\documentclass[a4paper, reqno, 11pt]{amsart}

\usepackage[left=2.5cm, right=2.5cm, top=3cm, bottom=3cm]{geometry}

\usepackage{amsaddr}
\usepackage{amssymb}
\usepackage{amsfonts}
\usepackage{mathrsfs}
\usepackage[T1]{fontenc}
\usepackage[USenglish]{babel}
\usepackage{varioref}
\usepackage{multirow}
\usepackage{array,multicol}
\usepackage{bigdelim}
\usepackage{bigstrut}
\usepackage{bm}
\usepackage{units}
\usepackage{bbold}
\usepackage{mathtools} 
\usepackage{extpfeil}
\usepackage{autobreak}
\usepackage{xparse}     
\usepackage{tabularray}
\usepackage{lscape}

\usepackage[font=small, labelfont=bf]{caption}
\usepackage[list=true, labelformat=brace, font=small, labelfont=bf]{subcaption}
\usepackage{graphicx}
\usepackage{csquotes}
\usepackage{longtable}
\usepackage{here}
\usepackage{color}

\usepackage[breaklinks=true
]{hyperref}
\hypersetup{pdftitle={F.Nill: The replacement number dynamics in SIR-type epidemic models I.}}

\usepackage[backend=biber,style=numeric]{biblatex}

\usepackage{enumitem}
\setlist[itemize]{align=parleft,left=0pt..1.5em}
\setlist{noitemsep}

\usepackage{url}

\def\customauthor{\empty}
\def\customdate{\empty}
\let\oldauthor\author
\renewcommand{\author}[1]{\def\customauthor{#1}}
\renewcommand{\date}[1]{\def\customdate{#1}}
\oldauthor{\customauthor\\\customdate}

\theoremstyle{definition}
\newtheorem{definition}{Definition}[section]

\theoremstyle{plain}
\newtheorem{theorem}[definition]{Theorem}
\newtheorem{lemma}[definition]{Lemma}
\newtheorem{corollary}[definition]{Corollary}
\newtheorem{proposition}[definition]{Proposition}

\theoremstyle{remark}
\newtheorem{remark}[definition]{Remark}

\numberwithin{equation}{section}

\NewDocumentCommand{\mybar}{ O{0.60} O{3pt} m }{
    \mathrlap{\hspace{#2}\overline{\scalebox{#1}[1]{\phantom{\ensuremath{#3}}}}}\ensuremath{#3}
}

\newcommand{\ncm}{\newcommand}
\ncm{\rncm}{\renewcommand}

\ncm{\lb}[1]{\label{#1}}

\rncm{\sec}{\setc{0}\section}
\ncm{\bsn}{\bigskip\noindent}
\ncm{\msn}{\medskip\noindent}
\ncm{\ssn}{\smallskip\noindent}
\ncm{\beq}{\begin{equation}}
\ncm{\beqnon}{\begin{equation*}}
\ncm{\eeq}{\end{equation}}
\ncm{\eeqnon}{\end{equation*}}
\ncm{\bea}{\begin{eqnarray}}
\ncm{\beanon}{\begin{eqnarray*}}
\ncm{\eea}{\end{eqnarray}}
\ncm{\eeanon}{\end{eqnarray*}}

\ncm{\ba}{\begin{array}}
\ncm{\ea}{\end{array}}
\ncm{\bpma}{\begin{pmatrix}}
\ncm{\epma}{\end{pmatrix}}
\ncm{\fns}{\footnotesize}

\DeclareMathOperator\sign{sign}
\DeclareMathOperator\reg{reg}

\DeclareMathOperator\Ad{Ad}

\DeclareMathOperator\diag{diag}

\DeclareMathOperator\age{age}

\DeclareMathOperator{\In}{In}
\DeclareMathOperator{\Out}{Out}
\DeclareMathOperator{\GL}{GL}

\ncm{\SSISS}{{\em SSISS }}
\ncm{\RND}{{\em RND }}

\ncm{\scenA}{\ensuremath{\mathrm{(A)}}}
\ncm{\scenB}{\ensuremath{\mathrm{(B)}}}
\ncm{\scenC}{\ensuremath{\mathrm{(C)}}}
\ncm{\scenD}{\ensuremath{\mathrm{(D)}}}
\ncm{\scenE}{\ensuremath{\mathrm{(E)}}}

\ncm{\scenAp}{\ensuremath{\mathrm{(A^+)}}}
\ncm{\scenBp}{\ensuremath{\mathrm{(B^+)}}}
\ncm{\scenCp}{\ensuremath{\mathrm{(C^+)}}}
\ncm{\scenDp}{\ensuremath{\mathrm{(D^+)}}}
\ncm{\scenEp}{\ensuremath{\mathrm{(E^+)}}}

\ncm{\scenAm}{\ensuremath{\mathrm{(A^-)}}}
\ncm{\scenBm}{\ensuremath{\mathrm{(B^-)}}}
\ncm{\scenCm}{\ensuremath{\mathrm{(C^-)}}}
\ncm{\scenDm}{\ensuremath{\mathrm{(D^-)}}}

\ncm{\scenXpm}{\ensuremath{\mathrm{(X_\pm)}}}

\newcommand{\RR}{\ensuremath{\mathbb{R}}}
\newcommand{\RRN}{\ensuremath{\mathbb{R}_{\geq 0}}}
\ncm{\NN}{\ensuremath{\mathbb{N}}}
\ncm{\ZZ}{\ensuremath{\mathbb{Z}}}
\ncm{\GG}{\ensuremath{\mathbb{G}}}
\rncm{\SS}{\ensuremath{\mathbb{S}}}
\ncm{\bone}{\ensuremath{\mathbb{1}}}

\newcommand{\bA}{ \mathbf{A}}
\newcommand{\bB}{ \mathbf{B}}
\newcommand{\bC}{ \mathbf{C}}
\newcommand{\bD}{ \mathbf{D}}
\newcommand{\bDb}{\bD_{\bbe}}
\newcommand{\bDelbe}{\mathbf{\Del}_{\bbe}}
\newcommand{\bE}{ \mathbf{E}}
\newcommand{\bF}{ \mathbf{F}}

\newcommand{\bJ}{ \mathbf{J}}

\newcommand{\bK}{ \mathbf{K}}

\newcommand{\bS}{ \mathbf{S}}
\newcommand{\bT}{ \mathbf{T}}
\newcommand{\bV}{ \mathbf{V}}
\newcommand{\bW}{ \mathbf{W}}

\newcommand{\bM}{ \mathbf{M}}
\newcommand{\bN}{ \mathbf{N}}
\newcommand{\bNb}{\bN_{\bbe}}
\newcommand{\bg}{ \mathbf{g}}

\newcommand{\bx}{ \mathbf{x}}
\newcommand{\by}{ \mathbf{y}}

\newcommand{\bp}{ \bm{p}}

\newcommand{\bbe}{ \bm{\beta}}
\ncm{\bemin}{\beta^{\min}}
\ncm{\bemax}{\beta^{\max}}
\ncm{\belo}{\beta^{L}}
\ncm{\behi}{\beta^{R}}
\ncm{\redN}{\textcolor{red}{\mathbf{N\!:\ }}}

\ncm{\Mbe}{\M_{\bbe}}
\ncm{\Mobe}{\E_{\bbe}}
\ncm{\Ebe}{\E_{\bbe}}
\ncm{\Fbe}{\F_{\bbe}}
\ncm{\Fbela}{\F_{\bbe,\la}}
\ncm{\sbe}{s_{\bbe}}
\ncm{\bmbe}{\bm{m}_{\bbe}}
\ncm{\mbe}{m_{\bbe}}
\ncm{\taube}{\tau_{\bbe}}
\ncm{\rhobe}{\bm{\rho}_{\bbe}}
\ncm{\bombe}{\bm{\om}_{\bbe}}
\ncm{\bOmbe}{\bm{\Om}_{\bbe}}
\ncm{\rhobei}{\rho_{\bbe,i}}
\ncm{\Tbe}{T_{\bbe}}
\ncm{\bTbe}{\bT_{\bbe}}
\ncm{\elbe}{\ell_{\bbe}}
\ncm{\ibe}{i_{\bbe}}
\ncm{\lbe}{\bm{l}_{\bbe}}
\ncm{\pibe}{\pi_{\bbe}}
\ncm{\cpibe}{\check{\pi}_{\bbe}}
\ncm{\psibe}{\psi_{\bbe}}
\ncm{\brbe}{\bra\bbe|}
\ncm{\vphibe}{\varphi_{\bbe}}
\ncm{\phibe}{\phi_{\bbe}}
\ncm{\Lbe}{L_{\bbe}}
\ncm{\bLabe}{\bLa_{\bbe}}
\newcommand{\bsi}{ \bm{\sigma}}
\ncm{\brsi}{\bra\bsi|}
\ncm{\Ps}{\P_{\bsi}}
\newcommand{\bga}{ \bm{\gamma}}

\newcommand{\bLa}{ \bm{\Lambda}}

\newcommand{\La}{ \Lambda}

\newcommand{\bDel}{ \bm{\Delta}}

\newcommand{\bom}{ \bm{\omega}}
\newcommand{\bOm}{ \bm{\Omega}}

\ncm{\pt}{\bp_\tau}

\newcommand{\bfe}{ \mathbf{e}}

\newcommand{\bzero}{\mathbf{0}}

\newcommand{\A}{\ensuremath{{\mathcal A}}}

\newcommand{\B}{\ensuremath{{\mathcal B}}}

\newcommand{\C}{\ensuremath{{\mathcal C}}}
\newcommand{\D}{\ensuremath{{\mathcal D}}}
\newcommand{\E}{\ensuremath{{\mathcal E}}}
\newcommand{\F}{\ensuremath{{\mathcal F}}}
\newcommand{\G}{\ensuremath{{\mathcal G}}}
\renewcommand{\H}{\ensuremath{{\mathcal H}}}
\newcommand{\K}{\ensuremath{{\mathcal K}}}

\newcommand{\X}{\ensuremath{{\mathcal X}}}

\newcommand{\U}{\ensuremath{{\mathcal U}}}
\newcommand{\V}{\ensuremath{{\mathcal V}}}
\newcommand{\T}{\ensuremath{{\mathcal T}}}

\newcommand{\Z}{\ensuremath{{\mathcal Z}}}
\rncm{\P}{\ensuremath{{\mathcal P}}}
\ncm{\PPe}{\P_\epsilon}

\rncm{\S}{\ensuremath{{\mathcal S}}}

\renewcommand{\L}{\ensuremath{{\mathcal L}}}
\ncm{\Labc}{\L_{a,b,c}}
\newcommand{\I}{\ensuremath{{\mathcal I}}}

\newcommand{\M}{\ensuremath{{\mathcal M}}}

\ncm{\Ins}{I_{\nu\sigma}}
\ncm{\hf}{h_f}
\ncm{\piP}{\pi_{\P}}
\ncm{\MB}{\M_{\B}}

\rncm{\O}{\mathcal{O}}

\ncm{\laseq}{{\em las}-equilibrium }
\ncm{\Qoo}{Q_{0,0}}
\ncm{\Qol}{Q_{0,1}}
\ncm{\VRND}{\mathcal{V}_{\rm RND}}
\ncm{\Bf}{\B_f}
\ncm{\eBf}{e\B_f}
\ncm{\Bfreg}{\B_f^{\reg}}
\ncm{\Me}{\M_\epsilon}

\ncm{\Pph}{\P_{\mathrm{phys}}}
\ncm{\Tph}{\T_{\mathrm{phys}}}
\ncm{\bTph}{\bar{\T}_{\mathrm{phys}}}
\ncm{\Iend}{\I_{\mathrm{end}}}
\ncm{\Ta}{\T_{\bm{a}}}
\ncm{\Aph}{\A_{\mathrm{phys}}}
\ncm{\Abio}{\A_{\mathrm{bio}}}
\ncm{\Abiop}{\A_{\mathrm{bio},+}}
\ncm{\Asoc}{\A_{\mathrm{soc}}}
\ncm{\Csoc}{\C_{\mathrm{soc}}}
\ncm{\Dsoc}{\D_{\mathrm{soc}}}
\ncm{\Asig}{\A_{\mathrm{split}}}
\ncm{\Csig}{\C_{\mathrm{split}}}
\ncm{\Dsig}{\D_{\mathrm{split}}}
\ncm{\Aone}{\A_{\mathrm{Model-1}}}
\ncm{\Atwo}{\A_{\mathrm{Model-2}}}
\ncm{\Asirs}{\A_{\mathrm{SIRS}}}
\ncm{\Asiso}{\A_{\mathrm{SIS_1}}}
\ncm{\Asist}{\A_{\mathrm{SIS_2}}}
\ncm{\Asisp}{\A_{\mathrm{SIS+}}}
\ncm{\Asism}{\A_{\mathrm{SIS-}}}
\ncm{\Asispm}{\A_{\mathrm{SIS\pm}}}
\ncm{\Dsis}{\D_{\mathrm{SIS}}}
\ncm{\Dsisp}{\D_{\mathrm{SIS+}}}
\ncm{\Dsism}{\D_{\mathrm{SIS-}}}
\ncm{\Dsispm}{\D_{\mathrm{SIS\pm}}}
\ncm{\Asisj}{\A_{\mathrm{SIS_j}}}
\ncm{\Aheth}{\A_{\mathrm{Heth}}}
\ncm{\Dheth}{\D_{\mathrm{Heth}}}
\ncm{\Kheth}{\K_{\mathrm{Heth}}}
\ncm{\Cph}{\C_{\mathrm{phys}}}
\ncm{\Cbio}{\C_{\mathrm{bio}}}
\ncm{\Cone}{\C_{\mathrm{Model-1}}}
\ncm{\Ctwo}{\C_{\mathrm{Model-2}}}
\ncm{\Csirs}{\C_{\mathrm{SIRS}}}
\ncm{\Dph}{\D_{\mathrm{phys}}}
\ncm{\bDph}{\bar{D}_{\mathrm{phys}}}
\ncm{\DBA}{\D_{AB}}
\ncm{\DAB}{\D_{AB}}
\ncm{\Dbio}{\D_{\mathrm{bio}}}
\ncm{\Dbionu}{\D_{{\mathrm{bio},\nu}}}
\ncm{\Dbioz}{\D_{{\mathrm{bio},0}}}
\ncm{\Done}{\D_{\mathrm{Model-1}}}
\ncm{\Dtwo}{\D_{\mathrm{Model-2}}}
\ncm{\DII}{\D_{\RN{2}}}
\ncm{\Dsirs}{\D_{\mathrm{SIRS}}}
\ncm{\Ksirs}{\K_{\mathrm{SIRS}}}
\ncm{\Kbio}{\K_{\mathrm{bio}}}
\ncm{\Kbionu}{\K_{{\mathrm{bio},\nu}}}
\ncm{\Lbio}{\L_{\mathrm{bio}}}
\ncm{\Gdil}{G_{\mathrm{dil}}}
\ncm{\GX}{G_{X}}
\ncm{\GI}{G_{I}}
\ncm{\GS}{G_{S}}
\ncm{\Gs}{G_{\sigma}}
\ncm{\Bs}{\B_{\sigma}}
\ncm{\tAph}{\tilde{\A}_{\mathrm{phys}}}
\ncm{\tAbio}{\tilde{\A}_{\mathrm{bio}}}
\ncm{\tAone}{\tilde{\A}_{\mathrm{Model-1}}}
\ncm{\tAtwo}{\tilde{\A}_{\mathrm{Model-2}}}
\ncm{\tAsirs}{\tilde{\A}_{\mathrm{SIRS}}}
\ncm{\TABC}{T(\bA,\bB,\bC)}
\ncm{\Tabc}{\TABC^{\leq 1}}

\ncm{\Vone}{\V^{(1)}}
\ncm{\Vtwo}{\V^{(2)}}
\ncm{\Vmodone}{\V^{\mathrm{Model\text{-}1}}}
\ncm{\Amodone}{\A^{\mathrm{Model\text{-}1}}}
\ncm{\Fmodone}{\F^{\mathrm{Model\text{-}1}}}
\ncm{\Fbemodone}{\Fbe^{\mathrm{Model\text{-}1}}}
\ncm{\Ebemodone}{\Ebe^{\mathrm{Model\text{-}1}}}
\ncm{\Fbeone}{\Fbe^{(1)}}
\ncm{\Ebeone}{\Ebe^{(1)}}
\ncm{\Fbetwo}{\Fbe^{(2)}}
\ncm{\Ebetwo}{\Ebe^{(2)}}
\ncm{\Ftwo}{\F^{(2)}}
\ncm{\Mtwo}{\M^{(2)}}
\ncm{\Vn}{\V^{(n)}}
\ncm{\Fben}{\Fbe^{(n)}}
\ncm{\Eben}{\Ebe^{(n)}}
\ncm{\Zone}{\Z_{1}}
\ncm{\Ztwo}{\Z_{2}}
\ncm{\Zthre}{\Z_{3}}
\ncm{\Zfour}{\Z_{4}}
\ncm{\Znull}{\Z_0}

\ncm{\Vmodtwo}{\V^{\mathrm{Model\text{-}2}}}
\ncm{\Fbemodtwo}{\Fbe^{\mathrm{Model\text{-}2}}}
\ncm{\Ebemodtwo}{\Ebe^{\mathrm{Model\text{-}2}}}

\ncm{\yph}{y_{\mathrm{phys}}}
\ncm{\Padm}{P_{\mathrm{adm}}}
\ncm{\Plow}{P_{\mathrm{low}}}
\ncm{\Phigh}{P_{\mathrm{high}}}
\ncm{\plow}{p_{\mathrm{low}}}
\ncm{\phigh}{p_{\mathrm{high}}}
\ncm{\Pc}{\P_{\mathrm{cut}}}
\ncm{\Reff}{X_{\mathrm{rep}}}
\ncm{\Reffo}{X_{\mathrm{rep,0}}^*}
\ncm{\Reffe}{X_{\mathrm{rep,end}}^*}
\ncm{\Ie}{I_{\mathrm{end}}^*}
\ncm{\Se}{I_{\mathrm{end}}^*}
\ncm{\dReff}{\dot{X}_{\mathrm{rep}}}
\ncm{\Tosc}{T_{\mathrm{osc}}}
\ncm{\Timm}{T_{\mathrm{imm}}}
\ncm{\Tinf}{T_{\mathrm{inf}}}
\ncm{\Thalf}{T_{\mathrm{half}}}

\ncm{\rvac}{r_{\mathrm{vac}}}

\newcommand{\al}{\alpha}
\newcommand{\la}{\lambda}
\newcommand{\be}{\beta}

\newcommand{\ga}{\gamma}

\newcommand{\Del}{\Delta}
\newcommand{\del}{\delta}
\newcommand{\si}{\sigma}
\newcommand{\ep}{\epsilon}
\newcommand{\vep}{\varepsilon}
\newcommand{\om}{\omega}
\newcommand{\Om}{\Omega}
\newcommand{\ka}{\kappa}

\newcommand{\vth}{\vartheta}

\ncm{\OP}{\Omega_{\PP,\ep}}
\ncm{\oG}{\omega_\G}
\ncm{\ome}{\omega_\ep}
\ncm{\phit}{\varphi_\tau}
\ncm{\p}{\psi}

\ncm{\Aal}{A_\alpha}
\ncm{\Bal}{B_\alpha}
\ncm{\sal}{\sigma_\alpha}

\rncm{\k}{\kappa}
\ncm{\an}{a_\nu}
\newcommand{\re}{{\,\hbox{$\textstyle\triangleright$}\,}}

\newcommand{\id}{{\rm id}}

\newcommand{\bra}{\langle}
\newcommand{\ket}{\rangle}
\def\End{\mbox{End}\,}

\def\Aut{\mbox{Aut}\,}

\def\cros{\,\raise1.9pt\hbox{$\scriptscriptstyle  > $}\!
          \raise1.5pt\hbox{$\scriptstyle\triangleleft$}\,}
\def\>cros{\cros}
\def\<cros{\,\raise1.5pt\hbox{$\scriptstyle\triangleright$}\!
           \raise1.9pt\hbox{$\scriptscriptstyle < $}\,}
\ncm{\veq}{{\scriptstyle\Vert}}

\ncm{\dR}{\partial_R}
\ncm{\dS}{\partial_S}
\ncm{\dI}{\partial_I}
\ncm{\dD}{\partial_D}
\ncm{\dN}{\partial_N}
\ncm{\dM}{\partial_M}
\ncm{\dX}{\partial_X}
\ncm{\dq}{\partial_q}
\ncm{\dx}{\partial_x}
\ncm{\dy}{\partial_y}
\ncm{\parH}{\partial H}
\ncm{\parHe}{\partial H_{\epsilon}}
\ncm{\parq}{\partial q}
\ncm{\parp}{\partial p}

\ncm{\rto}{\rightarrow}
\ncm{\mto}{\longmapsto}
\ncm{\lto}{\longrightarrow}
\ncm{\Lto}{\Longrightarrow}
\ncm{\lra}{\leftrightarrow}
\ncm{\LRA}{\Leftrightarrow}
\ncm{\LLRA}{\Longleftrightarrow}
\ncm{\LRa}{\Leftrightarrow}
\ncm{\LLRa}{\Longleftrightarrow}

\ncm{\tOP}{\tilde{\Omega}_{\PP,\ep}}
\ncm{\toe}{\tilde{\omega}_\ep}
\ncm{\tHe}{\tilde{H}_{\epsilon}}
\ncm{\tH}{\tilde{H}}
\ncm{\tV}{\tilde{V}}
\ncm{\tK}{\tilde{K}}
\ncm{\tOm}{\tilde{\Om}}
\ncm{\tA}{\tilde{\A}}
\ncm{\tM}{\tilde{\M}}
\ncm{\tI}{\tilde{\I}}
\ncm{\tP}{\tilde{\P}}
\ncm{\tbF}{\tilde{\bF}}
\ncm{\rt}{\tilde{r}_0}
\ncm{\tr}{\tilde{r}_0}
\ncm{\tga}{\tilde{\gamma}}
\ncm{\tal}{\tilde{\alpha}}
\ncm{\tGa}{\tilde{\Gamma}}
\ncm{\Nt}{\tilde{N}}
\ncm{\tHPe}{\tilde{H}_{\PP,\epsilon}}
\ncm{\tre}{\tilde{\rho}_{\epsilon}}
\ncm{\tq}{\tilde{q}}
\ncm{\tp}{\tilde{p}}
\ncm{\ta}{\tilde{a}}
\ncm{\tb}{\tilde{b}}
\ncm{\tbe}{\tilde{\beta}}
\ncm{\tc}{\tilde{c}}
\ncm{\td}{\tilde{d}}
\ncm{\tx}{\tilde{x}}
\ncm{\ty}{\tilde{y}}
\ncm{\tR}{\tilde{R}}
\ncm{\tep}{\tilde{\vep}}
\ncm{\etq}{e^{\tilde{q}}}
\ncm{\etp}{e^{\tilde{p}}}
\ncm{\ttau}{\tilde{\tau}}
\ncm{\trho}{\tilde{\rho}}
\ncm{\tthe}{\tilde{\theta}}

\ncm{\hH}{\hat{H}}
\ncm{\hV}{\hat{V}}
\ncm{\hK}{\hat{K}}
\ncm{\hKs}{\hat{K}_\sigma}
\ncm{\hHs}{\hat{H}_\sigma}
\ncm{\hvc}{\hat{v}_C}
\ncm{\hP}{\hat{\P}}
\ncm{\hA}{\hat{\A}}
\ncm{\hAph}{\hat{\A}_{\mathrm{phys}}}
\ncm{\hB}{\hat{\B}}
\ncm{\hC}{\hat{\C}}
\ncm{\hCph}{\hat{\C}_{\mathrm{phys}}}
\ncm{\hD}{\hat{\D}}
\ncm{\hphi}{\hat{\phi}}
\ncm{\crho}{\check{\rho}}
\ncm{\Reflat}{R_1^{\ \flat}}

\ncm{\ok}{\checkmark}
\ncm{\no}{\ding{55}}
\ncm{\s}{\mathsf{s}}
\ncm{\g}{\mathsf{g}}
\ncm{\h}{\mathsf{h}}
\ncm{\HIG}{H_\alpha}
\ncm{\Hal}{H_\alpha}
\ncm{\oIG}{\omega_\alpha}
\ncm{\HPLV}{H_{\mathrm {pLV}}}
\ncm{\omPLV}{\om_{\mathrm {pLV}}}
\ncm{\Hreg}{H^{\mathrm {reg}}}
\ncm{\omreg}{\om^{\mathrm {reg}}}
\ncm{\HPLVreg}{H_{\mathrm {pLV}}^{\mathrm {reg}}}
\ncm{\omPLVreg}{\omega_{\mathrm {pLV}}^{\mathrm{reg}}}
\ncm{\Halreg}{H_{\alpha}^{\mathrm {reg}}}
\ncm{\omalreg}{\omega_{\alpha}^{\mathrm {reg}}}
\ncm{\xnull}{x_{\mathrm {null}}}
\ncm{\Lt}{L_\tau}
\ncm{\Lmin}{L_{\min}}
\ncm{\dLt}{\dot{L}_\tau}
\ncm{\rv}{a_\mathrm{vac}}
\ncm{\dfe}{_\mathrm{dfe}}
\ncm{\en}{_\mathrm{end}}

\ncm{\DBo}{\D_{\Del B>0}}
\ncm{\DelB}{\Del B}

\newcommand{\Eqref}[1]{Eq. \eqref{#1}}

\ncm{\ulim}[1]{\underset{#1}{\lim}}
\ncm{\secref}[1]{Section \ref{#1}}
\ncm{\figref}[1]{Fig. \ref{#1}}

\newcommand{\0}{_{(0)}}

\newcommand{\minus}{\scalebox{0.75}[1.0]{$-$}}
\newcommand{\inv}{^{\minus 1}}
\ncm{\vsir}{V_{SIR}}
\ncm{\Vsir}{V_{SIR}}
\ncm{\hsir}{H_{SIR}}

\ncm{\zit}[1]{\autocite{#1}}
\ncm{\GHS}{\textsc{HGS }}
\ncm{\Upot}{$U\!$-potential}
\ncm{\QUpot}{quasi-\Upot}
\ncm{\UE}{{V^E}}
\ncm{\VE}{{V^E}}
\ncm{\alpm}{\upsilon_\pm}
\ncm{\alp}{\upsilon_+}
\ncm{\alm}{\upsilon_-}
\ncm{\vpm}{\upsilon_\pm}
\ncm{\xpm}{x_\pm}
\ncm{\qpm}{q_\pm}

\ncm{\vp}{\upsilon_+}
\ncm{\vm}{\upsilon_-}
\ncm{\vO}{v_{\O}}
\ncm{\vN}{v_N}
\ncm{\vt}{v_\tau}
\ncm{\ut}{u_\tau}
\ncm{\apm}{a_\pm}
\ncm{\bpm}{b_\pm}
\ncm{\upm}{u_\pm}
\ncm{\qp}{q_+}
\ncm{\qc}{q_c}
\ncm{\upl}{u_+}
\ncm{\xp}{x_+}
\ncm{\xc}{x_c}
\ncm{\epm}{\varepsilon_\pm}
\ncm{\fpm}{f_\pm}
\ncm{\Apm}{A_\pm}
\ncm{\Bpm}{B_\pm}
\ncm{\Dpm}{D_\pm}
\ncm{\Dp}{\vp}
\ncm{\Dc}{\Delta_c}

\ncm{\Spm}{S_\pm}
\ncm{\rSpm}{\rho S_\pm}
\ncm{\thpm}{\theta_\pm}

\ncm{\ntg}{\notag\\}
\ncm{\Ss}{S_{\textsl{sample}}}
\ncm{\Is}{I_{\textsl{sample}}}
\ncm{\Zs}{Z_{\textsl{sample}}}
\ncm{\Es}{E_{\textsl{sample}}}
\ncm{\Ns}{N_{\textsl{sample}}}
\ncm{\rhos}{\rho_{\textsl{sample}}}
\ncm{\gs}{\gamma_{\textsl{sample}}}
\ncm{\Zrge}{Z_{\rho,\gamma,E}}
\ncm{\Zmax}{Z_{\max}}
\ncm{\Qmax}{Q_{\max}}
\ncm{\el}{e^{\lambda}}
\ncm{\Ve}{V_{\epsilon}}
\ncm{\He}{H_{\epsilon}}

\ncm{\HPe}{H_{\PP,\epsilon}}
\ncm{\Emax}{E_{\max}}
\ncm{\rep}{\rho_{\epsilon}}
\ncm{\qe}{q_{\epsilon}}
\ncm{\expe}{\exp_{\epsilon}}
\ncm{\lne}{\ln_{\epsilon}}
\ncm{\Vpme}{V_{\pm,\ep}}
\ncm{\Vpe}{V_{+,\ep}}
\ncm{\Vme}{V_{-,\ep}}
\ncm{\qpme}{q_{\pm,\ep}}
\ncm{\qpe}{q_{+,\ep}}
\ncm{\qme}{q_{-,\ep}}
\ncm{\xpme}{x_{\pm,\ep}}
\ncm{\xpe}{x_{+,\ep}}
\ncm{\xme}{x_{-,\ep}}
\ncm{\qG}{q_\G}
\ncm{\pG}{p_\G}
\ncm{\ys}{y_2^*}
\ncm{\xs}{x_1^*}
\ncm{\vs}{v_2^*}
\ncm{\us}{u_1^*}
\ncm{\fl}{\varphi_\tau}
\ncm{\Gas}{\Gamma_\sigma}
\ncm{\tmp}{\age}
\ncm{\tImp}{\tau_{\I,\max}}

\begin{document}

\title[The replacement number dynamics I]
{The replacement number dynamics \\ in SIR-type epidemic models
\\
I: From SSISS to RND picture}
\author{Florian Nill}

\address{Department of Physics, Free University Berlin, Arnimallee 14, 14195 Berlin, Germany.}
\date{\small 2024-10-25}

\email{florian.nill@fu-berlin.de}

\begin{abstract}
In SIR-type epidemic models time derivative of prevalence $I$ can always be cast into the form 
$\dot{I}=(X-1)I$, where $X$ is the replacement number and recovery rate is normalized to one. Assuming 
$\dot{X}=f(X,I)$ for some smooth function $f$ defines a {\em replacement number dynamics} (RND). 
Choosing transmission coefficients $\be_1>\be_2$, any such system uniquely maps to an isomorphic {\em SSISS model}, i.e. an abstract SIR-type 3-compartment model.
Extending to negative values $\be_i<0$ takes care of demographic dynamics with compartment dependent birth and death rates. 
Fixing $f$ and varying $\be_i$ generates a family of isomorphic SSISS systems, the {\em SSISS fiber $\F(f)$}.
A symmetry group $\Gs$ acts freely and transitively on fibers 
$\F(f)$, so SSISS systems become a principal $\Gs$-fiber bundle over the space of RND systems.
Choosing two specific 6-parameter polynomials $f$ at most quadratic in $(X,I)$ covers a large class of models in the literature, 
with in total up to 17 (largely redundant!) parameters, including also reactive social behavior models.
Epidemiological admissibility conditions guarantee forward boundedness and absence of periodic solutions in all these models.

Part II of this work will prove existence and stability properties of endemic equilibria, which in RND picture simply boils down to analyzing zeros of the parabolas $f|_{I=0}$ and $f|_{X=1}$.
This will cover and extend well known results for a wide range of models by a unifying approach while also closing some open issues in the literature.

\end{abstract}

\maketitle

\footnotesize
\protect\begin{multicols}{2}
\tableofcontents
\protect\end{multicols}
\normalsize

\newpage

\section{Introduction}
\subsection{Motivation}
Building mathematical models to describe phenomena in natural sciences one typically encounters dynamic variables and external parameters. Within the model, values for external parameters are considered to be given from outside, like fundamental natural constants (speed of light $c$), parameters describing material or biological properties (spring constant 
$\kappa$, recovery rate $\ga$) or social behavior (contact rate $\be$). Naturally, reducing the number of essential parameters is always a goal to detect redundancies within parameter space and to simplify computations by unloading formulas. In the simplest case, a pure dimensional scale parameter may without loss be put equal to one by choosing dimensional units appropriately. 
More generally, a normalization program consists of finding appropriate coordinate transformations in combined 
parameter-and-dynamic-variable space such that the transformed system only depends on a maximally reduced subset of transformed parameters. 

Following this strategy, it has recently been shown by the author \cite{Nill_Redundancy} that constant per capita birth and death rates become redundant when looking at the dynamics of fractional compartment sizes $(S_1,S_2,I)$ in homogeneous%
\footnote{In this context ``homogeneous'' means the vectorfield driving the dynamics of absolute population sizes being a homogeneous function of order one.}
SIRS-type epidemic models with in general two susceptible and one infectious compartments and standard bilinear incidences 
$\be_iS_iI$. 
By shifting  model parameters appropriately, this holds for up to 9 demographic parameters including compartment dependent birth and death rates, vertical infection transmission, an $I$-dependent fraction of newborns being vaccinated and a resulting  time varying total population.

While standard positivity requirements remain untouched by these  shifts, the price to pay is that shifted transmission coefficients $\tbe_i$ may become negative. For example,
SI(R)S models with variable population and a non-susceptible $R$-compartment ($\be_2=0$) look like models with constant population and a {\em negative incidence rate 
$\tbe_2<0$}. 

This result leads to a unifying normalization prescription by always considering these models without vital dynamics and, instead, with two distinguished and possibly also 
negative incidence rates $\tbe_i\in\RR$. When normalized this way, seemingly different models in the literature become sub-cases of a common constant-population master model, whence isomorphic at coinciding shifted parameters. In this way various results in such models (like existence and stability of equilibrium points) could have been obtained from corresponding results in earlier papers. For convenience of the reader, Section \ref{Sec_SSISS-model} will give a short review on this matter.

\smallskip
In this work I will take the above results as starting point for a further normalization program. The basic idea is to find a suitable coordinate transformation best adapted to the general structure of such models. Like spherical coordinates for a particle moving in a spherically symmetric potential or 
``turning parameter space north'' to replace an 
external (say magnetic) force field $\bB=(B_1,B_2,B_3)$ by
$\tilde{\bB}=(0,0,|\bB|)$. 
Experience from these examples also motivates to surge for a symmetry group supporting these ideas. 

Before summarizing the plan of this paper, let me start with reviewing basic construction principles for 3-compartment SIR-type epidemic models. The time derivative of prevalence $I$ is given by the difference between ``In-flow'' and ``Out-flow'',
$$
\dot{I}=\In_I-\Out_I=(X-1)\Out_I.
$$
Speaking epidemiologically, $\In_I$ is the incidence flow, 
$\Out_I$ the recovery flow and $X:=\In_I/\Out_I$ defines the {\em replacement number}.
Standardly, one assumes the time of infectiousness to be distributed exponentially, with mean time 
$\tau_I = \ga\inv$, where $\ga>0$
is the recovery (better: removal) rate. This leads to the usual assumption $\Out_I=\ga I$, and subsequently to the more intuitive  description of  $X(t)$ as the expected number of secondary cases produced by a typical infectious individual during its mean time $\tau_I$
of infectiousness \autocite{Hethcote2000} (nowadays mostly called {\em effective reproduction number}). 

Next, denote $S_i$ the population densities in susceptible compartments $i$, $S_1+S_2+I=1$, and assume the standard incidence formula
$\In_I=\be_1S_1I+\be_2S_2I$, where $\be_i$ is the average number of effective contacts per unit of time (i.e. contacts leading to an
infection given the contacted was infectious) of a susceptible from $i$. 
This applies to diseases where the number of effective contacts per capita is independent of $N$. Also, $\be_i$ is assumed constant in time and independent of $S_i$. Thus we obtain
$$
\dot{I}=\ga(X-1)I,\qquad X:=(\be_1S_1+\be_2S_2)/\ga.
$$
Consequently, the nullclines of $I$ are given by 
$\{I=0\}\cup\{X=1\}$, which strongly suggests that $(X,I)$ should best be chosen as independent dynamic variables to study such systems. So, let us define the variable transformation 
$$
\vphibe: \bS=(S_1,S_2)\mapsto (X,I)=
(\tau_I\bra\bbe|\bS\ket, 1-S_1-S_2).
$$
Then, whatever the specific equation of motion for 
$\dot{S}_i$ might be, we will end up with a dynamical system of the form
\begin{equation}\begin{aligned}
\tau_I\dot{X}&=f(X,I),
\\ 
\tau_I\dot{I}&=(X-1)I,
\end{aligned}
\label{RND}
\end{equation}
for some function $f(X,I)$. Without loss, let us now put 
$\tau_I=\ga\inv=1$ and call a dynamical system \eqref{RND} on phase space $(X,I)\in\P:=\RR\times\RRN$ a {\em replacement number dynamics} (RND).

\subsection{Summary of results}
Given $f$ arbitrary smooth (for simplicity $f\in\C^\infty(\P)$), the natural question to ask is what is the abstract SIR-type epidemic model corresponding to the RND system \eqref{RND}? Conversely, given an epidemic model, provide a general formula determining $f$. Section \ref{Sec_replno} gives a complete answer. The abstract epidemic system always has the form
\begin{equation}\begin{aligned}
\dot{S}_i&=(\ga_i-\be_iS_i)I-\Om_i(\bS)S_i+\Om_j(\bS)S_j,\qquad
i\neq j\in\{1,2\},
\\
I&=1-S_1-S_2,
\end{aligned}
\label{SSISS}
\end{equation}
for some smooth functions $\Om_i$  and constants $\ga_i$ satisfying $\ga_1+\ga_2=\ga=1$. Let us call \eqref{SSISS} an abstract {\em SSISS dynamical system}. Note that the above description is redundant, i.e. for arbitrary scalar functions $K(\bS)$, replacing $\Om_i$ by $\Om_i+KS_j$, $j\neq i$, does not change the SSISS dynamics. Denote $[\bOm]$ the corresponding equivalence class, $\bF_{(\bbe,\bOm,\bga)}$ the vector field generating the SSISS dynamics \eqref{SSISS} and 
$\bV_f:=f\partial_X+(X-1)I\partial_I$ the RND vector field generating \eqref{RND}, where bold symbols are used for elements or maps taking values in $\RR^2$. In Section \ref{Sec_replno} we will provide explicit formulas to construct an affine bijection 
$
\phibe:([\bOm],\bga)\mapsto f
$
and its inverse, such that%
\footnote{Here, 
${\vphibe}_*\bF=D\vphibe\circ\bF\circ\vphibe\inv$ denotes the ``push-forward'' by $\vphibe$ on vector fields $\bF$.}
$$
{\vphibe}_*\bF_{(\bbe,\bOm,\bga)}=\bV_{\phibe([\bOm],\bga)}\,.
$$
Now, there are two major aspects to be observed. 

First, the RND vector field $\bV_f$ carries no information about 
$\bbe$. In other words, to reconstruct $\bS=\vphibe\inv(X,I)$ and $([\bOm],\bga)=\phibe\inv(f)$, one may pick 
$
\bbe\in\B:=\{(\be_1,\be_2)\in\RR^2\mid\be_1>\be_2\}
$
in principle arbitrarily%
\footnote{$\vphibe\inv$ and $\phibe\inv$ are ill defined for 
$\be_1=\be_2$. So, if necessary by permuting $1\leftrightarrow2$, we always assume $\be_1>\be_2$.}.
Thus, fixing $f$ and putting 
$\bF_{(\bbe,\bOm,\bga)}={\vphibe}_*\inv\bV_f$, 
we get a family $\F(f)$ of isomorphic SSISS systems - the
{\em SSISS fiber} of $f$ - paramatrized by 
$\bbe\in\B$ and mapping all to the same RND system $\bV_f$. Elements of this family are related by the group of 
linear coordinate transformations 
$$
\Gs:=\{\bg(\bbe',\bbe):=\varphi_{\bbe'}\inv\circ\vphibe\mid 
\bbe',\bbe\in\B\}.
$$ 
As shown in  Section \ref{Sec_Symmetry}, $\Gs$ consists of all $2\times2$-matrices with positive determinant acting from the left on column vectors $(S_1,S_2)^T$ and leaving $S_1+S_2$ invariant. In this way, the set of SSISS dynamical systems gets the structure of a principle $\Gs$-fiber bundle,
with fiber space $\B$ 
\footnote{Observe that $\Gs$ acts freely and transitively on 
$\B$ from the right.}
and base space given by the set of RND vector fields 
$\{\bV_f\}$.
This generalizes previous results for a class of 10-parameter SI(R)S models with $\be_2=0$ in \cite{Nill_SIRS}.%
\footnote{The scaling symmetry $G_S$ defined in  \cite{Nill_SIRS} is given by the subgroup of $\Gs$ leaving 
$\B\cap\{\be_2=0\}$ invariant.}

Second, for \Eqref{SSISS} to define an admissible epidemic system, we should at least require $\ga_i\geq0$ and 
$\Om_j|_{S_i=0}\geq0$, $j\neq i$. 
Clearly, this implies the {\em physical triangle} 
$\Tph:=\{\bS\in\RRN^2\mid S_1+S_2\leq1\}$ staying forward invariant. So, call $\bbe\in\B$ 
{\em compatible with $f$}, iff $\phibe\inv(f)$ is admissible in this sense, and
denote $\Bf\subset\B$ the subset of $f$-compatible $\bbe$. Then $f$ is called {\em admissible}, iff $\Bf\neq\emptyset$. 

Section \ref{Sec_quadratic} illustrates these notions for the special case of RND functions $f$ at most quadratic in $X$ (for short: {\em $X^2$-models}), where explicit formulas for $\Bf$ can be given. In SSISS picture, such models are specified by three functions $A_1(I)$, $A_2(I)$ and $B(I)$ such that the 
dynamics in \Eqref{SSISS} simplifies to
$$
\Om_i(\bS)=A_i(I)+B(I)\be_iS_j,\qquad i\neq j.
$$

Section \ref{Sec_R0} addresses the notion of a
{\em reproduction number} 
$R_0$ in the version of van den Driessche and Watmough \cite{Driesche_Watmough2000, Driesche_Watmough2002} within this formalism. 

Section \ref{Sec_periodic} uses methods from Busenberg-Driessche  \autocite{BusDries90} (see also \autocite{BusDries91, DerrickDriessche}), to show that admissible $X^2$-models do not have periodic solutions inside the physical triangle $\Tph$. By the Poincaré-Bendixson theorem this will help to prove stability properties of equilibrium points in Part II of this work \cite{Nill_SSISS_2}.

Section  \ref{Sec_examples-revisited} analyzes in detail two specific $X^2$-models. Model-1 is given by a 6-parameter polynomial $f=\sum_{i+j\leq 2}f_{ij}X^iI^j$, for which the
associated SSISS models $\phibe\inv(f)$ are specified by
8 parameters $(\al_i,\be_i,\ga_i,\theta_i,\delta)$ such that 
$$
A_i(I)=\al_i+\theta_iI, \qquad B(I)=\delta/(\be_1-\be_2).
$$ 
Putting $\theta_i=0$ and taking into account that all other parameters may also include demographic shifts, Model-1 covers a large class of prominent models in the literature with in total up to 14 parameters
\autocite{Hethcote1974, Hethcote1976, Hethcote1989,
BusDries90, DerrickDriessche, KorobWake, ORegan_et_al, Chauhan_et_al, Batistela_et_al, Had_Cast, KribsVel, LiMa2002, dOnofrio_et_al_2019, AvramAdenane2022, Nill_Omicron, Nill_SIRS}.
Here, $\del=\sign(1-f_{20})\,(\del_2-\del_1)$
where $\del_i=\nu_i-\mu_i$ is the birth-minus-death rate in compartment $i$. So, $\del\neq 0$ also covers models with differing mortality rates $\mu_1\neq\mu_2$, which in the above list had only been considered by \cite{BusDries90}%
\footnote{In \cite{BusDries90} only the case $\be_2=0$ is considered and \cite{AvramAdenane2022}  starts with 
$\mu_1\neq\mu_2$, but the bulk of results requires 
$\mu_1=\mu_2$.}.

If in addition $\theta_1>0$, this may be interpreted as an $I$-linear vaccination rate as in \cite{Nill_SIRS}, modeling the willingness of people to get vaccinated being dependent on published prevalence data. Following ideas of 
\autocite{dOnofrio_et_al_2007, dOnofrio_et_al_2008}, one may also let the portion of vaccinated newborns grow linearly with $I$, which leads to $\theta_1>0>\theta_2$.

Alternatively, putting 
$\al_1=0$, $\al_2=-\theta_2$ and $\theta_1\geq0\geq\theta_2$, we get a {\em social behavior model} (SBM), simulating people passing from $S_i$ to $S_j$ by changing their contact behavior depending on the values of $I$.

One may also consider more general {\em reactive vaccination models} (RVM), where the vaccination rate also contains a component
$\kappa XI$ being proportional to incidence. In the RND picture this leads to Model-2, which differs from Model-1 by the requirements $f_{20}=0$, $f_{21}=-\kappa<0$ and 
$f_{12}=-\kappa\be_2\leq0$.

The above examples motivate a stronger notion of so-called 
{\em $e$-admissibility}
(= epidemiologically admissible) in Section \ref{Sec_eadmissible}. Given $f$ in Model-1 or Model-2, then $\bbe\in\Bf$ is called 
{\em $e$-compatible} with $f$, iff
$\phibe\inv(f)$ satisfies $\theta_1\geq0\geq\theta_2$, and $f$
is called {\em $e$-admissible}, if such a $\bbe$ exists. Section 
\ref{Sec_eadmissible} gives a complete classification of $e$-admissible polynomials in Model-1 and Model-2.

Section \ref{Sec_conclusion} gives a summary of results and an outlook on part II of this work \cite{Nill_SSISS_2}, where existence and stability properties of equilibria in RND systems of type Model-1 or Model-2 are proven on the basis of admissibility and $e$-admissibility assumptions only, without relying on any SSISS model realization.

\subsection{Related literature} 
To give a very brief historical review on related literature, let me focus on deterministic and homogeneous SI(R)S-type 3-compartment models with standard bilinear incidence, which historically may be classified  along the two schemes

\begin{itemize}
\item[A)]
Constant vs. time-varying total population size $N$,
\item[B)]
Infection transmission only from $I$ to $S$ vs. also from $I$ to $R$ (in which case it makes sense to rename 
$S\equiv S_1$ and $R\equiv S_2$). 
\end{itemize}

Begging for understanding, the following overview is by far not exhaustive and concentrates on those papers being most related to the present work. 

\ssn
{\bf A)} Endemic models with constant population have first been constructed by adding a non-zero balanced birth and death rate to the classic SIR model of \autocite{KerMcKen}. As shown 
by \autocite{Hethcote1974} (see also
\autocite{Hethcote1976, Hethcote1989}), in this way already the simplest model without vaccination and loss of immunity shows a bifurcation from a stable disease-free equilibrium point (DFE) to a stable endemic scenario when raising the reproduction  number $R_0$ above one. Nowadays this is considered as Hethcote's {\em classic endemic model}.
Including linear vaccination and/or loss of immunity terms and optionally also considering recovery without immunity  one ends up with various types of constant population SI(R)S models without changing this picture, see for example \autocite{KorobWake, ORegan_et_al, Chauhan_et_al, Batistela_et_al, Nill_Omicron}. As shown in \cite{Nill_SIRS}, the true reason lies in the fact that due to symmetry concepts constant population SI(R)S models with up to 10 parameters all map to a normalized 2-parameter version of Hethcote's model.

Models with variable population $N(t)$ have mostly been studied under the assumption of constant per capita death rates, possibly an excess mortality in the $I$ and/or $R$ compartment and a constant (i.e. $N$-independent) birth flow landing in $S$. Plenty of examples can be found in the textbook by \cite{Martcheva}. Heuristically this may be justified by assuming that $N$ varies slowly on  characteristic epidemic time scales. But truly speaking, as already pointed out by \autocite{Mena-LorcaHeth}, this Ansatz rather models a constant immigration scenario. 

Instead, one may also consider per capita birth rates with a birth matrix $\nu_{ij}$ determining the fraction of newborns from compartment $i$ landing in compartment $j$. If birth rates and excess mortalities are constant, the dynamics of fractional compartment sizes is well known to stay independent of $N(t)$%
\footnote{This also holds for $N$-dependent death rates $\mu_i$, $i=1, 2, I$, as long as $\partial\mu_i/\partial N$ is independent of $i$, see e.g. \cite{LiMa2002, dOnofrio_et_al_2019} and Section \ref{Sec_birth+death}.}.
Apparently, this stream of models has been initiated by \autocite{BusDries90,BusDries91, DerrickDriessche, GaoHeth}%
\footnote{In \cite{GaoHeth} authors have also proposed an $N$-dependent birth rate to stabilize the total population size. In this case the dynamics of fractional compartment sizes also depends on $N$, see Section \ref{Sec_birth+death} and Remark 
\ref{Rem_N-dependence} for more details.}. 
A SIS-version  with varying population size has been analyzed by 
\autocite{LiMa2002}, a SIRS model with infection transmission also from outside in \autocite{Razvan} and the closed SIRS model without vaccination and varying population in
\cite{dOnofrio_et_al_2019}.
For generalizations to SEIR models see e.g.  \autocite{Greenhalgh,LiGraLiKa,SunHsieh,LuLu}.

\ssn
{\bf B)} 
A different approach to modeling partial and/or waning immunity consists of introducing a diminished incidence flow with rate 
$\be_R\equiv\be_2>0$ directly from $R\equiv S_2$ to $I$. This has presumably first been proposed in the so-called SIRI model of Derrick and van den Driessche \autocite{DerrickDriessche}. The authors also introduced an excess mortality $\Del\mu_I$ in compartment $I$, whence a time varying population size $N(t)$, but no vaccination nor immunity waning rates. 
For $R_0<1$, besides the expected locally asymptotically stable disease free equilibrium, they also found two endemic equilibria, one being a saddle and the other one also being  locally asymptotically stable.
Later Hadeler and Castillo-Chavez found the same phenomenon in their combined SIS/SIRS core group model with constant population, a constant vaccination rate and also two transmission rates $\be_i$ for $S\rto I$ and $R\rto I$ \autocite{Had_Cast}. 

Meanwhile, it is well known that models with infection incidents from several compartments may show a so-called {\em backward bifurcation} \autocite{Had_Dries}. This leads to coexistence of a bi-endemic scenario on top of the disease-free equilibrium, causing also hysteresis effects upon varying parameters. Apparently, a varying population size is not needed for this. In 
\autocite{KribsVel} the authors have improved and extended these results by adding also an immunity waning rate to the model of \autocite{Had_Dries}. More recently, in \autocite{AvramAdenane2022, AvramAdenane_et_al} the authors have given a thorough stability analysis of an eight parameter SIRS-type model by adding a varying population size to the model of \autocite{KribsVel} (apparently without being aware of that paper).

By now, applying the methods of \cite{Nill_Redundancy}, it becomes clear that the above classification schemes A) and B) become blurred. Instead,  it is more expedient to view all models as if they had constant population size and two distinguished and possibly also negative incidence rates 
$\tilde{\be}_i\in\RR$.

Closing this overview, I should also remark that one may  distinguish vaccinated and recovered people into separate compartments. This leads to 4-compartment models with backward bifurcation, see e.g.  \autocite{Arino_et_al, Yang_et_al}. 
Backward bifurcation has lately also been analyzed in SEIRS-type models for Covid-19 by considering two distinguished susceptible compartments. In \autocite{NadimChatto} the less susceptible compartment had been interpreted as an incomplete lockdown and in \autocite{Diagne_et_al} as an incomplete vaccination efficacy. 
Finally, backward bifurcation is also observed when considering $I$-dependent contact or recovery rates to model reactive behavior or infection treatment. However the list of papers on this topic over the last 20 years becomes too huge to be quoted at this place. 

\section{Homogeneous SIR-type models\label{Sec_SSISS-model} }
\subsection{The master model\label{the-model}}
\begin{figure}[ht!]
\centering
\includegraphics[width=0.4\textwidth]
	{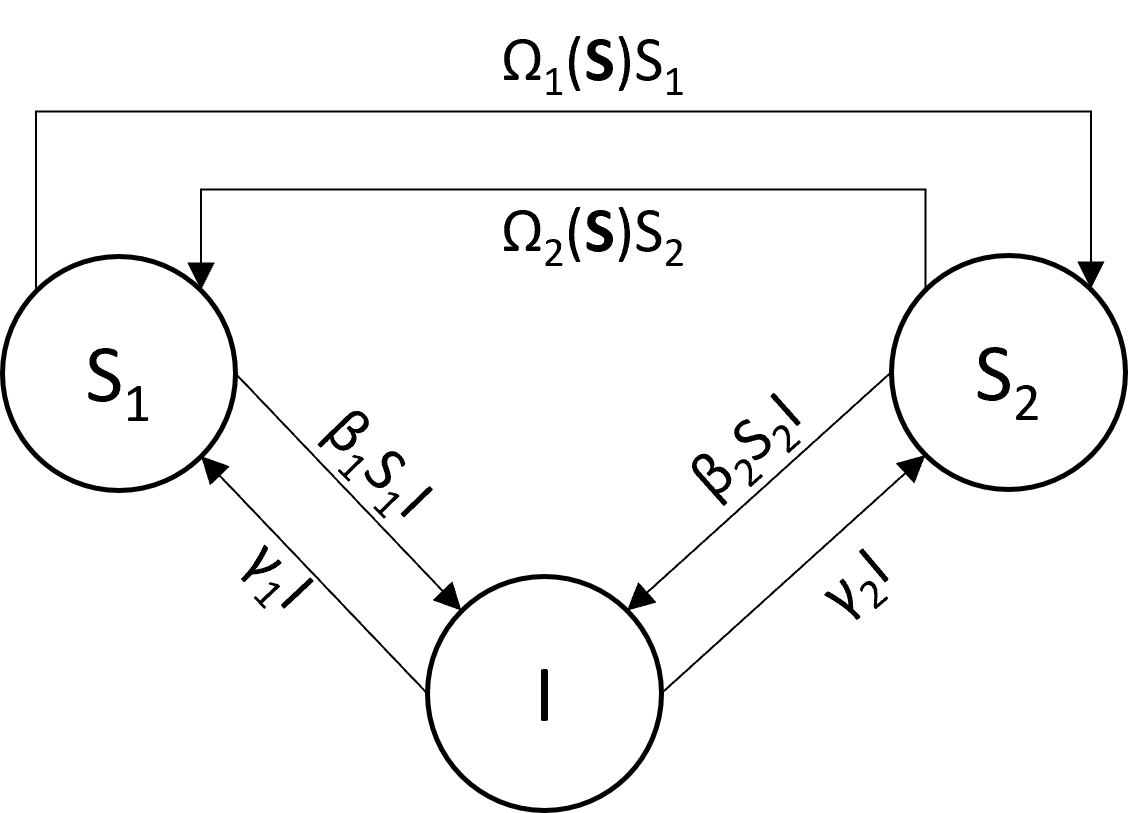}
	\caption{Flow diagram of the SSISS model.}
	\label{Fig_SSI-Flow}
\end{figure}

In this Section we define an abstract homogeneous SIR-type model consisting of three compartments with population densities
$(S_1,S_2,I)\in\RRN^3$ satisfying  $S_1+S_2+I=1$. Members of $I$ are infectious, members of $S_1$ are highly susceptible (socially active or not immune) and members of $S_2$ are less susceptible (partly immune or reducing contacts). Assuming standard incidence, the flow diagram between compartments is depicted in Fig. \ref{Fig_SSI-Flow}. 

Here, $\be_i$ denotes the number of effective contacts (i.e. contacts which would lead to an infection provided the contacted was infectious)
per unit time of a susceptible from $S_i$ and $\ga_i\geq 0$ denotes the recovery rates from $I\to S_i$, where we require 
$\ga:=\ga_1+\ga_2>0$.  The functions $\Om_i(\bS)$, 
$\bS:=(S_1,S_2)$, are assumed smooth (typically polynomial) and describe flows between $S_1$ and $S_2$ due to vaccination, loss of immunity and/or social behavior. Note that this description is redundant, since, for any scalar function $K(\bS)$, we may replace 
$(\Om_1,\Om_2)$  by $(\Om_1,\Om_2)+K(\bS)(S_2,S_1)$ without changing the dynamics. Figure \ref{Fig_SSI-Flow} leads to the dynamical system

\begin{align}
\begin{pmatrix}
\dot{S}_1\\ \dot{S}_2
\end{pmatrix}
&=\left[-I
\begin{pmatrix}
\be_1&0\\0&\be_2
\end{pmatrix}
+
\begin{pmatrix}
-\Om_1&\Om_2\\\Om_1&-\Om_2
\end{pmatrix}
\right]
\begin{pmatrix}
S_1\\ S_2
\end{pmatrix}
+I
\begin{pmatrix}
\ga_1 \\ \ga_2
\end{pmatrix}\,,
\label{dot_SS}
\\
\dot{I}&=\ga(X-1)I\,,\qquad\ga:=\ga_1+\ga_2>0\,,
\label{dot_I}
\end{align}
where dot denotes derivative w.r.t. time $t$. Here $X$ denotes the {\em replacement number} \autocite{Hethcote2000},  
\begin{equation}
X:=(\be_1S_1+\be_2S_2)/\ga.
\label{X}
\end{equation}

Note that putting $I:=1-S_1-S_2$, \Eqref{dot_I} becomes a consequence of \eqref{dot_SS}. In this way, consider \Eqref{dot_SS} as an abstract dynamical system on phase space 
\begin{equation}
\P_\si:=\{\bS\equiv(S_1,S_2)\in\RR^2\mid S_1+S_2\leq 1\},
\label{Psi}
\end{equation}
which is left invariant by the dynamics due to \eqref{dot_I}. Epidemiologically, we are restricted to initial conditions in the {\em physical triangle}
\begin{equation}
\Tph:=\RRN^2\cap\P_\si\,.
\label{phys_triangle}
\end{equation}
To guarantee $\Tph$ being forward invariant, we need 
$\dot{S}_i|_{S_i=0}\geq 0$. Since we want this to hold for all choices of recovery rates $\ga_i\geq 0$, we are naturally lead to impose the {\em admissibility constraints}
\begin{equation}
\ga_i\geq 0 \quad\land\quad\ga:=\ga_1+\ga_2>0
\quad\land\quad \Om_j|_{S_i=0}\geq 0,\ \ j\neq i.
\label{phys_constraint}
\end{equation}

Throughout,  the case 
$\be_1=\be_2$ will be ignored, since this makes notions like vaccination or loss of immunity obsolete. Mathematically, this case boils down to a pure SIS-model in the variables 
$(S:=S_1+S_2,I)$, which is explicitly solvable by separation of variables (see e.g. the textbook by \cite{Martcheva}). So, if necessary by permuting 
$1\leftrightarrow 2$, let's always assume $\be_1>\be_2$. 

\begin{remark}\label{Rem_variable_ga_i}
One might think of generalizing the system \eqref{dot_SS} by letting the splitting $\ga_i$ of recovered individuals between the two susceptible compartments also be a function of $\bS$ (or equivalently of $(X,I)$), while keeping $\ga=\ga_1+\ga_2$ constant. But this just amounts to a redefinition of 
$\Om_i$. To see this assume a decomposition
$$
\ga_1(\bS)=\ga_{10}+S_1\om_1(\bS)+S_2\om_2(\bS),\qquad
\ga_2(\bS)=\ga_{20}-S_1\om_1(\bS)-S_2\om_2(\bS),
$$
where $\ga_{10}+\ga_{20}=\ga$. Then putting $\ga_i=\ga_i(\bS)$ in \Eqref{dot_SS} would be equivalent to putting 
$\ga_i=\ga_{i0}$ and replacing 
$\Om_i$ by $\Om_i'$, where  
\begin{equation}
\Om_1'=\Om_1-I\om_1, \qquad \Om_2'=\Om_2+I\om_2.
\label{Om'}
\end{equation}
\end{remark}

\subsection{Redundancy of birth and death rates\label{Sec_birth+death}}
One may generalize the above model by including an up to 8-parameter demographic evolution dynamics with compartment dependent and constant per capita birth and death rates and time varying total population. 
Specifically, the following parameters are discussed in the literature.

\begin{table}[htbp]
\caption{Admissible demographic parameters}
\label{Tab_dem_par}
\begin{tabular}{lcp{0.75\textwidth}}
$\mu_i$ &:& 
		Mortality rate in $S_i$.
\\
$\mu_I$ &:& 
		Mortality rate in $I$.
\\
$\nu_i$ &:& 
		Rate of newborns from $S_i$. These are supposed 
		to be not infected.
\\
$\nu_I$ &:& 
		Rate of newborns from $I$.
\\
$\delta_i$ &:& $=\nu_i-\mu_i\,$.
\\

$\delta_I$ &:& $=\nu_I-\mu_I\,$.		
\\
$\delta$ &:& $=\delta_2-\delta_1\,$.
\\
$p_I$  &:& 
		Probability of a newborn from $I$ to be infected.
\\
$q_i$  &:& 
		Fraction of not infected newborns landing in
		$S_i$, $q_1+q_2=1$. For example, $q_2$ could be the fraction of not infected and vaccinated newborns.
\end{tabular}
\end{table}

Surprisingly, for a long time it had not been realized that in the dynamics of fractional variables all these demographic parameters become redundant.
Indeed, as has been shown just recently in \autocite{Nill_Redundancy} (see Appendix \ref{App_birth+death} for a brief review),  one still ends up with the dynamical system \eqref{dot_SS}-\eqref{dot_I}, with $(\be_i,\ga_i, \Om_i)$ replaced by 
$(\tilde{\be}_i,\tilde{\ga}_i, \tilde{\Om}_i)$, where
\begin{equation}
\begin{array}{rclrcl}
\tOm_1 &:=&\Om_1+q_2\nu_1+\delta_2S_2\,,\qquad\qquad\qquad &
\tOm_2 &:=&\Om_2+q_1\nu_2+\delta_1S_1\,, 
\\
\tga_1 &:=&\ga_1+q_1(1-p_I)\nu_I\,,& 
\tga_2 &:=&\ga_2+q_2(1-p_I)\nu_I\,,
\\
\tilde{\be}_1 &:=&\be_1+\delta_I-\delta_1\,,& 
\tilde{\be}_2 &:=&\be_2+\delta_I-\delta_2\,.
\end{array}
\label{tilde_parameters} 
\end{equation}
Birth and death rates being nonnegative, \Eqref{tilde_parameters} assures that if $(\Om_i,\ga_i)$ obey the admissibility constraints \eqref{phys_constraint}, then so do 
$(\tOm_i,\tga_i)$. On the other hand, for $\delta_I<\delta_i$, in particular when considering disease dependent excess mortality, the shifted transmission rates $\tbe_i$ may become negative. 
Also note that for $\del_1-\del_2>\be_1-\be_2$ we get 
$\tbe_1<\tbe_2$. In this case keeping the convention 
$\tbe_1>\tbe_2$ amounts to interchanging the compartments, 
$(\tilde{S}_1,\tilde{S}_2)=(S_2,S_1)$, which also implies that the formulas in \eqref{tilde_parameters} involve an additional permutation $1\leftrightarrow2$. \Eqref{tilde_parameters} now  motivates the following
\begin{definition}\label{Def_demogr_equiv}
SIR-type models \eqref{dot_SS} - \eqref{X} extended by birth and death rates as in \Eqref{SSISS+demogr} are called 
{\em $d$-equivalent} (demographically equivalent), if their tilde-parameters \eqref{tilde_parameters} coincide.
\end{definition}
Of course, along with the tilde system one also gets a time dependent total population $N(t)$,
$$
\dot{N}/N=\delta_1S_1+\delta_2S_2+\delta_II.
$$
So, unless fine tuned carefully, constant per capita birth and death rates in general imply the total population to approach $N(t)\to\infty$ or $N(t)\to 0$ exponentially as $t\to\infty$. 
However, since this happens on much larger time scales, it is still reasonable to analyze the decoupled system \eqref{dot_SS}
with tilde parameters in its own right.

Alternatively, as proposed by several authors, one may add a disease independent ($\equiv$ compartment independent) demographic correction term $d(N)$ to the constant per capita mortality rates $\mu_i$ and $\mu_I$. By choosing the function $d(N)$ appropriately one gets a finite limit population $N(t)\to N^*$ as $t\to\infty$. The nice thing to observe is that the derivation of \eqref{tilde_parameters} in Appendix \ref{App_birth+death} still holds for variable birth and death rates. Hence, any compartment independent correction term to the death rates drops out in tilde parameters, i.e. the dynamics for fractional variables still remains independent from $N(t)$. Examples are given in \cite{LiMa2002} and \cite{dOnofrio_et_al_2019}, see also Table \ref{Tab_examples}.

\begin{remark}\label{Rem_N-dependence}
Note that a similar correction term $b(N)$ to birth rates does show up in $\tOm_i$ and $\tga_i$, i.e. with this choice the tilde-system does couple to the $N(t)$-dynamics%
\footnote{This argument also applies to the popular method of choosing a constant (i.e. $N$-independent) number of newborns per unit of time, in which case $b(N)=b/N$.},
see e.g. \cite{GaoHeth, dOnofrio_et_al_2008}. The reason for this apparently unsymmetrical behavior is hidden behind our implicit assumption that the probability $q_i$ of an uninfected newborn to land in compartment $i$ does not depend on the compartment it stems from. If we drop this assumption and assume a diagonal birth matrix, then the demographic dynamics does indeed only depend on $\delta_i$ and $\delta_I$, see Appendix \ref{App_birth+death}.
\end{remark}

\subsection{Examples\label{Sec_examples}}
As the {\em standard example} consider the case where the functions $\Om_i$ are just constants, $\Om_i(\bS)=\al_i\geq 0$. Without demographic dynamics we have 
$\tbe_2=\be_2\geq0$,
$\al_1$ is interpreted as a vaccination rate and $\al_2$ as a loss-of-immunity rate. This example also includes the case of constant population and nonzero but balanced birth and death rates, $\delta_i=\delta_I=0$. In this case $\tbe_i=\be_i$, but 
positive birth rates $\nu_i$ and $\nu_I$ may also contribute to $\tal_i$ and $\tga_i$. So, this model is specified by 6 redundancy free tilde-parameters. If $\del_1=\del_2\neq0$, then possibly $\tbe_i<0$, but there are still just 6 tilde parameters, because effectively the dynamics only depends on
$\delta=\del_2-\del_1$. So, for $\tbe_2<0$ let's call this the {\em extended standard example}.

The {\em classic SIR model} is obtained from the standard example by keeping 
$\be_1>0$, $\ga_2>0$ and putting all other parameters to zero (in which case $S_1\equiv S$ and $S_2\equiv R$). If starting from SIR we turn on compartment independent birth and death rates $\nu=\mu>0$ and 
$q_1=1$, we get Hethcote's {\em classic endemic model}, which in tilde parameters differs from SIR only by $\tga_1=\tal_2=\nu>0$ \cite{Hethcote1974, Hethcote1976, Hethcote1989}, see also table \ref{Tab_examples} below.
So, with tilde parameters Hethcote's model looks like a 
{\em standard SIRS model}, 
which usually is defined by $\be_2=0$ and $\al_2>0$ (optionally also with vaccination and/or common birth=death rates, implying 
$\tal_1>0$ and split recovery rates $\tga_i>0$) \cite{KorobWake, ORegan_et_al, Nill_Omicron}. In fact, using tilde parameters and additional symmetry concepts, 
SIRS-type models with up to 10 parameters and constant population have recently been shown by the author to be all isomorphic to Hethcotes classic endemic model \cite{Nill_SIRS}. 

Finally, for $\be_2>0$ the  standard example has been used to demonstrate {\em backward bifurcation}, i.e. the appearance of bi-endemic scenarios at reproduction number $R_0<1$ \cite{Had_Cast, Had_Dries, KribsVel}.

Following and extending a proposal of \autocite{Nill_SIRS, Nill_Symm1}, the standard example may be extended to a {\em reactive vaccination model} (RVM) by letting the vaccination rate depend on prevalence $I$ and/or incidence 
$I(\be_1S_1+\be_2S_2)$.  This simulates the willingness of society to get vaccinated being influenced by published epidemic data. Hence we are lead to an Ansatz
\begin{equation}
\Om_1=\al_1+\theta_1 I+\kappa IX,\qquad \Om_2=\al_2,\qquad
\text{(RVM-Model)}
\label{RVM}
\end{equation}
with non-negative constants $\al_i,\theta_1,\kappa$.
For $\be_2=\kappa=0$ this becomes a SIRS model furnished with an $I$-linear vaccination rate \cite{Nill_SIRS}. 

In this context one should also quote  \autocite{dOnofrio_et_al_2007, dOnofrio_et_al_2008}, where the authors
use an information variable $M$ to 
model how information on current and past states of the disease influences decisions in families whether to vaccinate or not their children. In our formalism, reactive newborn vaccination could be modeled by letting $q_i$ be $I$-dependent, where $q_2$ should be increasing with $I$. Hence
$$
\begin{aligned}
q_1(I)	&=q_1(0)-I\Del q=q_1(1)+(1-I)\Del q,
\\
q_2(I)	&=q_2(0)+I\Del q=q_2(1)-(1-I)\Del q,
\end{aligned}
$$
with $q_1(I)+q_2(I)=1$, $q_i(I)\geq 0$ for all $I\in[0,1]$ and
$\Del q>0$. Putting $\om=\Del q(1-p_I)\nu_I\geq 0$, 
$1-I=S_1+S_2$ and passing to tilde parameters 
we get $\tbe_i$ as in \Eqref{tilde_parameters} and
\begin{equation}
\begin{array}{rclrcl}
\tOm_1 &:=&\Om_1+q_2(0)\nu_1+I\vth_1+\delta_2S_2\,,\qquad\qquad &
\tOm_2 &:=&\Om_2+q_1(0)\nu_2+I\vth_2+\delta_1S_1\,, 
\\
\vth_1	&:=&\nu_1\Del q-\om\,,&
\vth_2	&:=&-\nu_2\Del q +\om\,,
\\
\tga_1 &:=&\ga_1+q_1(1)(1-p_I)\nu_I\,,& 
\tga_2 &:=&\ga_2+q_2(1)(1-p_I)\nu_I\,.
\end{array}
\label{tilde+nu_vacc}
\end{equation}
Here we have used \Eqref{Om'} with 
$\om_1=\om_2=\om$. Note that one typically expects 
$\nu_i\geq\nu_I$ and therefore 
$\vth_1\geq0\geq\vth_2$. To see that the admissibility constraints \eqref{phys_constraint} are  still preserved for 
$\tOm_2$ observe 
$q_1(0)\nu_2+\vth_2=q_1(1)\nu_2+\om\geq0$. So adding $\Del q$ to Table \ref{Tab_dem_par} the number of redundant demographic parameters now increases from 8 to 9.

Finally, as a pure {\em social behavior model} (SBM) consider
\cite{Nill_Symm1}
\begin{equation}
\Om_1=\theta_1 I,\qquad \Om_2=\theta_2(I-1),\qquad
\theta_1\geq 0\geq\theta_2\qquad\text{(SBM-Model)}.
\label{SBM}
\end{equation} 
In this model, $S_1$ is interpreted as the ``freedom fraction'' = ``denier'' or ``careless'' or ``uneducated'' compartment, 
i.e. socially active susceptibles with a high number of contacts, and $S_2$ as the ``lock-down'' fraction = ``believer''
or ``careful'' or ``educated'' compartment, i.e. susceptibles with a reduced number of contacts. 
Driven by published epidemic data, in this model people change their behavior from $S_1$ to $S_2$ with rate 
$\theta_1I\geq 0$ and from $S_2$ to $S_1$ with rate 
$\theta_2(I-1)\geq 0$. 
Also, recovered individuals will have ``learned their lesson'' and stay in $S_2$, so $\ga_1=0$.

\begin{remark}
There are more elaborated social behavior models in the literature assuming contact rates being dependent on (the history of)  prevalence or incidence, see e.g. \cite{dOnofrio_et_al_2022} and references therein. In our approach we always restrict to $\be_i$ being constant.
\end{remark}

Let me now demonstrate in more detail how the use of tilde parameters covers a lot of seemingly different models in the literature by a unifying normalization prescription.  The following list is ordered chronologically (oldest first) and an updated version from  \autocite{Nill_Redundancy}. All models assume the rate of newborns to be compartment independent, 
$\nu_1=\nu_2=\nu_I=\nu$, and therefore $\delta_i-\delta_k=\mu_k-\mu_i$, $i,k\in\{1,2,I\}$.

\begin{longtable}{@{}lp{0.9\textwidth}@{}}
Heth &
Hethcotes classic endemic model 
\autocite{Hethcote1974, Hethcote1976, Hethcote1989} by putting $\nu=\mu_i=\mu_I>0$, $q_1=1$, $\beta_1>0$, 
$\ga_2>0$ and all other parameters vanishing.
\\
BuDr	&
The 7-parameter SIRS model with time varying population size in \autocite{BusDries90}, adding to Hethcote's model  an immunity waning rate $\al_2$ and allowing non-balancing mortality and birth rates $\nu\neq\mu_i\neq\mu_I$.
\\
SIRI	&
The 6-parameter SIRI model of \autocite{DerrickDriessche}, replacing the immunity waning rate $\al_2$ in \autocite{BusDries90} by the transmission rate $\be_2>0$ and also requiring $\mu_1=\mu_2$.
\\
SIRS	&
The 8-parameter constant population SI(R)S (i.e. mixed SIRS/SIS) model with two recovery flows $I\rightarrow S_1$ and 
$I\rightarrow S_2$. Hence 
$\nu=\mu_i=\mu_I$ and $\beta_2=0$.
\\
HaCa	&
The 6-parameter core system in \autocite{Had_Cast}, with  transmission and recovery rates $\be_i, \ga_i>0$, a vaccination term $\al_1>0$ and a constant population with balanced birth and death rates, $\nu=\mu_i=\mu_I>0$ and $q_1=1$. 
\\
KZVH	&
The 7-parameter vaccination models with constant popultauin of  \autocite{KribsVel}, adding an immunity waning rate $\al_2>0$ to the model of HaCa.
\\
LiMa	&
The 8-parameter SIS-model with vaccination and varying population size of \autocite{LiMa2002} keeping only 
$\ga_2=\be_2=0$ and assuming $\mu_1=\mu_2=d(N)$ for some suitable demographic function $d(N)$ and
$\mu_I=d(N)+\Del\mu_I$ with constant excess mortality 
$\Del\mu_I$.
\\
d'OmS	&
The 7-parameter SIRS-model without vaccination and with varying population size of \cite{dOnofrio_et_al_2019}, keeping 
$\al_1=\ga_1=\be_2=0$, a nonzero vertical infection probability $p_I$ and the same assumption on mortality rates as in LiMa above.
\\
AABH	&
The 8-parameter SIRS-type model analyzed by \autocite{AvramAdenane2022}, keeping only 
$\ga_1=q_2=0$ 
and  all other parameters positive.
The authors allow a varying population size by first discussing  the general case of all mortality rates being different and then concentrate on $\mu_1=\mu_2\neq \nu$ and excess mortality 
$\Del\mu_I>0$. 
\end{longtable}

\addtocounter{table}{-1} 
\begin{table}[htbp!]
\caption{\autocite{Nill_Redundancy} Mapping models in the literature
to the present choice of parameters.  The column $\#$ counts the number of free parameters in the original models.
Passing to fractional variables and tilde parameters, \Eqref{tilde_parameters},
$\#_{\mathrm{eff}}$ counts the number of effectively independent parameters.
}\label{Tab_examples}
$$
\begin{array}{l|cccccccccccc|c}
&\al_1&\al_2&\be_1&\be_2&\ga_1&\ga_2&
\nu&\mu_1&\mu_2&\mu_I&q_1&q_2&\#\\
\hline
\mathrm{Heth}&0&0&\ok&0&0&\ok&
\multicolumn{4}{c}{\nu=\mu_1=\mu_2=\mu_I}&1&0&3
\\\hline
\mathrm{BuDr}&0&\ok&\ok&0&0&\ok&\ok& \multicolumn{2}{l}{\mu_1\leq\mu_2}&\ok&1&0&7
\\\hline
\mathrm{SIRI}\footnotemark\addtocounter{footnote}{-1}
&0&0&\ok&\ok&0&\ok&\ok&
\multicolumn{2}{l}{\mu_1=\mu_2}&\mu_i+\Del\mu_I&1&0&6
\\\hline
\mathrm{SIRS}&\ok&\ok&\ok&0&\ok&\ok&
\multicolumn{4}{c}{\nu=\mu_1=\mu_2=\mu_I}&\ok&\ok&7
\\\hline
\mathrm{HaCa}&\ok&0&\ok&\ok&\ok&\ok&
\multicolumn{4}{c}{\nu=\mu_1=\mu_2=\mu_I}&1&0&6
\\\hline
\mathrm{KZVH}&\ok&\ok&\ok&\ok&\ok&\ok&
\multicolumn{4}{c}{\nu=\mu_1=\mu_2=\mu_I}&1&0&7
\\\hline
\mathrm{LiMa}&\ok&\ok&\ok&0&\ok&0& \ok&\multicolumn{2}{l}{\mu_i=d(N)}&\mu_i+\Del\mu_I&\ok&\ok&8
\\\hline
\mathrm{d'OMS}&0&\ok&\ok&0&0&\ok& \ok&\multicolumn{2}{l}{\mu_i=d(N)}&\mu_i+\Del\mu_I&1&0&7
\\\hline
\mathrm{AABH}\footnotemark\addtocounter{footnote}{1}
&\ok&\ok&\ok&\ok&0&\ok&\ok&\multicolumn{2}{l}{\mu_1=\mu_2\footnotemark}&\mu_i+\Del\mu_I&1&0&6
\addtocounter{footnote}{-3}
\\\hline
\end{array}
$$
\bigskip\bigskip
$$
\begin{array}{l|ccccccc|l}
&\tal_1&\tal_2&\tbe_1&\tbe_2&\tga_1&\tga_2&\delta&
\#_{\mathrm{eff}}
\\
\hline
\mathrm{Heth}&0&=\tga_1&=\be_1&0&=\tal_2&>0&0&3
\\\hline
\mathrm{BuDr}&0&\ok&\ok&<0&\ok&>0&\leq0&6 
\\\hline
\mathrm{SIRI}\footnotemark\addtocounter{footnote}{-1}
&0&=\tga_1&\ok&\ok&=\tal_2&>0&0&4
\\\hline
\mathrm{SIRS}&\ok&\ok&=\be_1&0&\ok&\ok&0&5
\\\hline
\mathrm{HaCa}&\ok&\ok&=\be_1&=\be_2&\ok&\ok&0&6
\\\hline
\mathrm{KZVH}&\ok&\ok&=\be_1&=\be_2&\ok&\ok&0&6
\\\hline
\mathrm{LiMa}&\ok&\ok&\ok&<0&>0&\ok&0&6
\\\hline
\mathrm{d'OMS}&0&\ok&\ok&<0&\ok&>0&0&5
\\\hline
\mathrm{AABH}\footnotemark
&\ok&\ok&\ok&\ok&>0&\ok&0&6
\\\hline
\end{array}
$$
\end{table}
\footnotetext{
${\mathrm{SIRI}}$ and ${\mathrm{AABH}}$ come in two versions with $1\leftrightarrow 2$ permuted, since the authors also allow $\be_1<\be_2$.
} 

Table \ref{Tab_examples} maps these examples to the present set of parameters and shows the number of effectively independent tilde parameters \eqref{tilde_parameters}. In summary, by looking at the ranges of tilde parameters in each model%
\footnote{Possibly disregarding boundary configurations 
$\tga_i=0$}, 
the following conclusions may be drawn
\cite{Nill_Redundancy}.

\begin{corollary} 

\begin{itemize}
\item[i)]
The variable population model of AABH
\autocite{AvramAdenane2022} is $d$-equivalent to the extended standard example%
\footnote{\label{FN_AABH} The bulk of results in Section 5 and 6 of AABH \autocite{AvramAdenane2022} assumes $\mu_1=\mu_2$, whence $\delta=0$.} and modulo $d$-equivalence includes all other models of Table \ref{Tab_examples}
$($BuDr only for $\mu_1=\mu_2)$.

\item[ii)]
The subcase $\min\{\be_1,\be_2\}<\Del\mu_I$ of AABH is 
$d$-equivalent to the SIS model of LiMa \autocite{LiMa2002}.

\item[iii)]
The subcase $\min\{\be_1,\be_2\}>\Del\mu_I$ of AABH  
is $d$-equivalent to the constant population model of KZVH \autocite{KribsVel}.

\item[iv)]
The subcase $\min\{\be_1,\be_2\} =\Del\mu_I$ of AABH is 
$d$-equivalent to the classic SIRS model.

\item[v)]
The models of HaCa \autocite{Had_Cast} and KZVH  \autocite{KribsVel} are $d$-equivalent.

\item[vi)]
The model of d'OMS \cite{dOnofrio_et_al_2019} is $d$-equivalent to the intersection in tilde-parameter space
of LiMa \autocite{LiMa2002} and BuDr \autocite{BusDries90}.
%

\end{itemize}
\end{corollary}
Apparently, none of these papers had been aware of the redundancy formulas \eqref{tilde_parameters}, which explains why the above relations have not been realized before and authors have struggled to prove equivalent results without noticing.
\section{The replacement number dynamics \label{Sec_replno}}
Unless noted explicitly, from now on we will always consider the demographically extended model with tilde-parameters \eqref{tilde_parameters} or \eqref{tilde+nu_vacc}. Yet, to unload the notation, the tilde above parameters will be dropped unless noted explicitly. So, put
$$
\B:=\{\bbe\equiv(\be_1,\be_2)\in\RR^2\mid\be_1>\be_2\}.
$$
Also, from now on let's choose time scale to be measured in units of $\ga\inv$, i.e. without loss put $\ga\equiv\ga_1+\ga_2=1$ (think of all rates of dimension $[time]\inv$ being divided by $\ga$). Recalling the admissibility constraints \eqref{phys_constraint} we then define
\begin{definition}\label{Def_SSISS_admissible}
The dynamical system \eqref{dot_SS} on phase space $\Ps$ 
\eqref{Psi} is called {\em admissible}, if 
$$
(\be_1,\be_2)\in\B,\qquad\ga_i\geq0,\qquad\ga_1+\ga_2=1,\qquad
\Om_j|_{S_i=0}\geq 0,\ \ j\neq i.
$$
\end{definition}

The basic idea of this work is to change dynamic variables
from $(S_1,S_2)$ to $(X,I)$. The motivation for this is given by \Eqref{dot_I}, which then simply becomes $\dot{I}=(X-1)I$. Since there must be a function $f(X,I)$ such that 
$\dot{X}=f$, in these variables the set of equilibrium points (EPs) of the dynamics is given by the zeros of $f|_{I=0}$ and $f|_{X=1}$. So, once we we have established the formulas to obtain $f$ from 
$(\be_i,\ga_i,\Om_i)$ and vice versa, the analysis of EPs and their stability properties will become substantially simplified.
This program will be pursued in part II of this work 
\autocite{Nill_SSISS_2}.

In this Section we will see, that expressing the abstract model \eqref{dot_SS}-\eqref{dot_I} in terms of variables $(X,I)$ should be considered as canonical also from a purely mathematical point of view. We start with providing a coordinate free formulation of the dynamics in Subsection \ref{Sec_SSISS-picture}. Expressing the equations of motion with respect to a so-called {\em compartment basis} one ends up with \Eqref{dot_SS}. This will be called the {\em SSISS picture}. 
In Subsection \ref{Sec_RND-picture} we change to variables 
$(X,I)$ to obtain an isomorphic representation of the dynamicas, called the {\em RND picture} (RND $=$ Replacement Number Dynamics). Subsection \ref{Sec_Symmetry} defines a symmetry group $\Gs$ acting freely and transitively on the space of compartment basises, such that basis dependent SSISS models map to the same RND model if and only if they are $\Gs$-equivalent. Subsection \ref{Sec_coordinates} provides explicit coordinate formulas for transforming back and forth between SSISS and RND picture.

\subsection{The SSISS picture\label{Sec_SSISS-picture}}
Denote $\bS\in\S\equiv\RR^2$ as column or ``ket'' vectors, $\S^*$ the dual of $\S$ as row or ``bra'' vectors and 
$\bra\cdot|\cdot\ket:\S^*\otimes\S\rto\RR$ the dual pairing.
Define $\bsi\equiv\brsi\in\S^*$ and $\bbe\equiv\brbe\in\S^*$ by
\begin{equation}
\brsi\bS\ket:=S_1+S_2,\qquad
\brbe\bS\ket:=\be_1S_1+\be_2S_2\equiv X.
\label{brasi+brabe}
\end{equation}
Then $\Ps:=\{\bS\in\S\mid \brsi\bS\ket\leq 1\}$ and 
$I(\bS):=1-\brsi\bS\ket$.
In what follows, $0\neq\bsi\in\S^*$ will be considered fixed arbitrarily while $\bbe\in\S^*$ is assumed varying. So, to unload the notation, the dependence on 
$\bsi$ will mostly be suppressed\footnote{By applying 
$\bg\in\Aut\S$ one may transform to any other nonzero 
$\bsi'\in\S^*$.}.
Given $0\neq\bsi\in\S^*$ put 
\begin{equation}
\B\equiv\Bs:=\{\bbe\in\S^*\mid (\bbe,\bsi) 
\text{ is an oriented basis in }\S^*\}.
\label{Bs}
\end{equation}
Now consider the set of matrix valued $C^\infty$-functions on $\Ps$ of the form
\begin{equation}
\M:= \{\bM:\Ps\to\End\S\mid\exists\bbe\in\S^*:
\bsi\bM = -I\bbe\}.
\label{M}
\end{equation}
Using the natural projection 
$$
\pi:\M\longrightarrow\S^*,\qquad
\bra\pi(\bM)|\bS\ket:=-I(\bS)\inv\brsi\bM(\bS)\bS\ket
$$
denote $\Mbe:=\pi\inv(\bbe)$ and $\MB:=\pi\inv(\B)$.
\begin{definition}\label{Def_SSISS}
By an abstract 
{\em SSISS model} $(\bM,\bga)$ we mean a dynamical system on $\Ps$ of the form 
\begin{equation}
\begin{aligned}
\dot{\bS}=\bF_{(\bM,\bga)}(\bS):=
\bM(\bS)\bS+I(\bS)\bga,
\\
\bM\in\MB,\quad\bga\in\S,\quad\brsi\bga\ket=1.
\end{aligned}
\label{dot-SSISS}
\end{equation}
\end{definition}
\bsn
Note that the definition \eqref{dot-SSISS} implies 
$\bF_{(\bM,\bga)}=\bF_{(\bM+\bK,\bga+\bm{k})}$ for all
$(\bK,\bm{k})\in\K$, where
\begin{equation}
\begin{aligned}
\K&:=\{(\bK,\bm{k})\in\M_0\times\ker\bsi\mid
\bK(\bS)\bS+I(\bS)\bm{k}=\bzero,\,\forall\bS\in\S\}=
\K_0\times\{\bzero\},
\\
\K_0&:=\{\bK\in\M_0\mid\bK(\bS)\bS=\bzero,\,\forall\bS\in\S\}
\end{aligned}
\label{K}
\end{equation}
So, adding $\bK\in\K_0$ to $\bM$ doesn't influence the dynamics. By convenient physics terminology let me call this a {\em ``gauge transformation''}. Hence, for fixed $\bbe\in\B$, the space of SSISS vector fields may be identified as 
$$
\Fbe:=\{\bF_{(\bM,\bga)}\mid\bM\in\Mbe\land \brsi\bga\ket=1\}=\Mbe/\K_0\times\bsi\inv(1).
$$
Given a SSISS model we call 
$\bfe_i\in\S$ a {\em compartment basis}, if w.r.t. this basis $\bsi=(1,1)$ and 
$\bbe=(\be_1,\be_2),\ \be_1>\be_2$.\footnote{One might more precisely call this an oriented compartment basis.} 
Given a compartment basis put 
\begin{equation}
\bDb:=-I\diag(\bbe)\equiv
-I(|\bfe_1\ket\be_1\bra\bfe_1| + |\bfe_2\ket\be_2\bra\bfe_2|).
\label{D}
\end{equation}
Then $\bDb\in\Mbe$ and putting
$\bM_\E:=\bM-\bD_{\pi(\bM)}$ we get $\bM_\E\in\E$ and
$\M\cong\S^*\oplus\E$, where
\begin{equation}
\E\equiv\M_0=\{\bM\in\M\mid\bsi\bM=0\}.
\label{M0}
\end{equation}
Since in a compartment basis any
$\bM_\E\in\E$ has the form 
\begin{equation}
\bM_\E=\bE(\bOm):=\begin{pmatrix}
-\Om_1&\Om_2\\\Om_1&-\Om_2
\end{pmatrix}
\label{E}
\end{equation}
for some uniquely determined 
$S^*$-valued function
$\bOm=(\Om_1,\Om_2)\in\C^\infty(\Ps,\S^*)$, we conclude that in such a basis
any SSISS model \eqref{dot-SSISS} looks like the system \eqref{dot_SS} with 
$\ga\equiv\ga_1+\ga_2=1$. 

\subsection{The RND picture\label{Sec_RND-picture}} Now, in place of a compartment basis, let's choose the {\em canonical basis} dual to $(\bbe,\bsi)$ in $\S$:
\begin{equation}
\bbe^\perp,\bsi^\perp\in\S,\qquad
\brbe\bbe^\perp\ket=\brsi\bsi^\perp\ket=0,\quad
\brbe\bsi^\perp\ket=\brsi\bbe^\perp\ket=1.
\label{dual}
\end{equation}
Putting $\P:=\{(X,I)\in\RR\times\RRN\}$, this provides a natural variable transformation 
\begin{equation}
\vphibe:\Ps\to\P,\qquad
\bS\mapsto(X=\brbe\bS\ket,\, I=1-\brsi\bS\ket),\qquad
\bbe\in\B
\label{phibe}
\end{equation}
with inverse given by
\begin{equation}
\bS={\vphibe}\inv(X,I)=X\,\bsi^\perp + (1-I)\,\bbe^\perp.
\label{vphibe_inv}
\end{equation}
Accordingly, for $(\bM,\bga)\in\Mbe\times\bsi\inv(1)$ and  
$\bbe\in\B$ the  transformed vector field is given by
$$
{\vphibe}_*\bF_{(\bM,\bga)}\equiv
D\vphibe\circ\bF_{(\bM,\bga)}\circ{\vphibe}\inv.
$$ 
By \Eqref{dot_I} and $\ga=1$ there must exist 
$f\in\C^\infty(\P)$ such that 
\begin{equation}
{\vphibe}_*\bF_{(\bM,\bga)}=\bV_f:=f\partial_X+(X-1)I\partial_I\,,
\label{Vf}
\end{equation}
or in other words,
\begin{equation}
\dot{X}=f(X,I),\qquad\dot{I}=(X-1)I.
\label{repl_dyn}
\end{equation}
\begin{definition}\label{Def_RND}
For $f\in\C^\infty(\P)$ the dynamical system \eqref{Vf}-\eqref{repl_dyn} on $\P$ is called a {\em replacement number dynamics} (RND). The set of such vector fields is denoted 
$\VRND:= \{\bV_f\mid f\in\C^\infty(\P)\}$.
\end{definition}
\bsn
Given $(\bM,\bga)\in\Mbe\times\bsi\inv(1)$, we now determine $f\in\C^\infty(\P)$. As an $\RR$-linear coset we have
\begin{equation}
\Mbe\times\bsi\inv(1)=(\bNb,\bbe^\perp)+(\E\times\ker\bsi),
\qquad \bNb:=-I|\bbe^\perp\ket\brbe,
\label{N}
\end{equation}
where $\E$ has been defined in \Eqref{M0}. Also, $\VRND$ naturally acquires a coset structure
$$
\VRND=\bV_0+\C^\infty(\P)\partial_X.
$$
Moreover, if $(\bM,\bga)=(\bNb,\bbe^\perp)$ then
$\dot{X}=0$, i.e.
$
{\vphibe}_*\bF_{(\bNb,\bbe^\perp)}=\bV_0.
$
To show that ${\vphibe}_*$ in fact provides an isomorphism of linear cosets,
\begin{equation}
\Fbe\cong\bF_{(\bNb,\bbe^\perp)}+
(\E\times\ker\bsi)/\K\cong\bV_0+\C^\infty(\P),
\label{coset_iso}
\end{equation}
we proceed in two steps.
First, for $\H:=\C^\infty(\P)^2\times\RR$ define 
the isomorphism $\bmbe:\H\to\E\times\ker\bsi$
\begin{align}
\bmbe(Q,P,\lambda)&:=
\left(\mbe(Q,P),\ \lambda\bsi^\perp\right)
\label{bmbe}
\\
\mbe(Q,P)&:=|\bsi^\perp\ket\left(Q\bra\bbe|+P\bra\bsi|\right)\circ\vphibe
\label{mbe}
\end{align}
and the epimorphism $\piP:\H\to\C^\infty(\P)$,
\begin{equation}
\piP(Q,P,\la)(X,I):=\la I+(1-I)P(X,I)+XQ(X,I).
\label{pi_P}
\end{equation}
The one-to-one relation between SSISS models 
and RND systems 
may then transparently be summarized by
\begin{proposition}\label{Prop_Repl_dyn}
{\rm (Replacement Number Dynamics I)} \\
Let $(Q,P,\la)\in\H$,
$(\bM,\bga)=(\bNb,\bbe^\perp)+\bmbe(Q,P,\la)$ and 
$f=\piP(Q,P,\la)$. Then
\begin{itemize}
\item[i)]
${\vphibe}_*\bF_{(\bM,\bga)}=\bV_f$\,.
\item[ii)]
$\bmbe(\ker\piP)=\K$\,.
\end{itemize}
\end{proposition}
Part i) of Proposition \ref{Prop_Repl_dyn} will follow from a more elaborated version, Theorem \ref{Thm_Repl_dyn} below. 
Part ii) says that 
gauge transformations 
$(\bK,0)\in\K$ leaving $\bF_{(\bM,\bga)}$ invariant precisely correspond to the ambiguity when expressing 
$f$ in terms of $(Q,P,\la)$.
More generally, this follows from

\begin{lemma}\label{Lem_bel_beta}
For $\bbe\in\B$ let 
$\lbe:\M\times\S\ni
(\bM,\bga)\mapsto(Q,P,\la)\in\H$ be given by 
\begin{equation}
(Q,P):=
\left(\bra\bbe|(\bM\circ\vphibe\inv)\bsi^\perp\ket,
\bra\bbe|(\bM\circ\vphibe\inv)\bbe^\perp\ket\right),\quad\la:=\brbe\bga\ket.
\label{QPla}
\end{equation}
Then $(\bNb,\bbe^\perp)\in\ker\lbe$,
$\lbe|_{\E\times\ker\bsi}=\bmbe\inv$ and
$\lbe(\K)=\ker\piP$.
\end{lemma}

\begin{proof}
The first two statements are obvious.
To prove the last statement 
use $\ker\piP=\{(Q,P,\la)\mid\la=0\,\land\,XQ=(I-1)P\}$ and
assume $(\bK,\bzero)\in\K\equiv\K_0\times\{0\}$. Putting 
$(Q,P,\la)=\lbe(\bK,\bzero)$ one concludes
$\la=0$, $(Q\bbe+P\bsi)\circ\vphibe=\bbe\bK$ and 
$$
\begin{aligned}
Q(X,I)X+P(X,I)(1-I)&
=Q(X,I)\bra\bbe|\bS\ket+P(X,I)\bra\bsi|\bS\ket
=\bra\bbe|\bK(\bS)\bS\ket=0,
\end{aligned}
$$
by definition of $\K_0$. Hence 
$\lbe(\K)\subset\ker\piP$. Conversely, if $(Q,P,0)\in\ker\piP$ put 
$$\bK:=
(Q\circ\vphibe)|\bsi^\perp\ket\brbe +
(P\circ\vphibe)|\bsi^\perp\ket\brsi\in\E.
$$
Then 
$\lbe(\bK,\bzero)=(Q,P,0)$ and 
$
\bK(\bS)\bS=\left[Q(X,I)X+P(X,I)(1-I)\right]\bsi^\perp=\bzero.
$
Hence $\bK\in\K_0$ and therefore 
$\ker\piP\subset\lbe(\K)$. 

\end{proof}
%
In the second step we now point out that the epimorphism $\piP$ in \eqref{pi_P} naturally splits. Being trivial, the proof of the following Lemma \ref{Lem_H_0} is omitted.

\begin{lemma}\label{Lem_H_0}
Denote 
$\bar{\H}:=\{(Q,P,\la)\in\H\mid\partial_X P=0\}$ and define 
$i_\P:\C^\infty(\P)\ni f\mapsto(Q_f,P_f,\la_f)\in\bar{\H}$ by
\begin{equation}
\begin{aligned}
Q_f(X,I)&:=X\inv(f(X,I)-f(0,I)),\\
P_f(I)&:=(1-I)\inv(f(0,I)-f(0,1)I),\\
\la_f&:=f(0,1).
\end{aligned}
\label{i_P}
\end{equation}
Then $i_\P$ is bijective and $\piP|_{\bar{\H}}={i_\P}\inv$. Hence 
$\H=\bar{\H}\oplus\ker\piP$, with projection 
$i_\P\circ\piP:\H\ni(Q,P,\la)\mapsto(\bar{Q},\bar{P},\la)\in\bar{\H}$, where $\bar{P}=P|_{X=0}$ and 
$\bar{Q}=Q+X\inv(1-I)(P-\bar{P})$. \qed
\end{lemma}
Next, denote 
\begin{equation}
\Ebe=\{\bM\in\E\mid\partial_X\bra\bbe|(\bM\circ\vphibe\inv)\bbe^\perp\ket\equiv\partial_XP=0\}
\label{Ebe}
\end{equation}
then $\bmbe(\bar{\H})=\Ebe\times\ker\bsi$ and, by part ii) of Proposition \ref{Prop_Repl_dyn}, the splitting 
$\H=\bar{\H}\oplus\ker\piP$ induces a splitting 
$$
\E=\Ebe\oplus\K_0.
$$
Since $\K_0\subset\E$ is the subspace of gauge transformation, we may identify 
$\pibe:\E\to\Ebe\cong\E/\K_0$ as the canonical projection to gauge equivalence classes. Similarly, we also identify
\begin{equation}
\Fbe\cong(\bNb+\Ebe)\times\bsi\inv(1).
\label{gaugefixing}
\end{equation}
\begin{definition}\label{Def_gaugefixing}
Define $\cpibe:\Mbe\to\bNb+\Ebe$ by 
$\cpibe\bM:=\bNb+\pibe(\bM-\bNb)$, then
$\bF_{(\bM,\bga)}=\bF_{(\cpibe\bM,\bga)}$. So, replacing $\bM$ by $\cpibe\bM$ may be considered as a ``gauge fixing'' condition, denoted as
{\em the canonical gauge}.
\end{definition}

\begin{theorem}\label{Thm_Repl_dyn}
{\rm (Replacement Number Dynamics II)} \\
For $\bbe\in\B$ put 
$\Lbe:=\piP\circ\lbe:\M\times\S\to\C^\infty(\P)$ and 
$\phibe:=\Lbe|_{(\bNb+\Ebe)\times\bsi\inv(1)}$. Then 
\begin{itemize}
\item[i)]
$\Lbe(\bNb,\bbe^\perp)=0$, $\Lbe(\E\times\ker\bsi)=C^\infty(\P)$ and 
$\ker\Lbe\cap(\E\times\ker\bsi)=\K$. Hence, $\Lbe$ induces an isomorphism of linear cosets 
$\Fbe\equiv(\Mbe\times\bsi\inv(1))/\K\cong\bV_0+\C^\infty(\P)\partial_X\equiv\VRND$.
\item[ii)]
For all $(\bM,\bga)\in\Mbe\times\bsi\inv(1)$ we have
${\vphibe}_*\bF_{(\bM,\bga)}=\bV_{\Lbe(\bM,\bga)}$, i.e.
\begin{equation}
\dot{X}=\la I+(1-I)P(X,I)+XQ(X,I).
\label{dot_X}
\end{equation}
\item[iii)]
Put $\bTbe:=\bmbe\circ i_P$. Then  
$\bTbe:\C^\infty(\P)\to\Ebe\times\ker\bsi$
defines a linear isomorphism satisfying
$\Lbe\circ\bTbe=\id_{\C^\infty(\P)}$. Hence,
for all $f\in\C^\infty(\P)$,
\begin{equation}\label{T_beta}
\begin{aligned}
\phibe\inv(f)&=(\bNb,\bbe^\perp)+\bTbe f,
\\ 
\cpibe&=\phibe\inv\circ\Lbe|_{\Mbe},
\\
{{\vphibe}_*}\inv \bV_f&= \bF_{\phibe\inv(f)}.
\end{aligned}
\end{equation}
\end{itemize}
\end{theorem}

\begin{proof}
Part i) follows from Lemma \ref{Lem_bel_beta}. To prove part ii) use \eqref{vphibe_inv} and \eqref{QPla} to compute
$$\dot{X}=\brbe\dot{\bS}\ket=\brbe\bM\bS\ket+I\brbe\bga\ket=
XQ+(1-I)P+\la I.
$$
Finally,
$\Lbe\circ\bTbe=\id_{\C^\infty(\P)}$ 
by Lemmas \ref{Lem_bel_beta} and  \ref{Lem_H_0}, proving also the other statements in iii).
\end{proof}

\subsection{The symmetry group $\Gs$\label{Sec_Symmetry}}
Beware that $\bV_f\in\VRND$ itself carries no information about 
$\bbe\in\B$. So, fixing 
$f\in\C^\infty(\P)$ and varying $\bbe\in\B$, the reconstruction $\bF_{(\bM,\bga)}={{\vphibe}_*}\inv \bV_f$ yields different SSISS dynamical systems 
$\bF_{(\bM,\bga)}\in\Fbe$, 
which therefore must be isomorphic. In fact, putting
\begin{equation}
\bg(\bbe',\bbe):={\varphi_{\bbe'}}\inv\circ\vphibe:\Ps\longrightarrow\Ps,\qquad \bbe',\bbe\in\B,
\label{g(b'b)}
\end{equation}
we get a groupoid of linear coordinate transformations, 
$\bg(\bbe'',\bbe')\bg(\bbe',\bbe)=\bg(\bbe'',\bbe)$, obeying 
$$
\bg(\bbe',\bbe)_*\Fbe=\F_{\bbe'}.
$$
As will be shown now, these isomorphisms are in fact induced by a natural symmetry group $\Gs$,
$$
\Gs:=\{\bg\in\Aut\S\mid\det\bg>0\,\land\,\bsi\bg=\bsi\}.
$$
Clearly, $\Gs$ leaves $\Ps$ and $\bsi\inv(1)$ invariant, which allows to define a left $\Gs$-action 
\begin{equation}
\re : \Gs\times\M\times\bsi\inv(1)\longrightarrow
\M\times\bsi\inv(1),\qquad 
\bg\re(\bM,\bga):=(\Ad\g\circ\bM\circ\bg\inv,\bg\bga)
\label{action}
\end{equation}
such that 
$$
\bg_*\bF_{(\bM,\bga)}=\bF_{\bg\re(\bM,\bga)},\qquad\forall\bg\in\Gs.
$$
To see that $\bg\re$ indeed maps SSISS systems to SSISS systems,
use $I\circ\bg=I$ and if $\bM\in\M_{\bbe}$ then 
$\Ad\g\circ\bM\circ\bg\inv\in\M_{\bbe\bg\inv}$.
Since one also easily checks $\bg\re\K=\K$,
$\Gs$ maps SSISS systems to SSISS systems and all SSISS systems within one $\Gs$-orbit are isomorphic
$$
\bg_*\Fbe=\F_{\bbe\bg\inv}\,.
$$
Recalling the definition of $\B\equiv\Bs$ in \eqref{Bs}, standard linear algebra yields
\begin{lemma}\label{Lem_BGs}
The group $\Gs$ acts freely and transitively on 
$\B$ from the right.
\end{lemma}
Also obviously $\vphibe\circ\bg=\varphi_{\bbe\bg}$ and therefore
$$
\bbe=\bbe'\bg\Longleftrightarrow\bg=\bg(\bbe',\bbe),\qquad
\forall\bbe,\bbe'\in\B,
$$
where $\bg(\bbe',\bbe)$ has been defined in \eqref{g(b'b)}.
Also note $\bg\re\Ebe=\E_{\bbe\bg\inv}$.

In summary, SSISS systems on $\Ps$ map to the same replacement number dynamics \eqref{repl_dyn} on 
$\P$, if and only if they are $\Gs$-equivalent:
\begin{corollary}\label{Cor_Phi}
Define $\Phi:\F_\B\longrightarrow\VRND\times\B$ by
\begin{equation}
\Phi(\bF_{(\bM,\bga)}):=({\vphibe}_*\bF_{(\bM,\bga)},\bbe),\qquad\bbe:=\pi(\bM),
\label{Phi}
\end{equation}
\begin{itemize}
\item[i)]
If $\Phi(\bF_{(\bM,\bga)})=(\bV_f,\bbe)$, then 
$\Phi(\bg_*\bF_{(\bM,\bga)})=(\bV_f,\bbe\bg\inv)$, for all 
$\bg\in\Gs$.
\item[ii)]
If $\Phi(\bF_{(\bM,\bga)})=(\bV_f,\bbe)$ and 
$\Phi(\bF_{(\bM',\bga')})=
(\bV_f,\bbe')$, then there uniquely exists 
$\bg\in\Gs$ such that 
$\bbe'=\bbe\bg\inv$ and 
$\bF_{(\bM',\bga')}=\bg_*\bF_{(\bM,\bga)}$.
\end{itemize}
\end{corollary}

Corollary \ref{Cor_Phi} implies that under the $\Gs$-action 
$\re$ the space of SSISS vector fields 
$\F_\B$ is isomorphic via $\Phi$ to the trivial principal $\Gs$-bundle $\VRND\times\B$ with $\Gs$-fiber $\B$ and base space
$\VRND\cong\F_\B/\Gs$.
This generalizes the finite dimensional setting in \cite{Nill_SIRS}. 
\begin{definition}\label{Def_SSISS-fiber}
Denote $\F(f):=\Phi\inv(f\times\B)\subset\F_\B$. Then $\F(f)$ is called {\em the SSISS fiber} of $f$.
\end{definition}
It is important to realize that $\Gs$ does not preserve admissibility properties like the constraints \eqref{phys_constraint}. In fact, in general not even the physical triangle $\Tph\subset\Ps$ \eqref{phys_triangle} is conserved. To see this, for 
$\bbe=(\be_1,\be_2)$ w.r.t. some compartment basis, consider the image of $\Tph$ under 
$\vphibe$, 
\begin{equation}
\vphibe(\Tph)=\T(\bbe):=\{(X,I)\in\P\mid
0\leq I\leq 1\,\land\, \be_2(1-I)\leq X\leq\be_1(1-I)\}.
\label{T(beta)}
\end{equation} 
In other words, $\T(\bbe)\subset\P$ is given by the triangle with corners $(X,I)\in\{(\be_1,0),\,(\be_2,0),\,(0,1)\}$.
Now clearly 
$\T(\bbe)\subset\T(\bbe')\Leftrightarrow
[\be_2,\be_1]\subset[\be_2',\be_1']$, and so we conclude:
\begin{lemma}\label{Lem_g.Tph}
Let $\bg=\bg(\bbe',\bbe)\in\Gs$. Then $\bg\Tph\subset\Tph$ if and only if $\be_2'\leq\be_2$ and $\be_1\leq\be_1'$.
\end{lemma}
\begin{proof}
We have $\varphi_{\bbe'}(\bg\Tph)=\T(\bbe)$ and
$\varphi_{\bbe'}(\Tph)=\T(\bbe')$.
\end{proof}
Some more details on $\Gs$ can be found in Appendix \ref{App_Gs.X^2}.

\subsection{Coordinate expressions\label{Sec_coordinates}}
From now on we consider a fixed compartment basis 
$\bfe_i\in\S$, such that 
with respect to such a basis the following coordinate expressions for the transformations between SSISS and RND picture are easily verified. 
\begin{align}
\B&=\{(\be_1,\be_2)\in\RR^2\mid\be_1>\be_2\},
\\
\Gs&=\{\bg\in GL(\RR^2, +)\mid g_{1j}+g_{2j}=1\},
\\
\bg(\bbe,\bbe')&=\frac{1}{\be_1-\be_2}
\begin{pmatrix}
\be_1'-\be_2\,,& \be_2'-\be_2\\
\be_1-\be_1'\,, & \be_1-\be_2'
\end{pmatrix}
\label{g_matrix}
\end{align}
\begin{align}
\bsi^\perp&=\frac{1}{\be_1-\be_2}
\begin{pmatrix}
1\\-1
\end{pmatrix},
&
\bbe^\perp&=\frac{1}{\be_1-\be_2}
\begin{pmatrix}
-\be_2\\\be_1
\end{pmatrix},
\label{dual-coordinates}\\
|\bsi^\perp\ket\brbe &=\frac{1}{\be_1-\be_2}
\begin{pmatrix}
\be_1&\be_2\\-\be_1&-\be_2
\end{pmatrix},
&
|\bsi^\perp\ket\brsi &=\frac{1}{\be_1-\be_2}
\begin{pmatrix}
1&1\\-1&-1
\end{pmatrix},
\\
\bga&=\frac{1}{\be_1-\be_2}
\begin{pmatrix}
\la-\be_2\\\be_1-\la
\end{pmatrix},
&\la&=\be_1\ga_1+\be_2\ga_2,
\label{ga(la)}
\\
P&=\be_1\Om_2+\be_2\Om_1+\be_1\be_2I,
&Q&=-(\Om_1+\Om_2+(\be_1+\be_2)I),
\label{PQ(Om)}
\\
\Om_1&=\frac{1}{\be_2-\be_1}(P+\be_1Q+\be_1^2I),&
\Om_2&=\frac{1}{\be_1-\be_2}(P+\be_2Q+\be_2^2I),
\label{Omi}
\\
\bNb&=\frac{I}{\be_1-\be_2}
\begin{pmatrix}
\be_1\be_2&\be_2^2\\
-\be_1^2& -\be_1\be_2
\end{pmatrix}
\notag\\&=
\bDb+I\bE(\bDelbe),\hspace{-20pt}&
\bDelbe&=
\frac{1}{\be_1-\be_2}(-\be_1^2,\be_2^2).
\label{N-D}
\end{align}
Here, we have used the definitions \eqref{QPla} together with  $\bM=\bDb+\bE\circ\bOm$ in \Eqref{PQ(Om)}
and the definitions  \eqref{D}, \eqref{E}, \eqref{N} in \Eqref{N-D}. 
\Eqref{vphibe_inv} now becomes
\begin{equation}
S_i=\frac{X - (1-I)\be_j}{\be_i-\be_j},\qquad j\neq i.
\label{Si(X,I)}
\end{equation}
From now on, all examples will be presented in 
canonical gauge, i.e. without loss we replace $(Q,P)$ by 
$(\bar{Q},\bar{P})$. Accordingly, for $f\in\C^\infty(\P)$ define 
$\bOmbe(f)\in\C^\infty(\Ps,\S^*)$ and $\bga_{\bbe}(f)\in\S$ by Eqs. \eqref{Omi} and \eqref{ga(la)}, respectively, with 
$(Q,P,\la)$ replaced by 
$(Q_f,P_f,\la_f)\circ\vphibe$, see Lemma \ref{Lem_H_0}. In particular 
$\bOmbe(0)=I\bDelbe$ and $\bNb=\bDb+\bE\circ\bOmbe(0)$. 
Part iii) of Theorem \ref{Thm_Repl_dyn} may now be restated as
\begin{equation}
\phibe\inv(f)=(\bNb,\bbe^\perp)+\bTbe f
=\left(\bDb+\bE\circ\bOmbe(f),\,\bga_{\bbe}(f)\right),
\label{phibe_inv}
\end{equation}
In this way we identify 
$\Fbe\cong(\bNb+\Ebe)\times\bsi\inv(1)=
(\bDb+\Ebe)\times\bsi\inv(1)$.
Also note that by \Eqref{PQ(Om)} 
$\be_1\Om_2+\be_2\Om_1=P-\be_1\be_2 I$.
Since $P_f$ only depends on $I$, we arrive at

\begin{corollary}
In a compartment basis the canonical gauge is characterized by the property that $\be_1\Om_2+\be_2\Om_1$ only depends $S_1+S_2$ $($equivalently on $I)$.
\end{corollary}

\section{RND models quadratic in $X$\label{Sec_quadratic}%
}
This section demonstrates how the above formalism applies to the specific class of models where $f$ is at most quadratic in $X$, 
\begin{equation}
f(X,I)=f_0(I)+f_1(I)X+f_2(I)X^2.
\label{X^2}
\end{equation}
For short, let us call these {\em $X^2$-Models}. As we will see in Section \ref{Sec_examples-revisited}, all
examples presented in Subsection \ref{Sec_examples} are indeed of this type.
Below, subsection \ref{Sec_X^2} shows how $X^2$-models are characterized in SSISS picture. Subsection \ref{Sec_admissible_1} translates the admissibility constraints of Definition \ref{Def_SSISS_admissible} from SSISS picture into an admissibility condition for $f$ and a compatibility condition for $\bbe$ in RND picture. Subsection \ref{Sec_R0}
defines the reproduction number and its relation to admissibility in RND picture. 
Finally, changing again to SSISS picture, subsection \ref{Sec_periodic} shows that admissible $X^2$-models do not admit periodic solutions.

\subsection{$X^2$-Models in SSISS picture\label{Sec_X^2}}
Denote
$\Vtwo\subset\VRND$ the sub-coset of RND vector fields
$\bV_f$, where $f$ is at most quadratic in $X$ as in \Eqref{X^2}.
The associated family of SSISS systems in canonical gauge is denoted by
\begin{equation}
\begin{aligned}
\Fbetwo	&:={\vphibe}_*\inv\Vtwo=(\bNb+\Ebetwo)\times\bsi\inv(1)
=(\bDb+\Ebetwo)\times\bsi\inv(1),
\\
\Ebetwo	&:=\{\bM\in\Ebe\mid\partial_X^2\bra\bbe|(\bM\circ\vphibe\inv)\bsi^\perp\ket\equiv\partial_X^2 Q=0\}.
\end{aligned}
\label{Fbetwo}
\end{equation}
From \Eqref{i_P} we get $f=\la_f I+(1-I)P_f+XQ_f$, where
\begin{equation}
\la_f=f_0(1), 
\qquad
P_f(I)=(1-I)\inv(f_0(I)-\la I),
\qquad
Q_f(X,I)=f_1(I)+Xf_2(I).
\label{laP0Q0}
\end{equation}
Now replace
$X=\be_i(1-I)+ (\be_j-\be_i)S_j,\ j\neq i$, in \Eqref{laP0Q0} and use  \Eqref{Omi} with $(Q,P)$ replaced by 
$(Q_f\circ\vphibe,P_f\circ\vphibe)$. 
Then $\bOmbe(f)=(\Om_1,\Om_2)$ is given by
\begin{align}
\Om_i&=A_i(I)+B(I)\be_iS_j,\qquad j\neq i,
\label{Ai+Bi}
\\
A_i(I)&=\frac{1}{\be_j-\be_i}
\left[P_f(I)+\be_if_1(I)+\be_i^2(I+(1-I)f_2(I))\right],\qquad j\neq i,
\label{Ai}
\\
B(I)&=f_2(I),\label{B(I)}
\end{align}
\begin{corollary}\label{Cor_SSISS-X^2}
SSISS systems $\phibe\inv(f)$ corresponding to $X^2$-models $f$ are of the form
\begin{equation}
\dot{S_i}=(-\be_i S_i+\ga_i)I-A_i(I)S_i+A_j(I)S_j-
(\be_i-\be_j)B(I)S_1S_2,
 \quad j\neq i,\quad\ga_1+\ga_2=1\,.
 \label{dot_Si_X^2}
\end{equation}
\end{corollary}
Thus, considering 
$(A_1, A_2,B,\ga_1,\ga_2)\in\C^\infty(\RRN)^3\times\bsi\inv(1)$ as affine coordinates in $\Fbetwo$, we may write 
$(A_1, A_2,B,\ga_1,\ga_2)=(\bOm_f,\bga_f)=\phibe\inv(f)$.
The inverse assignment 
$f\equiv(P_f,f_1,f_2,\la)=\phibe(A_1, A_2,B,\ga_1,\ga_2)$ is then given by
\begin{align}
P_f&=\be_1A_2+\be_2A_1+\be_1\be_2(I+(1-I)B),
\label{P0}\\
f_1&=-\left[A_1+A_2+
(\be_1+\be_2)(I+(1-I)B)\right],
\label{f1}\\
f_2&=B,
\label{f2}
\\
\la&=\be_1\ga_1+\be_2\ga_2\,,
\label{la}
\end{align}
The interested reader is invited to check that the transformations $(A_1,A_2,B)\longleftrightarrow(P_f,f_1,f_2)$ given by Eqs. \eqref{Ai}-\eqref{B(I)} and \eqref{P0}-\eqref{f2},
respectively, indeed invert each other. 
Also, by Corollary \ref{Cor_Phi}, fixing $f$ and varying $\be_i$ while keeping $\be_1>\be_2$, provides a family of $\Gs$-eqivalent systems \ref{dot_Si_X^2}, i.e. the SSISS fiber $\F(f)$ in Definition \ref{Def_SSISS-fiber}. A formula for the action of
$\bg\in\Gs$ on parameters  $(\bbe,\bA,B,\bga)\in\Fbetwo$ is given in Proposition \ref{Prop_action} in Appendix \ref{App_Gs.X^2}.

\subsection{Admissibility in RND picture%
\label{Sec_admissible_1}}
Recalling Definition \ref{Def_SSISS_admissible},
admissibility of the SSISS system \eqref{dot_Si_X^2} is equivalent to  
\begin{equation}
\bbe\in\B\ \ \land\ \ \la\equiv \be_1\ga_1+\be_2\ga_2\in[\be_2,\be_1]\ \ \land\ \ A_i(I)\geq 0,\ \forall I\in[0,1].
\label{physA}
\end{equation}
To analyze the impact on $f$, we now define
\begin{definition}\label{Def_admissible}
Let $\bV_f$ be an RND system. Then $\bbe\in\B$ is called {\em $f$-compatible}, if the SSISS system ${\vphibe}_*\inv\bV_f$ is admissible, 
and $f$ is called {\em admissible}, if such a $\bbe$ exists, i.e. if $\Bf\neq\emptyset$, 
\begin{equation}
\Bf:=\{\bbe\in\B\mid \bbe \text{ is $f$-compatible}\}.
\label{Bf}
\end{equation}
Correspondingly, the triangle $\T(\bbe)$ (\Eqref{T(beta)}) is called $f$-compatible, if $\bbe\in\Bf$.
\end{definition}
Note that admissibility in the SSISS picture is not a $\Gs$-invariant property. In fact, even if $f$ is admissible, there will still exist $\bbe\in\B$, for which the SSISS system 
${\vphibe}_*\inv\bV_f$ is not admissible, whence
$\Bf\subsetneq\B$. For $X^2$-models this may be analyzed to quite some detail. First, rewrite \Eqref{Ai} in the form
\begin{align}
A_i(I)&=\ \hf(\be_i,I)/(\be_j-\be_i),\quad j\neq i,
\label{Ai=hf/(bj-bi)}
\\
\hf(z,I)&:=\ P_f(I)+zf_1(I)+z^2\left(I+(1-I)f_2(I)\right).
\label{hf}
\end{align}

Hence, for $\bbe\equiv(\be_1,\be_2)\in\Bf$ the condition $A_i\geq 0,\ i=1,2,$ implies that the family of at most quadratic functions 
$\hf(\cdot,I)$ must have a zero in the interval 
$z\in[\be_2,\be_1]$, for every $I\in[0,1]$. Noting that 
$\ga_1+\ga_2=1$ implies
$\ga_i\geq0\LRA\be_2\leq\la\leq\be_1$, one readily concludes
\begin{lemma}\label{Lem_Bf}
For $f$ given by \eqref{X^2} and $\hf$ given by 
\eqref{hf}, we have 
\begin{equation}
\begin{aligned}
\Bf	&=(\Bf^+\times\Bf^-)\cap\B,
\\
\Bf^+&:=
\{z\geq\la\mid \hf(z,I)\leq 0,\,\forall I\in[0,1]\},
\\
\Bf^-&:=
\{z\leq\la\mid \hf(z,I)\geq 0,\,\forall I\in[0,1]\}.
\end{aligned}
\label{Bfpm}
\end{equation}
\end{lemma}
A generalization of Lemma \ref{Lem_Bf} to arbitrary smooth functions $f\in\C^\infty(\P)$ will be given in Part II of this work.

If $\hf(z,I)$ is at most quadratic in $z$ as above, then $\Bf^\pm$ are intersections of (possibly empty or semi-infinite) closed intervals, hence themselves either empty or  possibly semi-infinite closed intervals. 
To be more precise denote $Z_-(I)\leq Z_+(I)$ the zeros of 
$\hf(\,\cdot\,,I)$ and put 
$$
\begin{aligned}
\I_s	&:=\{I\in[0,1]\mid\sign\left(I+(1-I)f_2(I)\right)=s\}
\\
\Z_+	&:=\{Z_+(I)\mid I\in\I_+\},
\\
\Z_-	&:=\{Z_-(I)\mid I\in\I_-\},
\\
\Z_0	&:=
\{Z_-(I)\mid I\in\I_+\}\cup\{Z_+(I)\mid I\in\I_-\}\cup
\{-P_f(I)/f_1(I)\mid I\in\I_0\land f_1(I)<0\}.
\end{aligned}
$$
By analyzing positive and negative ranges of parabolas or straight lines one readily concludes that $\Bf\neq\emptyset$ requires 
$\Z_-\leq\Bf^-\leq\Z_0\leq\Bf^+\leq\Z_+$ and
\begin{equation}
f_1(I)<0\ \text{or}\ f_1(I)=P_f(I)=0,\qquad\forall I\in\I_0.
\label{f1<0}
\end{equation}
Also note that $Z_-$ might be empty, whereas $1\in\I_+$ implies 
$\Z_+$ and $\Z_0$ to be nonempty. This leads to define
$$\begin{aligned}
\bemax_1	&:=\inf\Z_+,\\
\bemin_1	&:=\sup\{\la,\Z_0\},\\
\bemax_2	&:=\inf\{\la,\Z_0\},\\
\bemin_2	&:=
\begin{cases}
\sup\{\Z_-\},	&\Z_-\neq\emptyset,
\\
-\infty,		&\Z_-=\emptyset.
\end{cases}
\end{aligned}
$$
In summary, we get
\begin{proposition}\label{Prop_admissible_X^2}
{\em (Admissibility in $X^2$-models)}
Let $f(X,I)$ be at most quadratic in $X$. Necessary for 
admissibility of $f$ is \Eqref{f1<0}. If this holds, then $f$ is admissible if and only if 
$\bemin_2\leq\bemax_2\leq\bemin_1\leq\bemax_1$ and 
$\bemin_2<\bemax_1$. In this case 
\begin{equation}
\Bf^-=[\bemin_2,\bemax_2],\qquad\Bf^+=[\bemin_1,\bemax_1].
\label{be_min_max1}
\end{equation}
\end{proposition}

\subsection{The Reproduction Number\label{Sec_R0}}
In this subsection we take a closer look at the admissibility condition $A_i(I)\geq0$ at the boundary values $I=0$ and $I=1$ in general $X^2$-models \eqref{dot_Si_X^2}. Clearly, $A_i(0)\geq0$ implies the line segment 
$\Tph\cap\{I=0\}$ staying forward invariant under the disease free SSISS dynamics. When changing to RND picture observe
$f|_{I=0}=\hf|_{I=0}$ by definition \eqref{hf} and therefore, 
\begin{equation}
A_i(0)=f(\be_i,0)/(\be_j-\be_i)\geq0,\quad j\neq i, \quad
\forall (\be_1,\be_2)\in\Bf.
\end{equation}
Hence we conclude
\begin{equation}
f(\Bf^+,0)\leq 0\leq f(\Bf^-,0).
\label{f(Bf^pm,0)}
\end{equation}
Recalling the definition $\T(\bbe)=\vphibe(\Tph)$, this equivalently means that the base line
$$
\T(\bbe)\cap\{I=0\}=[\be_2,\be_1]\times\{0\}\subset\P
$$
is forward invariant under the flow of $\bV_f$, for all compatible
$\bbe\equiv(\be_1,\be_2)\in\Bf$. Also, there must exist $R_0\in[\bemax_2,\bemin_1]$, such that $f(R_0,0)=0$, i.e. 
$(X^*,I^*)=(R_0,0)$ is a {disease-free equilibrium point} of 
$\bV_f$. Under the additional regularity condition 
$\partial_Xf(R_0,0)<0$, $R_0$ becomes the {\em Reproduction Number} in the version of van den Driessche and Watmough \cite{Driesche_Watmough2000, Driesche_Watmough2002}.

To get the relation between $R_0$ and our admissibility bounds 
$\be_i^{\min/\max}$ put $f(X,0)=f_{00}+f_{10}X+f_{20}X^2$ and adapt our notation by renaming 
$\{Z_+(0),Z_-(0)\}\equiv\{R_0,R_1\}$, where
\begin{align}
R_0&:=\begin{cases}\displaystyle
\frac{1}{2f_{20}}\left(-f_{10}-
\sqrt{f_{10}^2-4f_{20}f_{00}}\,\right), & f_{20}\neq 0,
\\
-f_{00}/f_{10}, & f_{20}=0\ \land\ f_{10}\neq0,
\\
\text{undefined}, & \text{else},
\end{cases}
\label{R0}\\
R_1&:=\begin{cases}\displaystyle
\frac{1}{2f_{20}}\left(-f_{10}+
\sqrt{f_{10}^2-4f_{20}f_{00}}\,\right), & f_{20}\neq 0,
\\
\text{undefined}, & f_{20}=0,
\end{cases}
\label{R1}
\end{align}
Hence, if $f$ is admissible, then
\begin{gather}
R_0\in\Z_0\subset[\bemax_2,\bemin_1]\\
f_{20}>0\Rightarrow R_1\in\Z_+\geq\bemax_1, \\
f_{20}<0\Rightarrow R_1\in\Z_-\leq\bemin_2.
\end{gather}
Also, observing $\hf(z,1)=P_f(1)+f_1(1)z+z^2$, let us denote
\begin{equation}
Z_\pm\equiv Z_\pm(1):=
\frac{1}{2}\left(-f_{1}(1)\pm
\sqrt{f_{1}(1)^2-4P_f(1)}\,\right),
\label{Z+-}
\end{equation}
Then the following Corollary is an immediate consequence of 
Proposition \ref{Prop_admissible_X^2}.
\begin{corollary}\label{Cor_admissible_X^2}
Let $f(X,I)$ be at most quadratic in $X$ and assume $f$ admissible. 
\begin{itemize}
\item[i)]
Then necessarily $Z_\pm\in\RR$ and
\begin{gather}
(f_{20}\neq 0 \land f_{10}^2-4f_{20}f_{00}\geq 0)
\quad \text{or}\quad
(f_{20}=0\land f_{10}<0) \quad \text{or}\quad (f|_{I=0}=0).
\label{nec_for_adm1}
\end{gather}
\item[ii)]
If these conditions hold, put
\begin{equation}
\begin{aligned}
\belo_1&:=\begin{cases}
\max\{\la,R_0,Z_-\}, & f|_{I=0}\neq 0,
\\
\max\{\la,Z_-\}, & f|_{I=0}= 0,
\end{cases} 
&
\behi_1&:=\begin{cases}
\min\{R_1,Z_+\}, & f_{20}>0,
\\
Z_+, & f_{20}\leq 0,
\end{cases}
\\
\behi_2&:=\begin{cases}
\min\{\la,R_0,Z_-\}, & \,f|_{I=0}\neq 0,
\\
\min\{\la,Z_-\}, & \,f|_{I=0}= 0,
\end{cases} 
&
\belo_2&:=\begin{cases}
R_1, & \qquad\quad\ \  f_{20}<0,
\\
-\infty, & \qquad\quad\ \  f_{20}\geq 0.
\end{cases}
\end{aligned}
\label{be_hi_lo}
\end{equation}
Then 
\begin{equation}
\Bf^+\equiv[\bemin_1,\bemax_1]\subset[\belo_1,\behi_1],\qquad
\Bf^-\equiv[\bemin_2,\bemax_2]\subset[\belo_2,\behi_2].
\label{bemin_bemax_Model-X^2}
\end{equation}
\end{itemize}

\qed
\end{corollary}
\begin{proof}
This follows straight forwardly from Proposition \ref{Prop_admissible_X^2} and the identities
\begin{align}
\hf(z,1)&=(z-Z_-)(z-Z_+)
\label{hf(X,1)}
\\
\hf(z,0)=f(z,0)&=
\begin{cases}
f_{20}(R_0-z)(R_1-z),& f_{20}\neq 0,
\\
-f_{10}(R_0-z),		& f_{20}=0\,\land\, f_{10}<0,
\\
0,					& f|_{I=0}=0.
\end{cases}
\label{hf(X,0)}
\end{align}
\end{proof}

\begin{remark}\label{Rem_R0}
Under the conditions of Corollary \ref{Cor_admissible_X^2}, 
the disease-free equilibrium $(X^*,I^*)\dfe:=(R_0,0)$ is globally stable in 
$\X_f\times\{0\}\subset\P$,
$$
\X_f:=\begin{cases}
\RR,	&f_{20}=0,
\\
[\bemin_2,\bemax_1]\setminus\{R_1\},	&R_1\neq R_0
\\
[\bemin_2,\bemax_1],	&R_1=R_0
\end{cases}
$$
Also, unless $R_1=R_0$,
we have 
$\partial_Xf(R_0,0)<0$, so in this case $R_0$ indeed coincides with the reproduction number in the version of van den Driessche and Watmough \cite{Driesche_Watmough2000, Driesche_Watmough2002}.
Using \Eqref{Si(X,I)}, in SSISS picture the disease free equilibrium $\bS^*\dfe$
is obtained from $R_0$ by the master formula%
\footnote{Assuming $f_{I=0}\neq0$.}
\begin{equation}
{S^*\dfe}_{,i}=(R_0-\be_j)/(\be_i-\be_j),\ j\neq i.
\label{S*dfe}
\end{equation}
\end{remark}

\subsection{Absence of periodic solutions\label{Sec_periodic}}
Admissibility in the SSISS picture implies $\Tph$ to be forward invariant. Correspondingly, in the RND picture all $f$-compatible triangles 
$\T(\bbe)$ are forward invariant under the flow of $V_f$. Hence, by the Poincaré–Bendixson theorem, any forward flow line starting in $\T(\bbe)$ converges to an equilibrium point (EP) in $\T(\bbe)$, provided there are no periodic solutions (including homoclinic loops and heteroclinic cycles) in $\T(\bbe)$. 

So, the aim of this subsection is to prove absence of periodic solutions inside physical triangles for all admissible $X^2$-models. To do so, we stay in the SSISS picture.  First, adapting methods from Busenberg-Driessche  \autocite{BusDries90} (see also \autocite{BusDries91, DerrickDriessche}), we have the following variant of the classical Bendixson–Dulac Theorem.

\begin{lemma}\label{Lem_BusDries90} \autocite{BusDries90}
Let $\U\subset\RR^3$ be open, 
$\Tph:=\{\by\in\RR_{\geq0}^3\mid\sum_i y_i=1\}\subset\U$ and  $\bF:\U\rto\RR^3$ be smooth 
and satisfy $\sum_iF_i=0$ in $\U$. 
Assume there exists
$u:\U\setminus\partial\Tph\to\RR$ smooth, such that 
\begin{equation}
\Psi(\by):=\nabla\cdot(u\bF)(\by)\leq 0\,,
\qquad\forall\by\in\Tph\setminus\partial\Tph
\label{Dulac}
\end{equation}
and $\Psi(\by)<0$ for some $\by\in\Tph\setminus\partial\Tph$. Then in 
$\Tph\setminus\partial\Tph$ there exist no periodic solutions, homoclinic loops or oriented phase polygons of the dynamical system 
$\dot{\by}=\bF(\by)$.
\end{lemma} 
\begin{proof}
First observe that the flow of $\bF$ leaves the planes 
$\{\mathbf{1}\cdot\by=const\}\subset\RR^3$ invariant, where 
$\mathbf{1}:=(1,1,1)$. Putting $\bV:=\by\times u\bF$ we have 
$\bV\cdot\bF=0$ and 
$\mathbf{1}\cdot(\nabla\times\bV)|_{\Tph}=\Psi|_{\Tph}$, where the second identity easily follows from 
$\mathbf{1}\cdot\bF=0$. Now the claim follows by applying Stoke's Theorem to the vector field $\bV$ as in the proof of Theorem 4.1 of \autocite{BusDries90}.
\end{proof}

\begin{remark}\label{Rem_tildeF}
The condition \eqref{Dulac} looks like a 3-dimensional version of the  classical Bendixson-Dulac theorem and in particular much simpler than the original Busenberg-Driessche formulas in   
\autocite{BusDries90, BusDries91, DerrickDriessche}. The reason is that in these papers in the presence of compartment dependent birth and death rates the condition $\mathbf{1}\cdot\bF=0$ seems to be violated outside $\Tph$. But indeed, the methods of
\cite{Nill_Redundancy} (see also Appendix \ref{App_birth+death}) demonstrate, that one may always replace 
$\bF$ by $\tilde{\bF}$ obeying 
$\bF|_{\Tph}=\tilde{\bF}|_{\Tph}$ and $\mathbf{1}\cdot\tbF=0$ also outside $\Tph$.
\end{remark}

\begin{theorem}\label{Thm_periodic+X^2}
In $X^2$-models \eqref{X^2} there exist no periodic solutions, homoclinic loops or oriented phase polygons in the interior 
$\T(\bbe)\setminus\partial\T(\bbe)$ of any compatible triangle.
\end{theorem}
\begin{proof}
\bsn
To apply Lemma \ref{Lem_BusDries90} to the SSISS version of $X^2$-models put $\by=(S_1,S_2,I)$ and $u=1/(S_1S_2I)$ as in \autocite{BusDries90}. Then the lifted dynamics in $\RR^3$ is given by Eqs. \eqref{dot_Si_X^2}, supplemented by 
$$
\dot{I}=(\be_1S_1+\be_2S_2-1)I.
$$
Thus,
$\Psi(\by)=\partial (u\dot{S}_1)/\partial S_1+
\partial (u\dot{S}_2)/\partial S_2+
\partial (u\dot{I})/\partial I$, where 
$$
\frac{\partial (u\dot{S}_i)}{\partial S_i}=
-\frac{1}{S_i^2}\left(\frac{\ga_i}{S_j}+\frac{A_j}{I}\right),
\ j\neq i,\qquad
\frac{\partial (u\dot{I})}{\partial I}=0.
$$
Hence, admissibility implies $\Psi(\by)<0$ for all 
$\by=(S_1,S_2,I)\in\Tph\cap\RR_+^3$.
\end{proof}
Theorem \ref{Thm_periodic+X^2} does not exclude the possibility of homoclinic loops or oriented phase polygons connecting equilibrium points sitting in the boundary 
$\partial\T(\bbe)$. However, for so-called {\em $e$-admissible models} (see Section \ref{Sec_eadmissible}), a classification of  all equilibrium scenarios will be given in Part II of this work, showing that such a scenario can in fact not appear.

\section{Examples revisited\label{Sec_examples-revisited}}

In this Section we analyze two specific master $X^2$-models, called Model-1 and Model-2, resulting from polynomial RND functions $f(X,I)$ at most quadratic also in $I$. 
Being polynomial, parameter spaces in both models become finite dimensional. So, at fixed $\bbe\in\B$, the transformations $\phibe$ from SSISS to RND picture, Eqs. \ref{Ai+Bi} - \ref{B(I)}, with inverse $\phibe\inv$, Eqs. \eqref{P0} - \eqref{la}, become 
$\bbe$-dependent affine bijections between finite dimensional parameter spaces. In fact, as computed explicitly in Sections \ref{Sec_Model-1} and \ref{Sec_Model-2}, they both turn out to be 6-dimensional. 

As we will see, in SSISS picture these two models cover all epidemiological examples introduced in Section \ref{Sec_examples}.
The admissibility conditions in Proposition \ref{Prop_admissible_X^2} and Corollary \ref{Cor_admissible_X^2} are then reformulated as restrictions on parameter spaces of these models.

\subsection{Model-1\label{Sec_Model-1}}

Model-1 is defined as the sub-case of \eqref{X^2} where $f(X,I)$ is a polynomial in both variables of maximal degree 2 ,
\begin{equation}
\Vmodone:=\{\bV_f\in\VRND\mid 
f(X,I)=\sum_{i=0}^2\sum_{j=0}^{2-i}f_{ij}X^iI^j\}
\subset\Vtwo.
\label{model1}
\end{equation}
Naturally, $\Vmodone$ becomes a coset space over $\RR^6$ with affine coordinates $(f_{ij})_{i+j\leq 2}$. From \eqref{laP0Q0} we get 
$$
\la_f=f_{00}+f_{01}+f_{02},\qquad P_f=f_{00}-f_{02}I,\qquad
Q_f=f_{10}+f_{11}I +f_{20}X,
$$
and therefore
\begin{equation}
\hf(z,I)=\left[f_{00}+f_{10}z+f_{20}z^2\right]+
I\left[-f_{20}+f_{11}z+(1-f_{20})z^2\right].
\label{hf_Model-1}
\end{equation}
From Eqs. \eqref{Ai=hf/(bj-bi)}-\eqref{hf}
we conclude 
$A_i=\al_i+\theta_i I$, $B=f_{20}$ and the associated 8-parameter family of SSISS-System becomes
\begin{equation}
\dot{S_i}=(-\be_i S_i+\ga_i)I-(\al_i+\theta_i I)S_i+
(\al_j+\theta_j I)S_j + (-1)^i\delta S_1S_2,
 \qquad j\neq i.
\label{standard_model}
\end{equation}
Here, $\ga_i$ is given by \eqref{ga(la)} and
\begin{align}
\al_i&=\dfrac{f(\be_i,0)}{\be_j-\be_i} =
\frac{\be_i^2f_{20}+\be_i f_{10}+f_{00}}{\be_j-\be_i},
\qquad j\neq i
\label{alpha}
\\
\theta_i&=
\frac{\be_i^2(1-f_{20})+\be_if_{11}-f_{02}}{\be_j-\be_i},
\qquad j\neq i, 
\label{theta}
\\
\delta&=(\be_1-\be_2)f_{20}.
\label{delta}
\end{align}
Eqs. \eqref{Ai=hf/(bj-bi)}, \eqref{Z+-} and \eqref{hf(X,1)} then imply
\begin{equation}
\begin{aligned}
A_i(1)&\equiv\al_i+\theta_i
=\dfrac{\hf(\be_i,1)}{\be_j-\be_i}
=\frac{(\be_i-Z_+)(\be_i-Z_-)}{\be_j-\be_i},\qquad j\neq i,
\\
Z_\pm&=\frac{1}{2}\left(
-(f_{10}+f_{11})\pm
\sqrt{(f_{10}+f_{11})^2+4(f_{02}-f_{00})}\right)
\end{aligned}
\label{Zpm_Model-1}
\end{equation}
At fixed $\bbe\in\B$ denote 
$\Fbemodone:={\vphibe}_*\inv\Vmodone\subset\Fbetwo$. So, this is also a 6-dimensional coset space with coordinates 
$\Fbemodone=\{(\al_i,\theta_i,\ga_i,\delta)\in\RR^7\mid
\ga_1+\ga_2=1\}$. 
In this way we may write 
$(\al_i,\theta_i,\ga_i,\delta)=\phibe\inv((f_{ij}))$.
Applying \eqref{P0}-\eqref{la}, the inverse 
$\phibe(\al_i,\theta_i,\ga_i,\delta)$
is given by
\begin{align}
f_{00}&=\be_1\al_2+\be_2\al_1 +\be_1\be_2f_{20},
\label{f00}\\
f_{10}&=-\left(\al_1+\al_2+(\be_1+\be_2)f_{20}\right),\\
f_{11}&=-\left((\be_1+\be_2)(1-f_{20})+\theta_1+\theta_2\right),\\
f_{02}&=-\left(\be_1\be_2(1-f_{20})+\be_1\theta_2+\be_2\theta_1\right),
\\
f_{20}&=\delta/(\be_1-\be_2),
\label{f20}\\
f_{01}&=\la-f_{00}-f_{02},\qquad
\la:=\be_1\ga_1+\be_2\ga_2\,.
\label{f01}
\end{align}
For the special case $f_{20}=\delta=0$, relations \eqref{alpha}-\eqref{f01}
have first been obtained in \autocite{Nill_Symm1}.

To compare with our examples in Subsection \ref{Sec_examples}, first assume $\delta=0$.
If $\theta_i=0$ we recover the {\em standard example} \eqref{standard_model} (extended to include $\be_i<0$) and if 
$\theta_1>0=\theta_2$ also the RVM-model \eqref{RVM} with 
$\ka=0$ and an $I$-linear  {\em reactive vaccination} rate. 
Reintroducing tilde-parameters in the presence of birth and death rates, the case $\theta_1>0>\theta_2$ may be interpreted 
as a model with reactive vaccination of non-infected newborns, see \Eqref{tilde+nu_vacc}.
Putting $\al_1=0$ and 
$\theta_1\geq 0\geq\theta_2=-\al_2$ yields the {\em social behavior Model} (SBM) \eqref{SBM}. Finally, the interpretation of a non-zero $\delta$-term is readily obtained by 
writing \Eqref{standard_model} in terms of tilde parameters and comparing with \Eqref{tilde_parameters}, which yields
\begin{equation}
\delta=\del_2-\del_1=(\nu_2-\mu_2)-(\nu_1-\mu_1).
\label{delta_nu_mu}
\end{equation}
Thus, Model-1 covers exactly the superposition of all examples in Subsection 
\ref{Sec_examples} obeying $\kappa=0$ (i.e. excluding models with {\em incidence dependent} vaccination rate).
In particular, using Eqs. \eqref{R0} and \eqref{S*dfe} for $f_{20}=0$, one recovers the well known formulas for the standard example at
$\theta_i=\kappa=\delta=0$ and $\ga=1$ (see, e.g. \cite{AvramAdenane2022}),
\begin{equation}
\bS^*\dfe = 
\left(\frac{\al_2}{\al_1+\al_2},\frac{\al_1}{\al_1+\al_2}\right),
\qquad
R_0 =\frac{\be_1\al_2+\be_2\al_1}{\al_1+\al_2}=
-\frac{f_{00}}{f_{10}}.
\label{R0_at_del=0}
\end{equation}
Reintroducing the tilde above parameters for 
$f_{20}=\delta/(\tbe_1-\tbe_2)\neq 0$, one gets instead
\begin{align}
&R_0=
\notag\\
&\frac{1}{2f_{20}}\left[
\tal_1+\tal_2+(\tbe_1+\tbe_2)f_{20}-
\sqrt{(\tal_1+\tal_2+(\tbe_1+\tbe_2)f_{20})^2-
4f_{20}(\tbe_1\tal_2+\tbe_2\tal_1+
\tbe_1\tbe_2f_{20}})\right]
\label{R0_original_variables}
\end{align}
and $\bS^*\dfe$ may then be obtained from
\Eqref{S*dfe}. Recalling the definition of tilde parameters in  \Eqref{tilde_parameters}, this makes final formulas in terms of (in total up to 17) SSISS model and demographic parameters quite involved. So, here  the reformulation as a 6-parameter  RND dynamical system becomes extremely helpful.

\begin{remark}\label{Rem_delta}
At this point a discussion on epidemiological interpretations of non-zero ranges of $\delta$ is appropriate. Assuming compartment independent birth rates, $\delta<0$ is equivalent to 
$\mu_2>\mu_1$. Epidemiologically this assumption seems reasonable if one assumes $\ga_1=0$ and absence of vaccination, in which case $S_2\equiv R$ would consist of ``removed'' individuals only. Here, as already argued bv Busenberg and van den Driessche 1990 \cite{BusDries90}, ``removed'' may not only mean ``recovered'', but also ``still being ill while no longer being infectious'' due to isolation measures. This motivates to consider a disease-related excess mortality $\mu_2>\mu_1$ as in \cite{BusDries90}, whence $\del<0$. 

To obtain a model for $\delta>0$, consider a highly infectious but non-fatal childhood disease with permanent immunity after recovery or vaccination, hence $\ga_1=0$, $\be_1>\ga_2>0$, $\al_1\geq0$ and 
$\be_2=\al_2=0$. Denoting $S_1$ the population density of susceptible children, which are assumed below sexual reproduction, we conclude 
$\nu_1=0$ and $\mu_1>0$ due to infant mortality for other reasons. Similarly, $\nu_I=0$ and $\mu_I>0$.
Finally, $S_2$ being ``recovered children or adults'', one may assume 
$0<\mu_2\approx\nu_2$. Consequently, the following assumptions seem reasonable and lead to $0<f_{20}<1$,
$$
\tbe_1=\be_1+\mu_1-\mu_I> 0\ \land\
\tbe_2=\mu_2-\nu_2-\mu_I<0\ \Longrightarrow\
0<\delta=\nu_2-\mu_2+\mu_1<\tbe_1-\tbe_2.
$$

Finally, let us discuss the case $f_{20}\geq1$. Obviously  $f_{20}=1\Leftrightarrow\delta=\tbe_1-\tbe_2\Leftrightarrow
\be_1=\be_2$ and  $f_{20}>1$ seems to imply 
$\delta>\tbe_1-\tbe_2>0$. But, according to \Eqref{tilde_parameters}, $\tbe_1-\tbe_2=\be_1-\be_2+\delta$.
So, if we insist on 
denoting $S_1$ the higher susceptible compartment, i.e. 
$\be_1>\be_2$, then necessarily $\tbe_1-\tbe_2<\delta$. Hence, the case $f_{20}>1$ should rather be interpreted by
\begin{equation}
\delta\equiv\delta_2-\delta_1<\be_2-\be_1<0\quad
\Longrightarrow\quad\delta<\tbe_1-\tbe_2<0\quad
\Longrightarrow\quad f_{20}>1.
\label{f20>1}
\end{equation}
In other words, when choosing the convention 
$\tbe_1>\tbe_2$  also for $f_{20}>1$, then one has to keep in mind that in this case  $S_1$ and $S_2$ become interchanged, 
$(\tilde{S}_1,\tilde{S}_2)=(S_2,S_1)$. Correspondingly, in this case the relations \eqref{tilde_parameters} between tilde and original parameters involve an additional permutation of indices $1\leftrightarrow2$. In summary, rewriting Eqs. \eqref{f00} - \eqref{f01} in terms of tilde parameters we must put 
$\tilde{\del}=\sign(1-f_{20})\,\del$, which gives 
\begin{equation}
f_{20}=\frac{\tilde{\del}}{\tbe_1-\tbe_2}=
\frac{\del}{\be_1-\be_2+\del}\,,\qquad
\be_1-\be_2=|f_{20}-1|(\tbe_1-\tbe_2)\,.
\label{f20(del,be)}
\end{equation}
\end{remark}

\begin{remark}
Also note, that 
typically (i.e. disregarding huge excess mortalities) birth and death rates are much smaller than transmission rates $\be_i$. So, assuming coinciding birth rates $\nu_1=\nu_2$ and interpreting 
$\Del\mu=\mu_2-\mu_1>0$ as a disease induced excess mortality as above, the scenario \eqref{f20>1} would imply 
$\Del\mu>\be_1-\be_2$. For $\be_2=0$ this would be an absolutely lethal disease. In fact, in this case a large fraction of $S_2$ would consist of isolated, very sick and soon dying people. Reintroducing 
$\ga\inv$ as the mean waiting time in $I$ and 
$\mu_2>\Del\mu>\be_1=r_0\ga$ as the mortality rate in $S_2$, the total expected survival time after infection would be smaller than $\ga\inv(1+r_0\inv)$, where 
$r_0$ denotes the basic reproduction number.

On the other hand, for moderate excess mortalities 
$\Del\mu\ll\be_1$, the assumption $f_{20}>1$ would mean 
$\be_1\approx\be_2$, in which case 
``vaccination'' and ``loss-of-immunity'' concepts would loose their meaning. So, when discussing scenarios $f_{20}\geq1$, one should keep in mind that these only apply to extraordinary cases. In particular, the case 
$f_{20}=1\Leftrightarrow\be_1=\be_2$ will mostly be disregarded from now on.

\end{remark}

Next, we look at the admissibility condition 
\eqref{physA}. In Model-1 this becomes $\ga_i\geq0$, 
$\al_i\geq 0$ and $\al_i+\theta_i\geq 0$. 

\begin{proposition}\label{Prop_admissible_Model-1}
{\em (Admissibility in Model-1)}
Necessary for $f$ in Model-1 to be admissible are the conditions in Corollary \ref{Cor_admissible_X^2}i). If these conditions hold, then 
\begin{equation}
\Bf^+\equiv[\bemin_1,\bemax_1]=[\belo_1,\behi_1],\qquad
\Bf^-\equiv[\bemin_2,\bemax_2]=[\belo_2,\behi_2].
\label{bemin_bemax_Model-1}
\end{equation}
Thus, $f$ is admissible if and only if
\begin{equation}
\belo_2<\behi_1\ \land\ 
\belo_2\leq\behi_2\leq\belo_1\leq\behi_1.
\label{admiss_Model-1}
\end{equation}
\end{proposition}
\begin{proof}
In Model-1 $A_i(I)$ and $\hf(z,I)$ depend only linearly on $I$. Thus, it suffices to check admissibility conditions at $I=0$ and $I=1$. The claim follows immediately from Proposition \ref{Prop_admissible_X^2} and Corollary \ref{Cor_admissible_X^2}.
\end{proof}
Clearly, all examples in Table \ref{Tab_examples} provide admissible examples for Model-1. In addition, they also satisfy 
$\theta_i=0$, whence $A_i(0)=A_i(1)$.  By \Eqref{Ai=hf/(bj-bi)}, in RND picture for general $X^2$-models $f$ this condition becomes
$$
A_i(0)=A_i(1)\Longleftrightarrow\hf(\be_i,0)=\hf(\be_i,1).
$$
In  $X^2$-models 
$\hf(\cdot,I)$ are at most quadratic polynomials, so denote $Y_\pm$ the zeros of $\hf(\cdot,1)-\hf(\cdot,0)$. Then by \eqref{hf}
$$\begin{aligned}
\hf(z,1)-\hf(z,0)&\equiv
(1-f_{20})z^2+(f_1(1)-f_{10})z+P_f(1)-f_{00}
\\
&=(1-f_{20})(z-Y_+)(z-Y_-).
\end{aligned}
$$
\begin{corollary}\label{Cor_Y_pm}
In $X^2$-models $f$ denote $\theta_i:=A_i(1)-A_i(0)$ and assume 
$f_{20}\equiv f_2(0)\neq 1$.
\begin{itemize}
\item[i)]
Then 
$\theta_1=\theta_2=0$, if and only if $(\be_1,\be_2)=(Y_+,Y_-)$.
\item[ii)]
If $f$ is admissible, then the SSISS fiber $\F(f)$ contains an admissible system obeying $\theta_1=\theta_2=0$, if and only if 
$(Y_+,Y_-)\in\Bf$.
\end{itemize}
\end{corollary}
In Model-1 we get in consistency with \Eqref{theta}
\begin{equation}
Y_\pm=\frac{1}{2(1-f_{20})}\left(-f_{11}\pm\sign(1-f_{20})
\sqrt{f_{11}^2+4f_{02}(1-f_{20})}\right).
\label{Ypm_Model-1}
\end{equation}
Using Eqs. \eqref{f00}-\eqref{f01} with tilde parameters, the examples of Table \ref{Tab_examples} may now straight forwardly be transformed to sub-cases of Model-1 in RND picture, see 
Table \ref{Tab_examples_Model1}. The additional constraints in 
Table \ref{Tab_examples_Model1} correspond to the parameter restrictions in Table  \ref{Tab_examples}. In particular note 
that $\tal_1\equiv f(\tbe_1,0)/(\tbe_2-\tbe_1)=0$ implies 
$\tbe_1=Y_+\in\{R_0,R_1\}$ and $\tga_1=\tal_2$ implies 
$f(\tbe_2,0)=\la-\tbe_2$.

\begin{table}[htbp!]
\caption{Identifying Examples from Table \ref{Tab_examples} as sub-cases of Model-1. In all cases  
the admissibility constraints \eqref{admiss_Model-1} are understood. Moreover, $\theta_i=0$ implies $(Y_+,Y_-)=(\tbe_1,\tbe_2)\in\Bf$. 
}
\label{Tab_examples_Model1}

$$
\begin{array}{l|cccccc|l}
&f_{00}&f_{01}&f_{02}&f_{10}&f_{11}&f_{20}&
\mathrm{additional\ constraints}
\\
\hline
\mathrm{Heth}&>0&0&0&<0&<0&0&f_{00}=f_{10}f_{11}
\\\hline
\mathrm{BuDr}&\ok&\ok&\ok&\ok&\ok&\ok&Y_+\in\{R_0,R_1\}
\\\hline
\mathrm{SIRI}&>0&>0&<0&<0&<0&0&Y_+=R_0,\ 
Y_-=f_{10}f_{11}-f_{00}+f_{01}+f_{02}
\\\hline
\mathrm{SIRS}&>0&\ok&0&<0&<0&0&
\\\hline
\mathrm{HaCa}&>0&\ok&<0&<0&<0&0&
\\\hline
\mathrm{KZVH}&>0&\ok&<0&<0&<0&0&
\\\hline
\mathrm{LiMa}&\ok&\ok&\ok&<0&\ok&0&
\\\hline
\mathrm{d'OMS}&\ok&\ok&\ok&<0&\ok&0&Y_+=R_0
\\\hline
\mathrm{AABH}&\ok&\ok&\ok&<0&\ok&0&
\\\hline
\end{array}
$$
\end{table}
\begin{remark}
As is well known, $R_0>1$ is a threshold for endemic scenarios.
However, in systems with varying population size, authors do not always use the definition $R_0=X^*\dfe$ as the value of the replacement number at the disease free equilibrium. To clarify the relations let us look at two examples in Tables \ref{Tab_examples} and \ref{Tab_examples_Model1}. 

First, in 
\cite{LiMa2002} Li and Ma considered a SIS version of Model-1 with $\be_2=0$, a compartment independent birth rate $\nu$, no vertical transmission,
$N$-dependent death rates $\mu_1=\mu_2=\mu$ and an excess mortality $\mu_I=\mu+\Del\mu_I$. 
This leads to tilde parameters
$\tbe_1=\be_1-\Del\mu_I$, $\tbe_2=-\Del\mu_I$, 
$\tal_i=\al_i+q_j\nu$, $j\neq i$,  
$\tga\equiv\tga_1+\tga_2=\ga+\nu$ and $\del=0$. 
Using \eqref{R0_at_del=0} with tilde parameters and reintroducing the time$\inv$ scaling parameter 
$\tga=\ga+\nu\neq1$ this gives
$$
R_0=\tga\inv\left[\frac{\be_1(\al_2+q_1\nu)}{\al_1+\al_2+\nu}
-\Del\mu_I\right].
$$
Instead, Li and Ma have used the formula (see Theorem 3.2 in
\cite{LiMa2002}) 
$$
R_0'=\frac{\be_1(\al_2+q_1\nu)}{(\al_1+\al_2+\nu)(\Del\mu_I+\tga)}.
$$
Obviously $R_0\geq1\LRA R_0'\geq1$, where equalities correspond to each other.

Second, to get an example with $\delta\neq0$, consider Model-1 with  $\al_i=0$ for $i=1$ or $i=2$. In this case one readily computes 
\begin{equation}
f(X,0)=f_{20}(X-\be_i)(X-\frac{\al_j}{f_{20}}-\be_j),\quad
j\neq i,
\label{al_i=0}
\end{equation}
implying $\{R_0,R_1\}=\{\be_i,\be_j+\al_j/f_{20}\}$, where ``which-is-which'' is determined by 
$f_{20}(R_1-R_0)\geq0$. 
An example  would be the SIRS model of Busenberg and van den Driessche \autocite{BusDries90} (BuDr in Tables
\ref{Tab_examples} and \ref{Tab_examples_Model1}), with parameters
$\ga_1=\al_1=\be_2=q_2=0$, a compartment independent birth rate $\nu$ and no vertical transmission. \Eqref{tilde_parameters} gives $\tal_1=0$,
$\tal_2=\al_2+\nu$, 
$\tbe_1=\be_1+\mu_1-\mu_I$, $\tbe_2=\mu_2-\mu_I$ and again
$\tga=\ga+\nu\neq1$. So, this time we get%
$$
R_0=\frac{\tbe_1}{\tga}=\frac{\be_1+\mu_1-\mu_I}{\ga+\nu}
$$
Instead the authors in \autocite{BusDries90} defined 
$R_0':=\be_1/(\ga+\nu+\mu_I-\mu_1)$ as their threshold parameter (called $R_1$ in  \autocite{BusDries90}). Again, we have $R_0\geq 1\Leftrightarrow R_0'\geq 1$, where equalities correspond to each other. 

Once again, these examples demonstrate the usefulness to have a unifying approach relating various seemingly unrelated expressions in the literature.
\end{remark}

\begin{remark}
As an example of a non-admissible Model-1 consider 
the Lotka–Volterra predator–prey system
$$
\dot{x}=ax-bxy,\qquad
\dot{y}=-cy+dxy,
$$
where $(x,y)$ are the population densities of prey and predator, respectively, and $a,b,c,d>0$. By rescaling time assume without loss $c=1$ and put $(X,I)=(dx,y)$ to obtain Model-1 with 
$f=aX-bXI$. The only non-zero coefficients being 
$f_{10}=a$ and $f_{11}=-b$, this model would be admissible if and only if $a\leq0$.
Correspondingly, this model is well known to be isomorphic to a SIR model with 
$\be_2=\ga_1=\al_2=\theta_i=0$ and negative ``vaccination rate'' 
$\al_1=-a$, which in our formalism may be verified  by putting
$(\be_1,\be_2)=(b,0)$ and applying 
$\phibe\inv$ (Eqs. \eqref{standard_model} - \eqref{delta}) to this model.
\end{remark}

\subsection{Model-2\label{Sec_Model-2}}
Let us next look at the RVM-model with an incidence dependent vaccination rate, $\kappa>0$ in \eqref{RVM}. Following 
\eqref{RVM} and using 
$\kappa IX = \kappa I[\be_1(1-I)+(\be_2-\be_1)S_2]$ we get an $X^2$-model in SSISS picture as in \eqref{dot_Si_X^2} with 
\begin{equation}
A_1=\al_1+\theta_1I+\kappa\be_1I(1-I),
\qquad
A_2=\al_2,
\qquad
B=-\kappa I.
\label{RVM(A_i,B)}
\end{equation}
To obtain a model class staying closed under the $\Gs$-action, it is convenient to start with a more general Ansatz, staying also form invariant under permutation $1\leftrightarrow 2$, i.e.
\begin{equation}
A_i=\al_i+\theta_iI+\kappa_i\be_iI(1-I),\ i=1,2,
\qquad
B=-(\ka_1+\ka_2) I.
\label{A_i_Model-2}
\end{equation}
Applying again Eqs. \eqref{P0}-\eqref{la} leads to an isomorphic RND system $\bV_f$ 
with polynomial $f=\sum_{i,j}f_{ij}X^iI^j$ given by the 7 coefficients
\begin{align}
f_{00}&=\be_1\al_2+\be_2\al_1 ,
\label{f00_Mod2}\\
f_{02}&=-(\be_1\theta_2+\be_2\theta_1+\be_1\be_2),
\\
f_{01}&=\la-f_{00}-f_{02},\qquad
\la\equiv\be_1\ga_1+\be_2\ga_2\,,
\\
f_{10}&=-\left(\al_1+\al_2\right),\\
f_{11}&=\left(\ka_1\be_2+\ka_2\be_1-\be_1-\be_2-\theta_1-\theta_2\right),\\
f_{12}&=-\kappa_1\be_2-\ka_2\be_1,\\
f_{21}&=-(\kappa_1+\ka_2).
\label{f21}
\end{align}
When introducing a demographic dynamics
to this model, we want $IX$ to remain the true incidence. So in this model we only allow tilde parameters obeying 
$\tilde{\be}_i=\be_i\geq0$ in \eqref{tilde_parameters}, implying 
$\del_I=\del_1=\del_2$ and therefore $\delta=0$. This also implies $\tilde{B}(I)=B(I)=f_2(I)$, and therefore in this model we always restrict to 
$f_{20}\equiv f_2|_{I=0}=B(0)=0$. 

To obtain the inverse $\phibe\inv(f)$ of the parameter transformation 
\eqref{f00_Mod2} - \eqref{f21},
note that if we replace $f_{11}$ by $f_{11}+f_{12}$, the last two equations in Eqs. \eqref{f00_Mod2}-\eqref{f21} decouple and the first 5 are independent of 
$\ka_i$ and coincide with Eqs. \eqref{f00}-\eqref{f01} in Model-1 for $\delta=f_{20}=0$. Thus, the previous results in Model-1 immediately take over with $f_{20}=0$ and $f_{11}$ replaced by $f_{11}+f_{12}$.
\begin{align}
\al_i&=
\frac{1}{\be_j-\be_i}(\be_i f_{10}+f_{00}),
\quad j\neq i,
\\
\theta_i&=
\frac{1}{\be_j-\be_i}
\left(\be_i^2+(f_{11}+f_{12})\be_i-f_{02}\right)
\label{theta_model2}
\\
&=\frac{1}{\be_j-\be_i}(\be_i-Y_+)(\be_i-Y_-),
\qquad j\neq i,
\notag
\\
Y_\pm&=\frac{1}{2}\left(-(f_{11}+f_{12})\pm
\sqrt{(f_{11}+f_{12})^2+4f_{02}}\right).
\label{Ypm_Model-2}
\\ 
\al_i+\theta_i&=
\frac{1}{\be_j-\be_i}(\be_i-Z_+)(\be_i-Z_-),
\quad j\neq i,
\\  \displaybreak[0]
Z_\pm&=\frac{1}{2}\left(-(f_{10}+f_{11}+f_{12})\pm
\sqrt{(f_{10}+f_{11}+f_{12})^2+4(f_{02}-f_{00})}\right)
\label{Zpm_Model-2}
\\
\ka_i&=\frac{\be_i-\La}{\be_j-\be_i}f_{21},\quad j\neq i,
\label{kappa_i}\\
\La&:=\frac{f_{12}}{f_{21}}
=\frac{\ka_1\be_2+\ka_2\be_1}{\ka_1+\ka_2}\geq 0
\label{R_kappa}
\end{align}
So, $\ka_i\geq 0$ is equivalent to $\be_2\leq \La\leq\be_1$ and $\ka_2=0$ is equivalent to 
$\be_2=\La$. 
Since eventually we want  
$\ka\equiv\ka_1>0$ and $\ka_2=\theta_2=0$, whence $\be_2=Y_-$ by Corollary \ref{Cor_Y_pm}, we impose the constraints 
$f_{21}=-\ka<0$ and $Y_-=f_{12}/f_{21}$. Since we also want 
$\{\al_1,\al_2,\theta_1\}\geq 0$, Model-2 is now defined by the following parameter restrictions
$$
\Vmodtwo :=\{\bV_f\in\VRND\mid 
f(X,I)=\sum_{i,j=0}^2 f_{ij}X^iI^j,\ (f_{ij})\in\text{Model-2}\}
$$
\begin{equation}
\text{Model-2} :=
\left\lbrace
\begin{tblr}{Q[l,m]|Q[l,m]}
\SetCell[r=2]{m}
(f_{ij})\in\RR^9 &
f_{20}=f_{22}=0 
\,\land\,f_{21}\equiv-\ka<0\,\land\,f_{12}=-\ka Y_-
\\
& 
\,\land\,\{f_{02},f_{10},f_{11}+f_{12}\}\leq0\leq
\{f_{00},\la\equiv f_{00}+f_{01}+f_{02}\}
\end{tblr}
\right\rbrace
\label{signs_model2}
\end{equation}

Note that the constraints \eqref{signs_model2} imply
$Y_\pm\geq0$, $f_{10}\neq0\Rightarrow R_0\equiv -f_{00}/f_{10}\geq0$,  and $Z_\pm\in\RR\Rightarrow Z_\pm\geq0$. Moreover, plugging $f_{12}=-\ka Y_-$ into \eqref{Ypm_Model-2} one easily verifies
\begin{lemma}
In Model-2 we have the identities
\begin{align*}
Y_+&=(\ka-1)Y_--f_{11}\geq Y_-\geq0
\\
Y_-&=
\begin{cases}
(2-2\ka)\inv\left[
-f_{11}+\sign(\ka-1)\sqrt{f_{11}^2+4(1-\ka)f_{02}}\right],
& \ka\neq1,
\\
f_{02}/f_{11},	&\ka=1.
\end{cases}
\end{align*}
\end{lemma}
This leaves us with 6 independent parameters
$(f_{00},f_{02},f_{10},f_{11},\la,\ka\}$ and the constraints 
$0\leq Y_-\in\RR$ and $f_{12}=-\ka Y_-$. So, effectively  parameter space of Model-2 in RND picture is again 6-dimensional. In SSISS picture the requirement $\theta_2=0$ restricts to $\be_2=Y_-$, keeping the SSISS fiber of $f\in\text{Model-2}$ just 1-dimensional, in consistency with parameter space of the RVM-Model \eqref{RVM(A_i,B)} being 7-dimensional%
\footnote{After normalizing time scale such that $\ga=1$.}.

From \Eqref{hf} we also get 
\begin{equation}
\begin{aligned}
\hf(z,I)&=(zf_{10}+f_{00})+I(z^2 + (f_{11}+f_{12})z-f_{02})+
I(1-I)f_{21}z(z-f_{12}/f_{21})
\\
&=f_{10}(z-R_0)+I(z-Y_+)(z-Y_-)+\ka I(1-I)z(Y_--z)
\\
&=(1-I)f_{10}(z-R_0)+I(z-Z_+)(z-Z_-)+\ka I(1-I)z(Y_--z).
\label{hf_Model-2}
\end{aligned}
\end{equation}
where in case $f_{10}=0$ the last two lines hold with $f_{10}R_0$ replaced by $-f_{00}$. Putting $\be_2=Y_-$, this verifies
$A_i(I)=\hf(\be_i,I)/(\be_j-\be_i)$, $j\neq i$.
Following the lines of Proposition \ref{Prop_admissible_X^2} and Corollary \ref{Cor_admissible_X^2}, we now analyze conditions for $f$ in Model-2 being admissible. First recall the definitions 
 \eqref{be_hi_lo}, which in Model-2 become  
\begin{equation}
\belo_2=-\infty,\qquad\behi_2=\min\{\la,R_0,Z_-\},\qquad\belo_1=\max\{\la,R_0,Z_-\},\qquad\behi_1=Z_+.
\label{behilo_Model-2}
\end{equation}
\begin{proposition}\label{Prop_admissible_Model-2}
{\em (Admissibility in Model-2)}
Consider $f$ in Model-2 and assume $Z_\pm\in\RR$. Then
\begin{equation}
\bemin_2\leq 0\leq\min\{Y_-,R_0,\la\}\leq\bemax_2\leq\behi_2.
\label{Bf-Model-2}
\end{equation}
If $Y_-\leq\{R_0,\la\}\leq Z_+$ and $Z_+>0$, then
$[0,Y_-]\subset\Bf^-$ and $[\belo_1,Z_+]=\Bf^+$, hence $f$ is admissible.
\end{proposition}

\begin{proof}
\Eqref{Bf-Model-2} follows from Corollary \ref{Cor_admissible_X^2}, $\la\geq0$ and $\hf(z,I)\geq0$ for 
$0\leq z\leq\{R_0,Y_-\}$ and $I\in[0,1]$. Consequently, $Y_-\leq\{R_0,\la\}$ implies $[0,Y_-]\subset\Bf^-$ and $\hf(z,I)\leq 0$ for all 
$z\in[\belo_1,Z_+]$ and all $I\in[0,1]$.
\end{proof}
\begin{remark}
Eventually in the RVM Model we want
$\theta_2=0$, whence $\be_2=Y_-$. Recalling 
$\bemax_2\leq R_0\leq\bemin_1$, this motivates the condition 
$Y_-\leq R_0$ above.
Also note that by the above arguments 
$Y_-\leq R_0\Rightarrow Y_-\leq Z_-$. A more systematic analysis of the relations between $Y_\pm$, $Z_\pm$ and $R_0$ will be given in Lemma \ref{Lem_YZR} in Section \ref{Sec_eadm-scenarios}.
\end{remark}

\section{$e$-Admissibility\label{Sec_eadmissible}}
In this section we analyze additional requirements in Model-1  and Model-2, which arise on top of the admissibility constraints in Definition \ref{Def_SSISS_admissible} from purely  epidemiological arguments. Correspondingly, systems obeying these more restrictive requirements will be called {\em $e$-admissible} (= epidemiologically admissible). 

\subsection{The basic concept\label{Sec_eadm-conceprt}}

As a motivation consider  Model-1 in SSISS picture, \Eqref{standard_model}, where in all epidemiological examples one naturally wants
$\theta_1\geq 0\geq \theta_2$. In fact, vaccination rates 
$\theta_1I$ should increase as prevalence $I$ increases, modeling reactive vaccination of newborns leads to 
$\vth_1>0>\vth_2$ in \Eqref{tilde+nu_vacc} and in the social behavior model the flow from $S_1$ to $S_2$ should increases with $I$, while the flow from $S_2$ to $S_1$ should  increases with $1-I$. In Model-2 \eqref{RVM(A_i,B)} we even want $\theta_2=0$. 
Note that these conventions always rely on choosing $S_1$ to be the higher susceptible compartment. 

Now, as discussed in \Eqref{f20>1} and Remark \ref{Rem_delta}, the case 
$\delta<\be_2-\be_1<0$ would yield $\tbe_1<\tbe_2$. 
Our convention 
$\tbe_1>\tbe_2$ therefore involves an additional permutation $1\leftrightarrow2$ in the definition of tilde parameters  \eqref{tilde_parameters} or \eqref{tilde+nu_vacc}%
\footnote{As discussed in Remark \ref{Rem_delta}, this case is exceptional. In reality we will mostly have
$|f_{20}|\equiv|\tilde{\del}|/(\tbe_1-\tbe_2)\ll 1$.}.
Since $\delta\neq 0$ always implies some non-vanishing birth or death rate, hence the need to use tilde-parameters, 
we are lead to the following Definition.
\begin{definition}\label{Def_eadmissible}
\begin{itemize}
\item[i)]
Admissible Model-1 SSISS systems \eqref{standard_model} are called 
{\em $e$-admissible}, if 
\begin{equation}
\begin{aligned}
\theta_1&\geq 0\geq\theta_2,	& \text{if }
\delta<\be_1-\be_2,
\\
\theta_1&\leq 0\leq\theta_2,	& \text{if }
\delta>\be_1-\be_2.
\end{aligned}
\label{eadmiss_1}
\end{equation}
\item[ii)]
Admissible Model-2 SSISS systems \eqref{RVM(A_i,B)} are called $e$-admissible, if
\begin{equation}
\theta_1\geq0=\theta_2.
\label{eadmiss_2}
\end{equation}
\item[iii)]
RND polynomials $f$ in Model-1 or Model-2 are called $e$-admissible, if there exists $\bbe\in\Bf$ such that 
$\phibe\inv f$ is $e$-admissible. In this case 
$\bbe$ is called {\em $e$-compatible} with $f$. We denote 
$$
\eBf:=\{\bbe\in\Bf\mid\bbe\text{ is }e\text{-compatible with }f\}.
$$
\end{itemize}
\end{definition}

\begin{proposition}\label{Prop_eadmiss}
{\em ($e$-Admissibility)}
Let $f$ in Model-1 or Model-2 be admissible and 
$f_{20}\neq 1$. 
Then $f$ is $e$-admissible if and only if $Y_\pm\in\RR$ and one (hence both) of the following conditions 
\eqref{Ypm_eadm_strong} - \eqref{Ypm_eadm} hold
\begin{align}
Y_-&<\min\{Y_+,\bemax_1\}\quad\, \text{and}\quad
Y_-\leq\bemax_2\leq\bemin_1\leq Y_+.
\label{Ypm_eadm_strong}
\\
Y_-&<\min\{Y_+,\behi_1\}\qquad\text{and}\quad
 Y_-\leq\behi_2\leq\belo_1\leq Y_+.
\label{Ypm_eadm}
\end{align}
In this case $Y_-\in\Bf^-$ and
$\Bf^+\equiv[\bemin_1,\bemax_1]=[\belo_1,\behi_1]$ and
\begin{equation}
\eBf=
\begin{cases}
\{(\be_1,\be_2)\in\Bf\mid Y_-\leq\be_2<\be_1\leq Y_+\},	&
\text{Model-1},
\\
\{(\be_1,\be_2)\in\Bf\mid Y_-=\be_2<\be_1\leq Y_+\},	&
\text{Model-2}.
\end{cases}
\label{eBf}
\end{equation}
\end{proposition}
\begin{proof}
For $f_{20}\neq 1$ and $\bbe\in\Bf$ we have 
\begin{equation}
\theta_i=\frac{\hf(\be_i,1)-\hf(\be_i,0)}{\be_j-\be_i}=
\frac{(1-f_{20})(\be_i-Y_+)(\be_i-Y_-)}{\be_j-\be_i},\qquad j\neq i.
\label{theta_Y_pm}
\end{equation}
Hence, \Eqref{eadmiss_1} is equivalent to 
$Y_-\leq\be_2<\be_1\leq Y_+$ and \Eqref{eadmiss_2} to
$Y_-=\be_2<\be_1\leq Y_+$, for all $(\be_1,\be_2)\in\Bf$. So, $e$-admissibility of $f$ is equivalent to $f$ being admissible and the conditions in \eqref{Ypm_eadm_strong}.
Morover, in Model-1 
$\eqref{Ypm_eadm_strong} \Leftrightarrow \eqref{Ypm_eadm}$, since there $\bemax_2=\behi_2$, 
$\bemin_1=\belo_1$ and $\bemax_1=\behi_1$, see Propositon \ref{Prop_admissible_Model-1}. In Model-2 we have in general
$\Bf^+\subset[\belo_1,\behi_1]$ and 
$\Bf^-\subset(-\infty,\behi_2]$, so 
$\eqref{Ypm_eadm_strong}\Rightarrow\eqref{Ypm_eadm}$. 
We now show that if we already know 
$\Bf\neq\emptyset$ in Model-2, then also
$\eqref{Ypm_eadm} \Rightarrow \eqref{Ypm_eadm_strong}$.

So, assume   \eqref{Ypm_eadm}.
Then \Eqref{behilo_Model-2} yields
$Y_-\leq\{Z_-,R_0,\la\}$ and $0\leq Y_-<\behi_1=Z_+$. Moreover, the admissibility assumption $\Bf\neq\emptyset$ yields $Z_+=\behi_1\geq\bemax_1\geq\bemin_1\geq\belo_1\geq
\{Z_-,R_0,\la\}$. Thus, the conditions of Proposition \ref{Prop_admissible_Model-2} are fulfilled, so we get
$[0,Y_-]\subset\Bf^-$ and 
$[\belo_1,\behi_1]=[\bemin_1,\bemax_1]$. Whence \eqref{Ypm_eadm} implies \eqref{Ypm_eadm_strong}.

In these cases $Y_-\in\Bf^-$ in Model-2 as just argued and in Model-1 by Proposition \ref{Prop_admissible_Model-1}, which yields 
$\Bf^-=(-\infty,\behi_2]$ for $f_{20}\geq0$ and 
$\Bf^-=[R_1,\behi_2]$ for $f_{20}<0$. In the last case we are left to show $R_1\leq Y_-$. Using the conditions 
$Y_-\leq\behi_2\leq \{Z_-,R_0\}\leq\belo_1\leq Y_+$, this  follows from Corollary \ref{Cor_YZR} below.
\end{proof}

\begin{remark}
Observe that in Model-2, according to  Proposition \ref{Prop_admissible_Model-2}, the conditions $Z_+>0$ and 
$Y_-\leq\{R_0,\la\}\leq Z_+$ had been sufficient for 
admissibility. According to Proposition \ref{Prop_eadmiss} they now become necessary for $e$-admissibility.
\end{remark}

Of course, in admissible SSISS systems the special case 
$\theta_i=0$ is $e$-admissible. Vice versa,  Corollary \ref{Cor_Y_pm} shows that $e$-admissible SSISS Systems are always $\Gs$-equivalent to systems with 
$\theta_1=\theta_2=0$.

\begin{corollary}\label{Cor_Ypm_in_Bf}
For $f$ in Model-1 or Model-2 assume $f_{20}\neq 1$.\begin{itemize}
\item[i)]
If $(Y_+,Y_-)\in\Bf$, then $f$ is $e$-admissible and 
$(Y_+,Y_-)\in\eBf$.
\item[ii)]
If $f$ is $e$-admissible, then the SSISS fiber $\F(f)$ contains 
precisely one representative%
\footnote{Of course, uniqueness should be understood modulo $d$-equivalence, see Definition \ref{Def_demogr_equiv}.}
obeying $\theta_1=\theta_2=0$. This representative is given by $\bbe\equiv(\be_1,\be_2)=(Y_+,Y_-)$ and it is admissible, if and only if $Y_+\in\Bf^+$.
\qed
\end{itemize}

\end{corollary}

Anticipating Corollary \ref{Cor_YZR} below, if $f_{20}\leq0$ then $e$-admissibility already implies $(Y_+,Y_-)\in\Bf$. Hence,
for any  $e$-admissible SSISS model with parameters  
$\bx=(\al_i,\be_i,\ga_i,\theta_i,\delta\leq0)$ (Model-1) or 
$\bx=(\al_i,\be_i,\ga_i,\theta_1,\ka, \delta=0)$ (Model-2),
there 
exists a unique $\bg=\bg(\bbe',\bbe)\in\Gs$, such that the 
$\bg$-transformed system $\bx'$ satisfies 
$\theta'_1=\theta'_2=0$. Here, $\be'_i$ is given by
$Y_-=\be_2'\leq\be_2<\be_1\leq\be_1'=Y_+$, which by
Lemma \ref{Lem_g.Tph} also implies 
$\bg\Tph\subset\Tph$. 

A general formula for the action of $\Gs$
on $X^2$-systems in SSISS picture is given in Appendix \ref{App_Gs.X^2}.
Specifically, for $\delta=0$ we have $\delta'=0$, $\ka'=\ka$,
$\bga'=\bg\bga$ and
$$
\al'_i=\frac{(\be_j-\be_i)\al_i-(\be'_i-\be_i)(\al_1+\al_2)}{\be'_j-\be'_i},\qquad j\neq i.
$$
For example, the Social Behavior Model \eqref{SBM} with 
$\al_1=\ka=\delta=0$ and $\al_2=-\theta_2>0$ is isomorphic to a standard model with $(\be'_1,\be'_2)=(Y_+,Y_-)$ and
$$
\al'_1=\frac{\be'_1-\be_1}{\be'_2-\be'_1}\,\theta_2,\qquad
\al'_2=\frac{\be_1-\be'_2}{\be'_2-\be'_1}\,\theta_2.
$$
\begin{remark}
This also generalizes a result in \cite{Nill_SIRS}, where it has been shown, that in the SIRS model an $I$-linear vaccination term can always be put to zero by a suitable scaling transformation. In fact, the scaling transformations of $S\equiv S_1$ introduced there are given by the subgroup of 
$\Gs$ leaving $\be_2=0$ invariant.
\end{remark}

\subsection{Classification of scenarios\label{Sec_eadm-scenarios}}
In view of the square roots in the definitions of $Y_\pm, Z_\pm$ and $R_\nu,\,\nu=0,1$, the conditions for $e$-admissibility in Proposition \ref{Prop_eadmiss}  become rather involved when expressed in terms of the coefficients $f_{ij}$.  
It will, therefore, be very helpful to realize that the subset of $e$-admissible polynomials $f$ may be reparametrized by using either two of these three pairs of roots as independent parameters. 

Moreover, in Proposition \ref{Prop_eadmiss} the condition 
$Y_-\leq \{Z_-,R_0\}\leq Y_+$ appears necessary for $e$-admissibility. In Lemma \ref{Lem_YZR} and Table \ref{Tab_YZR} below it is shown, that this condition leaves us with seven disjoint scenarios. In particular, certain  configurations are not possible, which closes the proof that $e$-admissibility implies 
$Y_-\in\Bf^-$ in  Proposition \ref{Prop_eadmiss} and also 
$Y_+\in\Bf^+$ if $f_{20}\leq0$.
Finally, this approach will lead to an alternative characterization of $e$-admissibility in Theorem \ref{Thm_eadm}.

First observe that by definition \eqref{hf} $R_\nu,Y_\pm$ and $Z_\pm$ appear as zeros of corresponding parabolas $p_R, p_Y$ and $p_Z$, respectively, where\footnote{In Model-1 we have $f_{12}=f_{21}=0$ and in Model-2 $f_{20}=0$.}
\begin{equation}
p_R(x):=\hf(x,0)\equiv f(x,0), \qquad
p_Z(x):=\hf(x,1), \qquad p_Y(x):=\hf(x,1)-\hf(x,0).
\label{parabolas}
\end{equation}
\begin{align}
p_Z(x)&= 
x^2 +(f_{10}+f_{11}+f_{12})x +(f_{00}-f_{02})
&&=(x-Z_+)(x-Z_-)\,,
\label{pZ}\\
p_Y(x)&=(1-f_{20})x^2 +(f_{11}+f_{12})x-f_{02}
&&=(1-f_{20})(x-Y_+)(x-Y_-)\,,
\label{pY}\\
p_R(x)&=f_{20}x^2+f_{10}x+f_{00}=f(x,0)
&&=\begin{cases}
f_{20}(x-R_0)(x-R_1),	&f_{20}\neq0\,,
\\
f_{10}(x-R_0),			&f_{20}=0\,.
\end{cases}
\label{pR}
\end{align}
\begin{remark}\label{Rem_f20=0}
To simplify the discussion in the following, note that the definitions imply 
\begin{equation}
\lim_{f_{20}\to 0}f_{20}R_1=-f_{10}.
\label{f20=0}
\end{equation}
So, without always mentioning explicitly, in all formulas below in case $f_{20}=0$ the product $f_{20}R_1$ is to be replaced by 
$-f_{10}$ as an independent parameter. If also $f_{10}=0$, then truly $R_0$ would be undefined. Recalling that in this case admissibility would also require $f_{00}=0$, we simply ignore this and interpret \Eqref{pR} as 
\begin{equation}
f_{20}=f_{10}=0\ \Longleftrightarrow\ p_R\equiv f|_{I=0}=0\ \Longleftrightarrow\ Y_\pm=Z_\pm
\label{f(x,0)=0}
\end{equation}
\end{remark}
\begin{lemma}\label{Lem_YZR}
Let $p_Z, p_Y, p_R$ be given by the rhs of Eqs. \eqref{pZ}-\eqref{pR}, with
$Z_\pm, Y_\pm,R_\nu,f_{20},f_{10}$ real valued.
Assume $f_{20}\neq 1$ and
\begin{equation}
Z_-\leq Z_+,\qquad Y_-\leq Y_+,\qquad 
0\leq
\begin{cases}
f_{20}(R_1-R_0),	& f_{20}\neq 0,
\\
-f_{10},	& f_{20}=0.
\end{cases}
\label{-<+}
\end{equation}

\begin{itemize}
\item[i)]
Then $p_Z-p_Y=p_R$ if and only if the following relations hold
\begin{equation}
\begin{aligned}
Z_+Z_-+(f_{20}-1)Y_+Y_- &=
\begin{cases}
f_{20}R_0R_1,	&f_{20}\neq 0
\\
-f_{10}R_0,		&f_{20}=0
\end{cases}
\\
(Z_++Z_-)+(f_{20}-1)(Y_++Y_-) &=
\begin{cases}
f_{20}(R_0+R_1),	&f_{20}\neq 0
\\
-f_{10},		&f_{20}=0
\end{cases}
\end{aligned}
\label{p_Z-p_Y=p_R}
\end{equation}
\item[ii)]
If in addition to \Eqref{p_Z-p_Y=p_R} 
$Y_-\leq \{Z_-,R_0\}\leq Y_+$, then one of the scenarios in Table
\ref{Tab_YZR} must show up. Here, for given values of $f_{20}$, in each row the conditions in columns 3-6 are equivalent.
\end{itemize}\qed
\end{lemma}

\begin{table}[htpb!]
\caption{Disjoint scenarios for $Y_-\leq \{Z_-, R_0\}\leq Y_+$
and $f_{20}\neq1$ (Lemma \ref{Lem_YZR}). The equivalences of conditions in columns 3-5 are visualized in Table \ref{Tab_graphical}. Precisely one of these scenarios must show up, if $f$ is $e$-admissible (Theorem \ref{Thm_eadm} and Table \ref{Tab_eadmissible}).}
\label{Tab_YZR} 
\renewcommand{\arraystretch}{1.4}
\rotatebox{90}{
\begin{tabular}{|l|c||c|c|c|c|}
\hline
\multicolumn{2}{|c||}{Cases} &
\multicolumn{4}{c|}{
Given $f_{20}$, these conditions are equivalent.}
\\
\cline{1-6}
No. & $f_{20}$ & $(Y_\pm,Z_\pm)$ & 
$(Y_\pm,R_\nu)$ & $(Z_\pm,R_\nu)$ & All
\\
\hline
\hline
1a 											& 
$=0$											&
$Y_\pm=Z_\pm$								&
$f|_{I=0}=0$							&
$f|_{I=0}=0$									&
$Y_\pm=Z_\pm	\,\land\,f|_{I=0}=0$
\\
\hline

\multirow{2}{*}{1b}							&
\multirow{2}{*}{$=0$}						&
$Y_-\leq Z_-\leq Y_+\leq Z_+$					&
$Y_-\leq R_0\leq Y_+$						&
$Z_-\leq R_0\leq Z_+$						& 
$Y_-\leq Z_-\leq R_0\leq Y_+\leq Z_+$ 		
\\
											&&
$Y_+<Z_+\,\lor\,Y_-<Z_-$							&
$f_{10}<0$									&
$f_{10}<0$									&
$f_{10}<0$									
\\
\hline
2 											&
$<0$				 							&
$Y_-\leq Z_-\leq Y_+\leq Z_+$					&
$R_1\leq Y_-\leq R_0\leq Y_+$					&
$R_1\leq Z_-\leq R_0\leq Z_+$					&
$R_1\leq Y_-\leq Z_-\leq R_0\leq Y_+\leq Z_+$	
\\
\hline
3a 											&
$\in(0,1)$ 									&
\multirow{2}{*}{$Y_-\leq Z_-\leq Y_+\leq Z_+$}&
\multirow{2}{*}{$Y_-\leq R_0\leq Y_+\leq R_1$}&
$Z_-\leq R_0\leq Z_+\leq R_1$					&
$Y_-\leq Z_-\leq R_0\leq Y_+\leq Z_+\leq R_1$	
\\
\cline{1-2}\cline{5-6}
3b 											&
$>1$ 										&&&
$R_0\leq Z_-\leq R_1\leq Z_+$					&
$Y_-\leq R_0\leq Z_-\leq Y_+\leq R_1\leq Z_+$	
\\
\hline
4a										&
$\in(0,1)$ 									&
\multirow{2}{*}{$Y_-\leq Z_-\leq Z_+<Y_+$}	&
\multirow{2}{*}{$Y_-\leq R_0\leq R_1<Y_+$}	&
$Z_-\leq R_0\leq R_1<Z_+$					&
$Y_-\leq Z_-\leq R_0\leq R_1<Z_+<Y_+$								
\\
\cline{1-2}\cline{5-6}
4b 											&
$>1$ 										&&&
$R_0\leq Z_-\leq Z_+<R_1$					&
$Y_-\leq R_0\leq Z_-\leq Z_+< R_1<Y_+$			
\\
\hline
\end{tabular}
}
\renewcommand{\arraystretch}{1}
\end{table}
Lemma \ref{Lem_YZR} is proven in Appendix \ref{App_eadmissibility}.
The following Corollary closes the proof of $Y_-\in\Bf^-$ in Proposition \ref{Prop_eadmiss}.
\begin{corollary}\label{Cor_YZR}
For $f$ in Model-1 or Model-2 assume 
$Y_-\leq \{Z_-,R_0\}\leq Y_+$. Then  
$f_{20}<0\Rightarrow R_1\leq Y_-$ and 
$f_{20}\leq0\Rightarrow Y_+\leq Z_+$.
\end{corollary}
\begin{proof}
This can immediately be read off from Table \ref{Tab_YZR}.
\end{proof}

Modulo Remark \ref{Rem_f20=0}, using Eqs. 
\eqref{pZ} - \eqref{p_Z-p_Y=p_R} we now obtain three equivalent reparametrizations of $f$ as functions of 
$(\bW,\la,\ka,f_{20})$, where $\bW=(Y_\pm,Z_\pm)$ or 
$\bW=(Y_\pm,R_\nu)$ or $\bW=(Z_\pm,R_\nu)$. 

\begin{equation}
\begin{aligned}
f_{00}&=Z_+Z_-+(f_{20}-1)Y_+Y_- 	&&=f_{20}R_0 R_1
\\
f_{02}&=(f_{20}-1)Y_+Y_-			&&=f_{20}R_0 R_1-Z_+Z_-
\\
f_{10}&=(1-f_{20})(Y_++Y_-)-(Z_++Z_-)	&&=-f_{20}(R_0+R_1)
\\
f_{11}&=(f_{20}-1)(Y_+ +Y_-)-f_{12} 
	&&=f_{20}(R_0+R_1)-(Z_++Z_-)-f_{12}
\\
f_{01}&=\la-f_{00}-f_{02}		
\\
f_{12}&=-\ka Y_-					
\\
f_{21}&=-\kappa					
\\
f_{20}&=f_{20}					
\\
f_{22}&\equiv 0						
\end{aligned}
\label{XY-parameters}
\end{equation}
Here, we get Model-1 for $\ka=0$  and Model-2 for  
$f_{20}=0$, $\ka>0$, $\la\geq 0$, $Y_-\geq0$ and 
$f_{00}\geq0\geq f_{10}$. Also, the case $f|_{I=0}=0$ is covered by $f_{20}=0$ and $Y_\pm=Z_\pm$, in which case $R_\nu$ remain undefined.

\begin{remark}
The true reason for the identities \eqref{XY-parameters} lies  the fact that $\hf$ is at most quadratic in $x$ and $I$,
$$
\hf(x,I)=Ip_Z(x)+(1-I)p_R(x)+\ka I(1-I)x(Y_--x),
$$
and that $f$
may uniquely be reconstructed from $\hf$ and 
$\la\equiv f(0,1)$. Indeed, the definitions \eqref{laP0Q0} and \eqref{hf} imply
$$
f(X,I)=(1-I)\,\hf(\frac{X}{1-I}, I)+
\la I-\frac{X^2I}{1-I}.
$$
\end{remark}

\begin{remark}\label{Rem_f20=1}
Formally the case $\ka=0$ and $f_{20}=1$ could also be treated here by putting $(f_{20}-1)Y_+Y_-=(f_{20}-1)(Y_++Y_-)=0$. Note that this would imply $f_{11}=f_{02}=0$, whence 
$\theta_1=\theta_2=0$ by \Eqref{theta}.
\end{remark}

For $f_{20}\neq1$ $e$-admissible scenarios are now completely characterized as follows. Theorem \ref{Thm_eadm} generalizes previous results obtained for $f_{20}=0$ and $\ka=0$ in \cite{Nill_Symm1}.

\begin{theorem}\label{Thm_eadm}
{\em (Classification of $e$-admissible scenarios)}\\
Assume Eqs. \eqref{-<+} - \eqref{p_Z-p_Y=p_R} with 
$f_{20}\neq 1$ and let $f$ be given by \Eqref{XY-parameters},
where $\kappa=0$ in Model-1 and $f_{20}=0$, $\ka>0$ and 
$Y_-\geq0$ in Model-2. Then $f$ is $e$-admissible, if and only if the following conditions hold
\begin{itemize}
\item[i)]
$Y_-<Y_+$.
\item[ii)]
One of the scenarios in Table \ref{Tab_YZR} is realized.
\item[iii)]
In this scenario one (hence both) of the corresponding conditions in columns 3 and 4 of Table \ref{Tab_eadmissible} hold.
\end{itemize}
\end{theorem} 
\begin{table}[htpb!]
\caption{Assuming $Y_-<Y_+$, $f_{20}\neq1$ and one of the scenarios in Table \ref{Tab_YZR}, this table provides two equivalent conditions for $e$-admissibility
(Theorem \ref{Thm_eadm}).}
\label{Tab_eadmissible} 
\renewcommand{\arraystretch}{1.4}
\begin{tabular}{|l|c||c|c|}
\hline
\multicolumn{2}{|c||}{Scenario} &
\multicolumn{2}{c|}{Equivalent conditions for $e$-admissibility.}
\\
\hline
No. & $f_{20}$ & $\la$ & $Y_\pm$
\\
\hline\hline
1a,b											&
$=0$											&
\multirow{3}{*}{$Y_-\leq\la\leq Y_+$}			&
\multirow{3}{*}{$Y_\pm\in\Bf^\pm$}			
\\
\cline{1-2}
2 											&
$<0$				 							&&
\\
\cline{1-2}
3a,b											&
$>0$ 										&&
\\
\hline
4a											&
$\in(0,1)$ 									&
$Y_-\leq\la\leq R_1\,\land\,Y_-<R_1$			&
$\Bf\neq\emptyset\,\land\,Y_-\in\Bf^-$
\\
\cline{1-3}
4b											&
$>1$											&
$Y_-\leq\la\leq Z_+\,\land\,Y_-<Z_+$			&
$Y_+>\behi_1>Y_-$				
\\
\hline
\end{tabular}
\renewcommand{\arraystretch}{1}
\end{table}
\begin{remark}
Due to $f_{20}=0$, in Model-2 we are restricted to Scenarios 1a,b. In these cases the sign restrictions for Model-2 in
\Eqref{signs_model2} follow from $Y_-\geq0$ and Table \ref{Tab_YZR}.
\end{remark}
\begin{proof}
Throughout, in all formulas below omit $R_0$ in Scenario 1a.

First assume $f$ $e$-admissible, whence 
$\emptyset\neq\eBf\subset\Bf$. 
Applying Proposition \ref{Prop_eadmiss} we get 
$Y_-<\min\{Y_+,\behi_1\}$, $Y_-\in\Bf^-$ and 
$Y_-\leq\behi_2\leq\{R_0,Z_-,\la\}\leq\belo_1\leq Y_+$. By Lemma \ref{Lem_YZR}, one of the scenarios in Table \ref{Tab_YZR} must be realized.
By inspection, in scenarios 1a,b, 2 and 3a,b this implies $Y_+<\behi_1$, whence 
$Y_+\in\Bf^+$. Hence, in these scenarios the conditions of columns 3 and 4 in Table \ref{Tab_YZR} are both satisfied.

Scenarios 4a,b require $f_{20}>0$, hence they only show up in Model-1, in which case $\Bf^-=(-\infty,\behi_2]$ and 
$\Bf^+=[\belo_1,\behi_1]$, see Proposition \ref{Prop_admissible_Model-1}. 
Now assume scenario 4a, implying $Y_-<\behi_1=R_1<Y_+$ by inspection and
$\la\leq\belo_1\leq\behi_1=R_1$. So, again the conditions of columns 3 and 4 in Table \ref{Tab_YZR} are both satisfied.
Similarly, in scenario 4b we have $Y_-<\behi_1=Z_+<Y_+$ and
$\la\leq\belo_1\leq\behi_1=Z_+$, which again verifies columns 3 and 4 in Table \ref{Tab_YZR}.
Thus, we have proven i), ii) and the conditions in columns 3 and 4 of Table \ref{Tab_eadmissible}.

Conversely, assume now i)-ii). Then clearly column 3 of Table \ref{Tab_eadmissible} follows from column 4. Indeed, if non-empty then $\Bf^-\leq\la\leq\Bf^+$ by definition and
$\la\leq\belo_1\leq\behi_1=R_1<Z_+<Y_+$ in scenario 4a and 
$\la\leq\belo_1\leq\behi_1=Z_+<R_1<Y_+$ in scenario 4b.
So we are left to show that i), ii) and column 3 of Table \ref{Tab_eadmissible} imply $e$-admissibility.

Given these assumption, inspection of Table \ref{Tab_YZR} verifies
$\belo_2\leq Y_-\leq\behi_2\equiv\min\{R_0,Z_-,\la\}$ in all scenarios. By Propositions \ref{Prop_admissible_Model-1} and
\ref{Prop_admissible_Model-2}, respectively, in both models and all scenarios we conclude $Y_-\in\Bf^-$. Moreover, inspection of 
Scenarios 1a,b, 2, or 3a,b verifies 
$\max\{R_0,Z_-,\la\}\equiv\belo_1\leq Y_+\leq\behi_1\leq Z_+$, whence also $Y_-<Z_+$. Propositions \ref{Prop_admissible_Model-1} and \ref{Prop_admissible_Model-2} then imply $\Bf^+=[\belo_1,\behi_1]\ni Y_+$. Hence, in these scenarios $(Y_+,Y_-)\in\Bf$ and by Corollary \ref{Cor_Ypm_in_Bf} $f$ is $e$-admissible.

We are left with scenarios 4a,b, which may only show up in Model-1. In Scenario 4a we conclude 
$\max\{R_0,Z_-,\la\}\equiv\bemin_1\leq R_1=\bemax_1<Y_+$ and 
$Y_-<R_1=\bemax_1$, hence $(Y_-,R_1)\in\Bf$. Similarly, in Scenario 4b
$\max\{R_0,Z_-,\la\}\equiv\bemin_1\leq Z_+=\bemax_1<Y_+$ and 
$Y_-<Z_+=\bemax_1$, hence $(Y_-,Z_+)\in\Bf$. Finally, as already argued, $Y_-\in\Bf^-$, so in both scenarios $f$ is admissible,
$Y_-\leq\bemax_2\leq\bemin_1<Y_+$ and 
$Y_-<\min\{Y_+,\bemax_1\}$. By Proposition \ref{Prop_eadmiss} this assures $e$-admissibility of $f$.
\end{proof}
\subsection{Examples\label{Sec_eadm_examples}}
To get a more intuitive insight into this classification let us fill these scenarios with epidemiological examples.
According to Corollary \ref{Cor_Ypm_in_Bf} and Table \ref{Tab_eadmissible}, all admissible SSISS models of type Model-1 \eqref{standard_model} or Model-2 \eqref{RVM(A_i,B)} with 
$\theta_1=\theta_2=0$ must fall into one of Scenarios 1a,b, 2 or 3a,b. Scenarios 1a,b are obtained if $\del_1=\del_2$, hence for all examples in Table \ref{Tab_examples} obeying $\del=0$ and also for Model-2. Scenario 1a additionally requires 
$f|_{I=0}=0$, hence a continuum of disease-free equilibria like for example in the classic SIR model.
For $\del_2>\del_1$ we get Scenario 3a, for 
$0<\del_1-\del_2<\be_1-\be_2$ Scenario 2 and for 
$0<\be_1-\be_2<\del_1-\del_2$ Scenario 3b,
see \eqref{f20>1} and the discussion in Remark \ref{Rem_delta}
\footnote{Also note that the case $\del_1-\del_2=\be_1-\be_2$ would give $\tbe_1 = \tbe_2$, which has been disregarded throughout.}.
So, the model of a childhood disease discussed in Remark \ref{Rem_delta} provides an example of Scenario 3a and the model of BuDr \cite{BusDries90} belongs to Scenario 2, if 
$0<\mu_2-\mu_1<\be_1$ and to Scenario 3b, if 
$0<\be_1<\mu_2-\mu_1$ (recall $\del_1-\del_2=\mu_2-\mu_1>0$ in BuDr).%

In Scenarios 4a,b we have $Y_+>\bemax_1$, so they necessarily require $\tilde{\theta}_1/(1-f_{20})>0$. 
In fact, keeping $\tilde{\theta}_2=0$ by choosing 
$\tbe_2=Y_-$%
\footnote{Recall $Y_-\in\Bf^-$ by Proposition \ref{Prop_eadmiss}, so this stays consistent with admissibility.},
\Eqref{theta_Y_pm} implies 
$$
\tilde{\theta}_1=(Y_+-\tbe_1)(1-f_{20}).
$$
Here 
$\tilde{\theta}_1=\theta_1$ for $f_{20}<1$ and 
$\tilde{\theta}_1=\theta_2$ for $f_{20}>1$, see Remark \ref{Rem_delta}.
Hence, Scenario 4a may be obtained from 3a (e.g. the childhood disease) by adding an $I$-linear vaccination rate and keeping all other parameters invariant. 
$$
Y_+>R_1\ \Longleftrightarrow\ 
\theta_1=\tilde{\theta}_1>(R_1-\tbe_1)(1-f_{20})
$$
To see this use that $R_0$ and $R_1$ are the zeros of 
$f|_{I=0}$, whence independent of $\tilde{\theta}_i$, and that admissibility guarantees 
$$
\tbe_1\leq\bemax_1=\behi_1=
\begin{cases}
Z_+\leq R_1	&\text{in Scenario 3a},
\\
R_1<Z_+	&\text{in Scenario 4a}.
\end{cases}
$$
Thus, increasing $\tilde{\theta}_1$ as above keeps 
$R_1-\tbe_1>0$ and $\tbe_2=Y_-\leq R_0$ invariant and increases 
$Y_+, Z_+$ from $Y_+\leq Z_+\leq R_1$ to
$R_1<Z_+<Y_+$. Hence, $(\tbe_1,\tbe_2)$ stays in $\Bf$ and  we land in Scenario 4a, while $e$-admissibility is assured by Table 
\ref{Tab_eadmissible}.

In principle, the same strategy could be applied to go from Scenario 3b to 4b by lowering $\tilde{\theta}_1$ from 
$\tilde{\theta}_1=0$ to
$\tilde{\theta}_1<0$. However, $f_{20}>1$ implies the permutation $1\leftrightarrow2$, whence
$\tilde{\theta}_1=\theta_2$, which in Section \ref{Sec_examples} had been introduced as a social behavior parameter to model a transition flow from $S_2$ to $S_1$ 
with rate $\theta_2(I-1)S_2>0$. The interpretation was to describe people increasing their contact behavior as prevalence $I$ goes down%
\footnote{For simplicity, in this discussion we disregard the possible interpretation of $\theta_1>0>\theta_2$ describing reactive vaccination rates of newborns \eqref{tilde+nu_vacc}.}. 
So, in this picture the two transmission coefficients $\be_1>\be_2$ are due to different contact behaviors in society rather than a consequence of vaccination vs. loss-of-immunity effects.

In this light, it seems more appropriate to stay with the pure social behavior model (SBM) to begin with. This model had been defined in Section \ref{Sec_examples} by
$\be_1>\be_2\geq0$, $\ga_2>0$, 
$\theta_1\geq0\geq\theta_2$ and $\ga_1=\al_1=\al_2+\theta_2=0$.
Adding demographic dynamics to this model, it is natural to assume newborns from $S_i$ to land in $S_i$ and non-infected newborns from infected mothers to land in the ``careful'' compartment $S_2$. Thus, we assume a birth matrix
$\nu_{ji}=\nu\del_{ji}$ and $\nu_{iI}=\nu\delta_{i2}$ for 
$i\neq j =1,2$, implying $\tal_i=\al_i$, $\tthe_i=\theta_i$,
$\tga_1=\ga_1=0$ and 
$\tbe_i=\be_i+\mu_i-\mu_I$, see Appendix \ref{App_birth+death}
(again, if $f_{20}>1$ then tilde parameters get an additional permutation $1\leftrightarrow 2$). This model may now in principle appear in all scenarios of Theorem \ref{Thm_eadm}.

\begin{corollary}\label{Cor_SBM}
Consider the social behavior model $($SBM$)$ with birth and death rates as described above and put $\Del\mu:=\mu_2-\mu_1$. Then 
\begin{equation}
\Del\mu=(\tbe_1-\tbe_2)f_{20}\sign(f_{20}-1)
=\frac{(\be_1-\be_2)f_{20}}{f_{20}-1}
\label{f20(Delmu,be)}
\end{equation}
and this model appears in all seven $e$-admissible Scenarios as shown in Table \ref{Tab_SBM}.
\end{corollary}

\begin{table}[htpb!]
\caption{$e$-Admissible scenarios of the social behavior model
with demographic dynamics (Corollary \ref{Cor_SBM}). Here $\Del\mu:=\mu_2-\mu_1$, 
$\theta_1\geq0\geq\theta_2$, $\ga_1=\al_1=\al_2+\theta_2=0$ and $\tbe_i=\be_j+\mu_j-\mu_I$, where $j=i$ if $f_{20}<1$ and $j=3-i$ if  $f_{20}>1$.}
\label{Tab_SBM} 
\begin{tblr}{
colspec = {|Q[l,m]|Q[c,m]||Q[c,m]|Q[c,m]|Q[c,m]|Q[c,m]|Q[c,m]|
Q[c,m]|Q[c,m]|}, 
}
\hline
\SetCell[c=2]{c}Scenario	&&
\SetCell[c=2]{c} SSISS picture	&&
\SetCell[c=5]{c}RND picture		&&&&
\\
\hline
No. & $f_{20}$ & $\Del\mu$ & $\theta_i$	&
$R_0$ & $R_1$ & $Z_+$	& $Z_-$	& $\la$
\\
\hline
\hline
1a 								& 
\SetCell[r=2]{c} $=0$			&
\SetCell[r=2]{c}$\Del\mu=0$		&
$\theta_2=0$						&
\SetCell[c=2]{c}$f|_{I=0}=0$		&&
\SetCell[r=6]{c}$\tbe_1+\theta_1$	&
\SetCell[r=6]{c}$\tbe_2$			&
\SetCell[r=6]{c}$\tbe_2$
\\
\hline
1b							&&&
$\theta_2<0$					&
$\tbe_1$						&
{$\displaystyle \lim_{f_{20}\to 0}-f_{20}R_1$ \\ 
$=f_{10}=\theta_2$}
&&&	
\\
\hline
2							&
$<0$							&
{$0<\Del\mu$ \\ $<\be_1-\be_2$}	&&
$\tbe_1$						&
$\tbe_2-\theta_2/f_{20}$		&&&
\\
\hline
3a							&
\SetCell[r=3]{c}$\in(0,1)$	&
\SetCell[r=3]{c}$\Del\mu<0$	&
{$\theta_2\leq$\\$\Del\mu-f_{20}\theta_1$}	&
$\tbe_1$						&
$\tbe_2-\theta_2/f_{20}$		&&&
\\
\hline
\SetCell[r=2]{l}
4a							&&&
{$\Del\mu>\theta_2>$\\$\Del\mu-f_{20}\theta_1$}	&
$\tbe_1$						&
$\tbe_2-\theta_2/f_{20}$		&&&
\\
\hline
						&&&
$\theta_2\geq\Del\mu$	&
$\tbe_2-\theta_2/f_{20}$	&
$\tbe_1$					&&&
\\
\hline
3b												&
\SetCell[r=3]{c}$>1$								&
\SetCell[r=3]{c}{$\Del\mu>$\\$\be_1-\be_2$\\$>0$}	&
{$f_{20}\theta_1\geq$ \\ $\Del\mu-\theta_2$}		&
\SetCell[r=3]{c}$\tbe_2$							&
\SetCell[r=3]{c}$\tbe_1-\theta_2/f_{20}$			&
$\tbe_2+\theta_1$								&
$\tbe_1$											&
\SetCell[r=3]{c}$\tbe_1$
\\
\hline
\SetCell[r=2]{l} 
4b							&&&
{$\Del\mu<f_{20}\theta_1$ \\ $<\Del\mu-\theta_2$}		&&&
$\tbe_2+\theta_1$						&
$\tbe_1$									&
\\
\hline
							&&&
$f_{20}\theta_1\leq\Del\mu$	&&&
$\tbe_1$						&
$\tbe_2+\theta_1$			&
\\
\hline
\end{tblr}
\end{table}

\begin{table}[htpb!]
\caption{Reconstructing social behavior models in SSISS picture
from arbitrary $e$-admissible Model-1 parameters as specified in Theorem \ref{Thm_eadm}. In Scenarios 1b, 2a,b and 3a,b the solution is unique and in Scenarios 4a,b there are exactly two solutions (unless $R_0=R_1$ or $Z_+=Z_-$, respectively). 
}
\label{Tab_SBM_inv} 
\begin{tblr}{
colspec = {|Q[l,m]|Q[c,m]||Q[c,m]|Q[c,m]|Q[c,m]|Q[c,m]|Q[c,m]|}, 
}
\hline
\SetCell[c=2]{c}Scenario	&&
\SetCell[c=5]{c}SBM model in SSISS picture &&&&
\\
\hline
No.		& $f_{20}$	& $\Del\mu$	& 
$\tbe_1$	& $\tbe_2$	& $\theta_1$	& $\theta_2$ 
\\
\hline
\hline
1a 								& 
\SetCell[r=2]{c}	$=0$				&
\SetCell[r=6]{c}\rotatebox{90}{$f_{20}(\tbe_2-\tbe_1)$}	&	
$Z_+-\theta_1$					&
\SetCell[r=6]{c}	$Z_-$			&
$\in[0,Z_+-Z_-)$					&
$0$								
\\
\hline
1b				&&&
$R_0$			&&
$Z_+-R_0$		&
$f_{10}$			
\\
\hline
2								&
$<0$								&&
$R_0$							&&
$Z_+-R_0$						&
$f_{20}(Z_--R_1)$	
\\
\hline
3a							&
\SetCell[r=3]{c}$\in(0,1)$	&&
$R_0$						&&
$Z_+-R_0$					&
$f_{20}(Z_--R_1)$
\\
\hline
\SetCell[r=2]{l}	
4a					&&&
$R_0$				&&
$Z_+-R_0$			&
$f_{20}(Z_--R_1)$
\\
\hline
					&&&
$R_1$				&&
$Z_+-R_1$			&
$f_{20}(Z_--R_0)$	
\\
\hline
3b						&
\SetCell[r=3]{c}$>1$		&
\SetCell[r=3]{c}\rotatebox{90}{$f_{20}(\tbe_1-\tbe_2)$}	&
$Z_-$					&
\SetCell[r=3]{c}$R_0$	&
$Z_+-R_0$				&
$f_{20}(Z_--R_1)$		
\\
\hline
\SetCell[r=2]{l} 
4b						&&&
$Z_-$					&&
$Z_+-R_0$				&
$f_{20}(Z_--R_1)$		
\\
\hline
						&&&
$Z_+$					&&
$Z_--R_0$				&
$f_{20}(Z_+-R_1)$		
\\
\hline
\end{tblr}
\end{table}

\begin{proof}
Assuming compartment independent birth rates yields $\Del\mu=-\delta$, hence \eqref{f20(Delmu,be)} follows from \eqref{f20(del,be)}. Equivalence of columns 2 and 3 in Table \ref{Tab_SBM} is an immediate consequence. Column 9 follows from $\la=\tbe_1\tga_1+\tbe_2\tga_2$ and $\tga_1=\ga_1=0$ in rows 1-5 and $\tga_2=\ga_1=0$ in rows 6-7.
The remaining formulas in Table \ref{Tab_SBM} follow from the identities 
($i\neq j\in\{1,2\}$)
$$
\tal_i=\tal_j+\tthe_j=0\ \Longrightarrow\
\begin{cases}
Z_+ 		&\hspace{-7pt}= \max\{\tbe_j,\tbe_i+\tthe_i\},
\\
Z_- 		&\hspace{-7pt}= \min\{\tbe_j,\tbe_i+\tthe_i\},
\\
f|_{I=0}	&\hspace{-7pt}=(X-\tbe_i)(f_{20}(X-\tbe_j)+\tthe_j).
\end{cases}
$$
\end{proof}

Observe that Table \ref{Tab_SBM} defines the map 
$(\Del\mu,\tbe_1,\tbe_2,\theta_1,\theta_2)\mapsto
(f_{20},R_0,R_1,Z_+,Z_-)$
from SSISS picture to RND picture, subdivided into 9 parameter ranges as specified in columns 3 and 4. Comparing with the target ranges in column 5 of Table \ref{Tab_YZR}, the restricted maps are easily checked to be 
one-to-one in each row of Table \ref{Tab_SBM}, except scenario 1a. The inverse assignments are displayed in Table \ref{Tab_SBM_inv}. 
Thus, we also conclude

\begin{corollary}\label{Cor_SBM_inv}
As shown in Table \ref{Tab_SBM_inv},
any $e$-admissible Model-1 is isomorphic to an SBM
\footnote{To be precise, ``SBM'' means $d$-equivalence class of social behavior models (see Definition \ref{Def_demogr_equiv}).}.
This SBM is unique
for Scenarios 1b, 2, 3a,b and there are exactly two%
\footnote{Unless $R_0=R_1$ or $Z_+=Z_-$, respectively.}
in Scenarios 4a,b.
\end{corollary}

\section{Summary and Outlook\label{Sec_conclusion}}

In this paper we have introduced the notion of a 
{\em replacement number dynamics} (RND) given by a two dimensional dynamical system of the form
$$
\dot{I}=(X-1)I,\qquad\dot{X}=f(X,I),\qquad
(X,I)\in\RR\times\RRN\,.
$$
Choosing two transmission coefficients $\be_1>\be_2$, any such system uniquely maps to an isomorphic {\em SSISS model}, i.e. an 
abstract SIR-type model with two susceptible and one infectious compartment and population densities $(S_1,S_2,I)$ satisfying  $S_1+S_2+I=1$ and
$$
\dot{S}_i=(\ga_i-\be_iS_i)I-\Om_iS_i+\Om_jS_j,
\qquad j\neq i\in\{1,2\},\ \be_1>\be_2.
$$
Here, a normalized recovery rate $\ga=\ga_1+\ga_2=1$ is understood and the ambiguity 
$\Om_i\mapsto \Om_i+KS_j$, $j\neq i$, 
is removed by requiring $\be_1\Om_2+\be_2\Om_1$ to be only $I$-dependent (``canonical gauge condition'').
Given $\bbe=(\be_1,\be_2)$, the one-to-one relation 
$\phibe:(\bOm,\bga)\mapsto f$ between SSISS picture and RND picture is determined by identifying $X$ as replacement number in the SSISS system via the standard incidence formula,
$$
X=\be_1S_1+\be_2S_2.
$$
By the methods of \cite{Nill_Redundancy}, extending this correspondence to negative values $\be_i<0$ takes care of incorporating demographic dynamics with up to 9 parameters including compartment dependent birth and death rates, vertical transmission and an $I$-linear vaccination rate of newborns.

Fixing $f$ and varying $\be_i$ generates a family of isomorphic SSISS systems, the {\em SSISS fiber $\F(f)$}.
The subgroup $\Gs\subset\GL(\RR^2,+)$ of $2\times2$-matrices acting on $(S_1,S_2)$ and leaving $S_1+S_2$ invariant acts freely and transitively on SSISS fibers, such that the set of 
$\C^\infty$-SSISS systems becomes a principle $\Gs$-fiber bundle with base space given by the family of RND systems with $f\in\C^\infty(\RR\times\RRN)$.

Physical and epidemiological considerations require SSISS model parameters to obey certain {\em admissibility constraints}. It turns out that these are not conserved by the $\Gs$-action.
So, given $f$ we call $\bbe$ {\em compatible with $f$}, if 
$\phibe\inv(f)$ defines an admissible SSISS system. Denoting 
$\Bf$ the set of $f$-compatible values $\bbe$, then $f$ is called admissible if $\Bf\neq\emptyset$.

Admissibility in particular guarantees the {\em physical triangle} 
$$\Tph=\{(S_1,S_2,I)\in\RRN^3\mid S_1+S_2+I=1\}$$
to stay forward invariant under the SSISS dynamics. 
The images of $\Tph$ under the transformations from SSISS to RND picture define triangles $\T(\bbe)$ with corners 
$\{(\be_2,0), (\be_1,0), (0,1)\}$ in $(X,I)$-space. Consequently, in RND picture $\T(\bbe)$ stays forward invariant for all $\bbe\in\Bf$.

If $\bbe\in\Bf$ and $f$ is at most quadratic in $X$, we have shown that there exist no periodic solutions
homoclinic loops or oriented phase polygons in the interior 
$\T(\bbe)\setminus\partial\T(\bbe)$. Moreover, in such ``$X^2$-models'' forward invariance of the line segment 
$\T(\bbe)\cap\{I=0\}$ implies the existence of a unique 
{\em disease-free equilibrium} 
$(X^*,I^*)\dfe=(R_0,0)\in\T(\bbe)$ obeying
$f(X,0)=(R_0-X)\rho(X)$ and $\rho(R_0)\geq0$. In SSISS picture $R_0$ becomes the {\em reproduction number} in the version of van den Driessche and Watmough \cite{Driesche_Watmough2000, Driesche_Watmough2002}. By construction, $R_0$ must be $\Gs$-invariant and 
$R_0\in[\be_2,\be_1]$ for all $\bbe\in\Bf$.

We have then studied in more detail two specific classes of RND functions $f$, called Model-1 and Model-2, given by 6-parameter polynomials at most quadratic in $X$ and $I$,
$$
f=\sum_{i,j=0}^2 f_{ij}X^iI^j. 
$$
Conditions for admissibility in these models are completely specified by determining $\Bf$ in terms of the coefficients $f_{ij}$ only. 
In SSISS picture Model-1 covers and extends a large class of prominent SI(R)S-type models in the literature, with in total up to 17 parameters%
\footnote{Including the above mentioned 9 demographic parameters.}.
In addition to standard vaccination and loss-of-immunity parameters, these also include parameters describing {\em reactive vaccination} or {\em reactive contact behavior}
in response to published prevalence data $I$. Reactive vaccination rates proportional to {\em incidence} are described in Model-2.

Finally, on top of admissibility we have introduced the notion of a so-called
{\em $e$-admissibility} condition by requiring reactive behavior parameters $\theta_i$ to obey
$\theta_1\geq0\geq\theta_2$, which turns out to be a natural epidemiological constraint.
Following the same philosophy as before, $f$ in Model-1 or Model-2 is called 
$e$-admissible, if the SSISS fiber $\F(f)$ contains an $e$-admissible representative. We conclude with a complete classification of $e$-admissible polynomials $f$, again without relying on any SSISS model description. As a corollary, SSISS models in a large $e$-admissibility class become isomorphic to ``conventional'' models (i.e. with $\theta_i=0$) by applying a suitable $\Gs$-transformation. Conversely, for any $e$-admissible polynomial $f$ the SSISS fiber $\F(f)$ contains at least one $e$-admissible representative given by a pure reactive  social behavior model.

In summary, this paper clarifies the connection between an
abstract class of SSISS dynamical systems and the apparently  simpler world of RND systems. Epidemiological admissibility conditions on the SSISS model side are transformed into parameter restrictions on RND functions $f$. 
In this way RND systems with two types of quadratic 6-parameter polynomials $f$
cover a large class of models in the literature, possibly including demographic dynamics and/or reactive social behavior.
Pre-images under this reduction induce a partitioning of SSISS
systems into equivalence classes under $\Gs$-transformations (SSISS fibers $\F(f)$) and demographic equivalence.

\smallskip
Let me close with a short outlook to Part II of this work \cite{Nill_SSISS_2}, where studying existence and stability properties of equilibrium points (EP) will considerably simplify in RND picture.
In fact, we just have to look at the zeros of two parabolas, $f|_{I=0}$ (disease-free EP) and 
$f|_{X=1}$ (endemic EP), while taking care that these lie in some $f$-compatible triangle 
$\T(\bbe)$. Global and/or local stability properties of these EPs will then follow from absence of periodic solutions 
(Poincaré–Bendixson theorem) and eigenvalues of the Jacobian.

An endemic EP in 
$\T(\bbe)$ is given by a solution of $f(1,I^*)=0$ obeying $0<I^*<1$ and $\be_2\leq(1-I^*)\inv\leq\be_1$. For $R_0>1$ the existence will be a simple consequence of forward invariance of $\T(\bbe)$. In Model-1 or Model-2 endemic EPs must lie in the smallest $f$-compatible triangle $\T(\bbe)$. Generically, the second zero of $f|_{X=1}$ will lie outside all $f$-compatible triangles, so it will be nonphysical in SSISS picture, i.e. $S_1^*<0$ or $S_2^*<0$ or $I^*<0$. 
However, unexpectedly, there are special cases allowing also 
{\em bi-endemic scenarios} for $R_0>1$.
In these cases the second EP lies in the boundary
$\partial\T(\bbe)$, corresponding in SSISS picture to $S_1^*=0$ or $S_2^*=0$. 

For $f_{20}=0$ this will provide a unifying and geometrically transparent approach covering most results on existence and stability of endemic EPs in the above mentioned class of models,
while simultaneously also applying to models simulating prevalence and/or incidence dependent reactive social behavior.
For $f_{20}\neq0$ this approach will close previously open cases like the model of AABH  \autocite{AvramAdenane2022} for differing mortality rates, $\mu_1\neq\mu_2$, or the model of BuDr  \cite{BusDries90} extended to a positive transmission rate $\be_2>0$.

For $R_0\leq1$ one may derive bounds in parameter space to verify and extend the well-known appearance of endemic bifurcation and bi-endemic scenarios in models with two susceptible compartments  
\autocite{DerrickDriessche, Had_Cast, Had_Dries, AvramAdenane2022, AvramAdenane_et_al}. 
In addition, there will also be unexpected scenarios with just one endemic EP%
\footnote{The second EP lies outside $\Tph$, so this is not a bifurcation point.} 
in $\partial\T(\bbe)$ obeying $S_2^*=0$ and $I^*=1-S_1^*>0$.

\smallskip
As a final message, in this Part I explicit formulas for the transition from SSISS to RND picture may take some hard getting used to. But, in turn the analysis of equilibrium points and their stability will substantially simplify in Part II. Also, the derivation of these formulas stays valid, if the functions 
$\Om_i$ are explicitly time-dependent, as long as $\be_i$ and 
$\ga=\ga_1+\ga_2$ stay constant. This suggests, that passing to RND picture might also be useful for models treating for example pulse vaccination or time-delay methods.

\bsn
{\bf Acknowledgments:} I would like to thank Florin Avram for encouraging interest and useful discussions.

\printbibliography

\appendix
\section{Birth and death rates\label{App_birth+death}}
This Appendix shortly reviews results from \autocite{Nill_Redundancy}, showing that constant per capita birth and death rates are redundant if one accepts possibly negative transition rates $\be_i$. To simplify formulas denote 
$S_0:=I$ and put $\SS_i:=NS_i$ the population in compartment $i$, where $N(t)$ is the time varying 
total population, $N=\sum_i\SS_i$. Let $\mu_i$ be the mortality rate in compartment $i$ and $\nu_{ik}$ the rate of newborns from compartment $k$ landing in compartment $i$. Here, it is essential to assume that newborns from susceptibles are not infected, i.e. $\nu_{01}=\nu_{02}=0$. Also put 
$\nu_k=\sum_i\nu_{ik}$ and  $\delta_k=\nu_k-\mu_k$. Then the SSISS dynamics with birth and death rates is given by
\begin{equation}
\begin{aligned}
\dot{\SS}_i&=(-\be_iS_i+\ga_i)\SS_0-\Om_i\SS_i+\Om_j\SS_j-\mu_i\SS_i+\sum_{k=0}^2\nu_{ik}\SS_k,
\quad i,j\in\{1,2\},\ j\neq i,
\\
\dot{\SS}_0&=
\left(\be_1S_1+\be_2S_2+\nu_{00}-\mu_0-\ga_1-\ga_2\right)\SS_0
\\
\dot{N}/N&=\sum_{k=0}^2\delta_k S_k
=\delta_i(1-S_j-S_0)+\delta_jS_j+\delta_0S_0
\\
&=(\nu_i-\mu_i)+(\delta_j-\delta_i)S_j+(\delta_0-\delta_i)S_0,\qquad i,j\in\{1,2\},\ j\neq i.
\end{aligned}
\label{SSISS+demogr}
\end{equation}
Here, by assumption of homogeneity in the variables 
$(\SS_0,\SS_1,\SS_2)$, 
$\Om_i$ may be considered as a function of the fractional sizes $S_1$ and $S_2$ only. Now, for $i,j\in\{1,2\}$ and $j\neq i$ use 
$\dot{S}_i=\dot{\SS}_i/N-S_i\dot{N}/N$ and 
$\nu_i=\nu_{ii}+\nu_{ji}$,
to obtain
$$
\begin{aligned}
\dot{S_i}&=(-\tbe_iS_i+\tga_i)S_0-\tOm_iS_i+\tOm_jS_j,
&& i,j\in\{1,2\},\ j\neq i,
\\
\tOm_i	&=\Om_i+\nu_{ji}+\delta_jS_j,
&& i,j\in\{1,2\},\ j\neq i,
\\
\tbe_i	&=\be_i+\delta_0-\delta_i
\\
\tga_i	&=\ga_i+\nu_{i0}
\end{aligned}
$$
Apparently, this coincides with \Eqref{tilde_parameters}. 

Observe that the above derivation also holds, if birth and/or death rates depend on $N$. In this case one typically assumes 
$d\mu_k/dN$ to be independent of the compartment $k$, so this dependency drops out in tilde parameters. In case of $N$-dependent birth rates this argument only holds, if we also assume off-diagonal elements of the birth matrix to be independent of $N$.

\section{The $\Gs$-action\label{App_Gs.X^2}}

This Appendix provides an explicit formula for the action of 
$\bg\in\Gs$ on the sub-class of $X^2$-models in SSISS picture,
$\Fbetwo=(\bDb+\Ebetwo)\times\bsi\inv(1)$, see \Eqref{Fbetwo}.

Let us first gather some more structural details on $\Gs$. As one easily verifies using Lemma \ref{Lem_BGs},
$\bg(\bbe',\bbe)=\bg(\tilde{\bbe}',\tilde{\bbe})$ for
$\{\bbe',\bbe,\tilde{\bbe}',\tilde{\bbe}\}\subset\B$,
if and only if there exists $(a,b)\in\RR_+\times\RR$ such that $\tilde{\bbe}=a\bbe+b\bsi$ and $\tilde{\bbe}'=a\bbe'+b\bsi$. Next, $\Gs$ leaves $\ker\bsi$ invariant, hence 
$\bg\bsi^\perp=\chi_\bg\bsi^\perp$, where
$\chi:\Gs\ni\bg\mapsto\chi_\bg$
defines a linear character of $\Gs$. 
Pick $\bbe\in\B$ such that $\bra\bbe|\bsi^\perp\ket=1$. Then any other $\bbe'\in\B$ is of the form
$\bbe'=a\bbe+b\bsi$, $a\neq0$. Hence,
$\chi(\bbe,\bbe')\equiv\chi_{\bg(\bbe,\bbe')}=
\bra\bbe|\bg(\bbe,\bbe')\bsi^\perp\ket=a$.
Thus, viewing $\RR$ as an additive group and putting
$j_{\bbe}:\RR\ni b\mapsto\bg(\bbe,\bbe+b\bsi)\in\Gs$, then $j_{\bbe}$ provides a group monomorphism satisfying 
$j_{\bbe}(\RR)=\ker\chi$.
Also,
$s_{\bbe}:\RR_+\ni a\mapsto\bg(\bbe,a\bbe)\in\Gs$ provides a  monomorphism satisfying 
$\chi\circ s_{\bbe}=\id_{\RR_+}$. In summary, we have a split short exact sequence
$$
\RR\xlongrightarrow{\ j_{\bbe}\ }\Gs\xtofrom[s_{\bbe}]{\chi}\RR_+
$$
and therefore $\Gs$ is isomorphic to the semidirect product
$
\Gs\cong\RR\cros\RR_+.
$
Also, using Eqs. \eqref{g_matrix} and \eqref{dual-coordinates}, one easily verifies
\begin{equation}
\chi(\bbe,\bbe')\equiv\chi_{\bg(\bbe,\bbe')}=
\frac{\be_1'-\be_2'}{\be_1-\be_2}
\label{chi}
\end{equation}

Now let's see how $\bg\equiv\bg(\bbe',\bbe)\in\Gs$ acts on 
$\Fbetwo$. Clearly, 
$\bg_*\Fbetwo=\F_{\bbe'}^{(2)}$ and
by \eqref{action}
$g_*\bF_{(\bM,\bga)}=\bF_{(\bg\re\bM,\bg\bga)}$,
where $\bg\re\bM:=\Ad\bg\circ\bM\circ\bg\inv$. 
Hence, we have to analyze, how this action operates on $\bM=\bDb+\bE\circ\bOm$, where $\bOm\in\C^\infty(\Ps,\S^*)$ is of the form
$$
\begin{aligned}
\bOm(\bS)&=\bA(I)+B(I)\bLabe(\bS),
\qquad I\equiv 1-\bra\bsi|\bS\ket,
\\
\bA(I)&=(A_1(I),A_2(I)),
\\
\bLabe(\bS)&=(\be_1S_2,\be_2S_1),
\end{aligned}
$$
see Eqs. \eqref{Ai+Bi} and \eqref{dot_Si_X^2}.
First, for 
$\bom=(\om_1,\om_2)\in\S^*$ use
\begin{equation}
\bE(\bom)=
\begin{pmatrix}
-\om_1,&\om_2\\\om_1,&-\om_2
\end{pmatrix}
=
\begin{pmatrix}
1\\-1
\end{pmatrix}
(\om_1,\om_2)\bJ,
\qquad\bJ:=\begin{pmatrix}
-1,&0\\0,&1
\end{pmatrix},
\label{EJ}
\end{equation} 
Hence, $\bg\bE=\chi_\bg\bE$, where $\chi_\bg\in\RR_+$
is the linear character \eqref{chi} of $\Gs$. Putting 
\begin{equation}
\bar{\bg}:=\chi_\bg\bJ\bg\inv\bJ
\label{bar(g)}
\end{equation}
we conclude
$\Ad\bg\circ\bE=\bE\circ R_{\bar{\bg}}$, where 
$R_{\bar{\bg}}:S^*\to S^*$ denotes right multiplication by 
$\bar{\bg}$. So,
\begin{equation}
\bg\re(\bE\circ\bOm)=\bE\circ R_{\bar{\bg}}\circ\bOm\circ\bg\inv.
\label{gE(w)}
\end{equation}
Next, use $\bDb=\bNb-I\bE(\bDelbe)$ by \eqref{N-D},
$\bg\re\bNb=\bN_{\bbe'}$ and $\Gs$-invariance of $I$ to conclude
\begin{equation}
\bg\re\bDb=
\bD_{\bbe'}+I\,\bE(\bm{\Del}_{\bbe'}-\bDelbe\,\bar{\bg}).
\label{gbDb}
\end{equation}
Finally, denote $\Mtwo\subset\M_\B$ the foliation
$$
\begin{aligned}
\Mtwo&=\cup_{\bbe\in\B}(\bDb+\Ebetwo)
\cong\B\times\C^\infty(\RRN)^3.
\end{aligned}
$$

\begin{proposition}\label{Prop_action}
Using $(\bbe,\bA,B)\in\B\times\C^\infty(\RRN)^3$ as coordinates on $\Mtwo$, the action of $\bg\in\Gs$ on $\Mtwo$ is given by
$\bg\re(\bbe,\bA,B)=(\bbe',\bA',B)$, where 
$\bbe'=\bbe\bg\inv$ and 
$$
\bA':=\bA\bar{\bg}+(I+(1-I)B)(\bDel_{\bbe'}-\bDelbe\bar{\bg}),
\qquad\bar{\bg}:=\chi_\bg\bJ\bg\inv\bJ
$$
\end{proposition}

\begin{proof}
By \eqref{gE(w)}, \eqref{gbDb} and $I\circ\bg\inv=I$ we are left to show 
$$
R_{\bar{\bg}}\circ\bLabe\circ\g\inv=
\bLa_{\bbe'}+(1-I)(\bDel_{\bbe'}-\bDelbe\bar{\bg}).
$$
Using the definition of $\bDelbe$ in \eqref{N-D}, the formula for $\chi_\bg$ in \eqref{chi} and the identity $\bra\bbe|\bS\ket=(1-I)\be_i+(\be_j-\be_i)S_j$ for $i=1$ or $i=2$ and  $j\neq i$, one computes
$$
\begin{aligned}
\bLabe(\bS)&=\frac{\bra\bbe|\bS\ket}{\be_1-\be_2}\bbe\bJ-(1-I)\bDelbe
\\
(R_{\bar{\bg}}\circ\bLabe\circ\g\inv)(\bS)
&=\frac{\bra\bbe'|\bS\ket}{\be'_1-\be'_2}\bbe'\bJ
-(1-I)\bDelbe\,\bar{\bg}.
\end{aligned}
$$
\end{proof}

\begin{remark}
Let $\bOmbe(f)$ be given by Eqs. \eqref{Ai+Bi}-\eqref{B(I)}.
Then it takes some lengthy calculations to check that Proposition \ref{Prop_action} is consistent with
$$
g(\bbe',\bbe)\re(\bDb+\bE\circ\bOmbe(f))=
\bD_{\bbe'}+\bE\circ\bOm_{\bbe'}(f).
$$
\end{remark}

\section{Proof of of Lemma \ref{Lem_YZR}\label{App_eadmissibility}} 

Part i) of Lemma \ref{Lem_YZR} is obvious by comparing the polynomial coefficients of 
$p_Z-p_Y$ and $p_R$. 
To prove part ii) we first check that the list of scenarios in Table \ref{Tab_YZR} is complete. The conditions in ii) allow two basic case distinctions, $Y_-\leq Z_-\leq Y_+\leq Z_+$ (scenarios 1a,b, 2, 3a,b) or $Y_-\leq Z_-\leq Z_+<Y_+$ (scenarios 4a,b). We first show that the second case can only show up if $f_{20}>0$. Indeed, using $p_Z(Y_+)=p_R(Y_+)$, in this case we conclude
$$
0<(Y_+-Z_+)(Y_+-Z_-)=
\begin{cases}
f_{20}(Y_+-R_0)(Y_+-R_1), &f_{20}\neq 0,
\\
f_{10}(Y_+-R_0)\leq 0, 	&f_{20}=0.
\end{cases}
$$
Recalling $R_1\leq R_0$ for $f_{20}<0$ and $f_{10}\leq 0$ for 
$f_{20}=0$, we get a contradiction for $f_{20}\leq 0$. Hence columns 2 and 3 in Table \ref{Tab_YZR} characterize a complete set of scenarios. To prove that given column 2 the conditions in columns 3 - 5 are equivalent we rely on graphical arguments. The key observation is that the identity 
$p_Z-p_Y=p_R$ implies the following intersection scenarios
$$
p_Y(R_{\nu})=p_Z(R_{\nu}),\qquad
p_Y(Z_\pm)=-p_R(Z_\pm),\qquad
p_Z(Y_\pm)=p_R(Y_\pm).
$$
Graphically, these are visualized in Table \ref{Tab_graphical}.
Analytically they lead to the following equations.
\begin{align}
(1-f_{20})(R_\nu-Y_+)(R_\nu-Y_-)&=
(R_\nu-Z_+)(R_\nu-Z_-),
\label{p3=0}
\\
(1-f_{20})(Z_\pm-Y_+)(Z_\pm-Y_-)&=
\begin{cases}
-f_{20}(Z_\pm-R_0)(Z_\pm-R_1), &f_{20}\neq 0,
\\
f_{10}(R_0-Z_\pm), 	&f_{20}=0\,\land\,f_{10}<0,
\end{cases}
\label{p1=0}
\\
(Y_\pm-Z_+)(Y_\pm-Z_-)&=
\begin{cases}
f_{20}(Y_\pm-R_0)(Y_\pm-R_1), &\quad f_{20}\neq 0,
\\
f_{10}(Y_\pm-R_0), 	&\quad f_{20}=0\,\land\,f_{10}<0.
\end{cases}
\label{p2=0}
\end{align}
To conclude the equivalence of inequalities in each row of Table \ref{Tab_YZR} from the above equations requires a lengthy and tedious chain of arguments. Hence a graphical proof by visiting Table \ref{Tab_graphical} shall do. Here the case 1a is omitted, since it is obvious.
\qed

\begin{longtblr}[
caption = {Proof of Lemma \ref{Lem_YZR}. The parabolas obey 
$p_Z-p_Y=p_R$, so in each row the three figures are equivalent.
The No. corresponds to the scenario in Table \ref{Tab_YZR}.},
entry		= {Intersecting parabolas},
label 		= {Tab_graphical},
note{$\dag$}	= {In case $f_{20}=0$ (1b) replace $f_{20}R_1$ by 
$-f_{10}$.},
note{$\dag\dag$}	= {In these cases the parabolas would not necessarily have to intersect given the ordering of their zeros alone.},
]{
colspec = {|Q[l,b]|Q[c,m]|Q[c,m]|Q[c,m]|}, 
rowhead = 2, 
}
%
\hline
\SetCell[c=4]{c}
In each column two pairs of zeroes$^\dag$ are ordered as in Table \ref{Tab_YZR}.
\\
\hline
No. & $(Y_\pm,Z_\pm)$ & 
$(Y_\pm,R_\nu)^\dag$ & $(Z_\pm,R_\nu)^\dag$ 
\\
\hline
\hline
1b											&
{\\ \includegraphics[scale=.5]{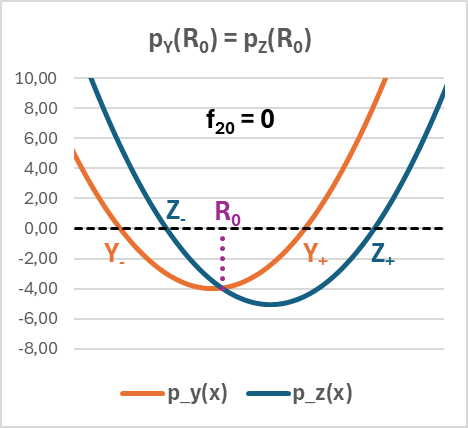}}	&
{\\ \includegraphics[scale=.5]{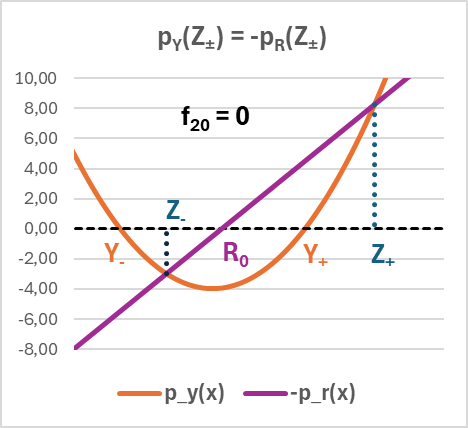}}	&
{\\ \includegraphics[scale=.5]{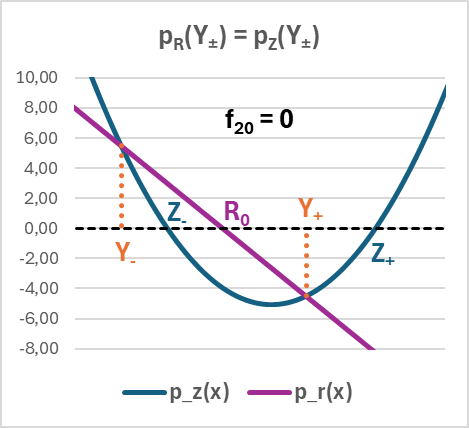}}	


\\
\hline
2 											&
{\\ \includegraphics[scale=.5]{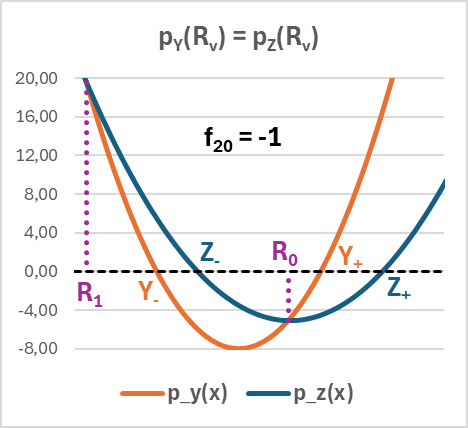}}		&
{\\ \includegraphics[scale=.5]{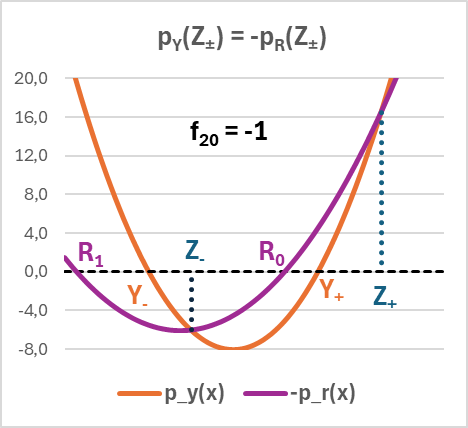}}		&
{\\ \includegraphics[scale=.5]{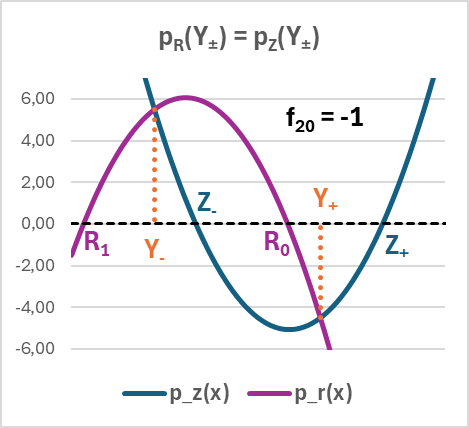}}							
\\
\hline
3a 											&
{\\ \includegraphics[scale=.5]{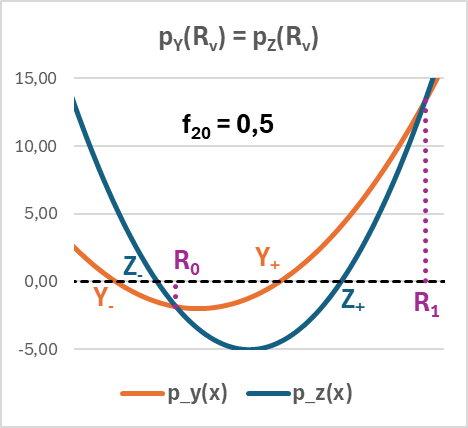}}	&
{\\ \includegraphics[scale=.5]{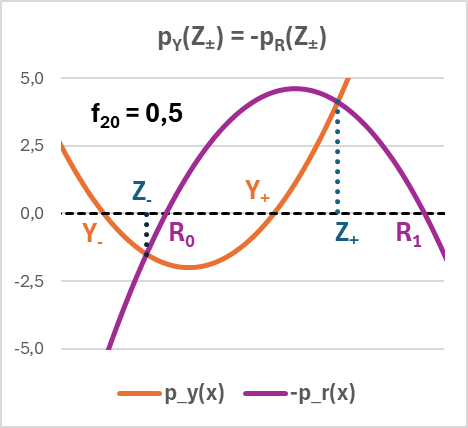}}	&
{\\ \includegraphics[scale=.5]{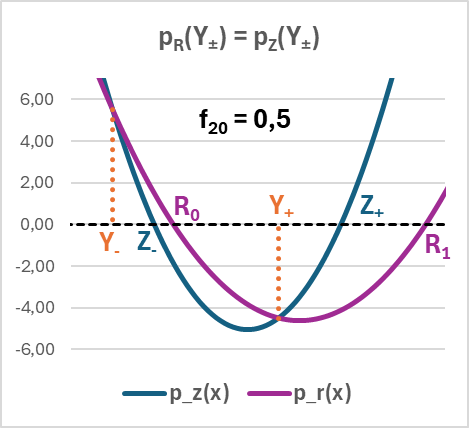}}			
\\
\hline
3b 											&
{\\ \includegraphics[scale=.5]{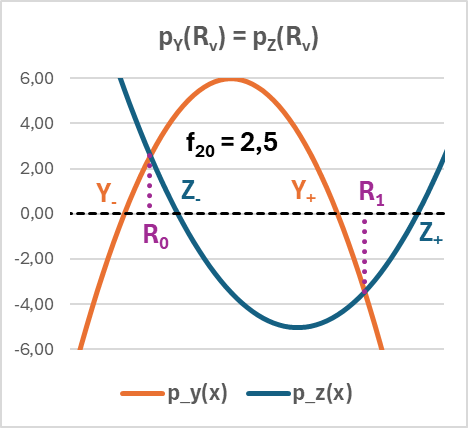}}	&
{\\ \includegraphics[scale=.5]{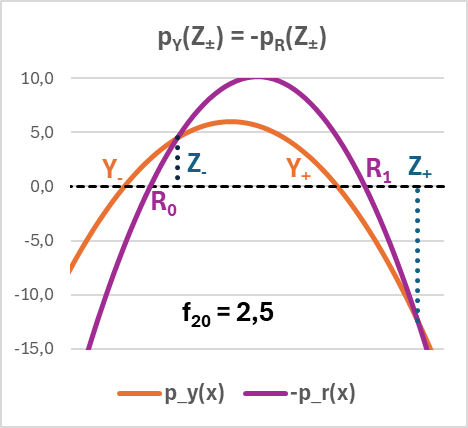}}	&
{\\ \includegraphics[scale=.5]{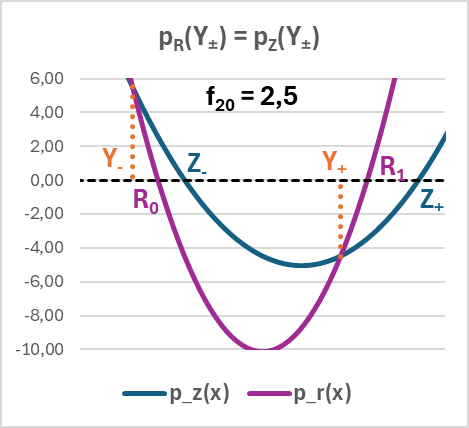}}	
\\
\hline	
4a											&
{\TblrNote{$\dag\dag$} \\ \includegraphics[scale=.5]{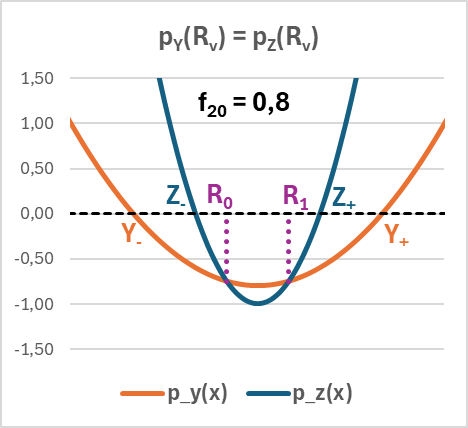}}	&
{\\ \includegraphics[scale=.5]{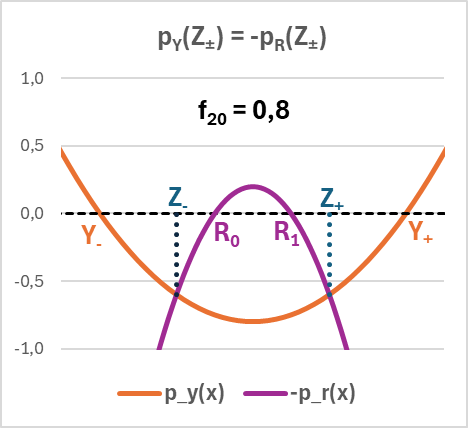}}	&
{\\ \includegraphics[scale=.5]{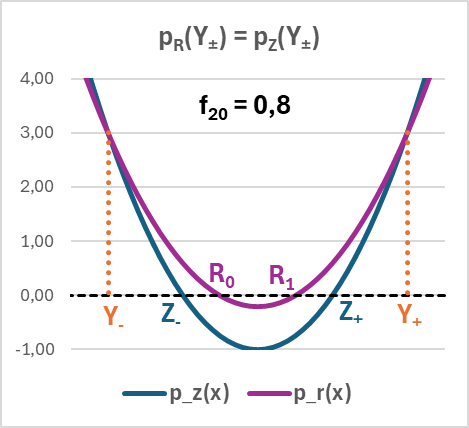}}						
\\
\hline
4b 											&
{\\ \includegraphics[scale=.5]{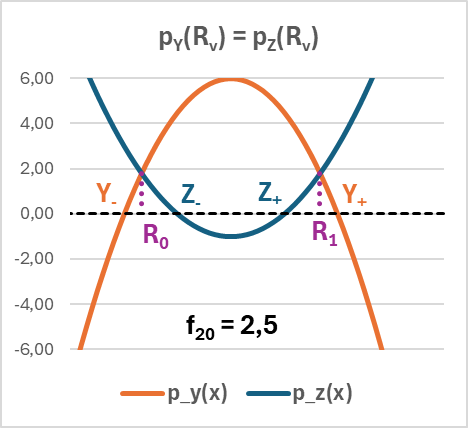}}	&
{\TblrNote{$\dag\dag$}\\ \includegraphics[scale=.5]{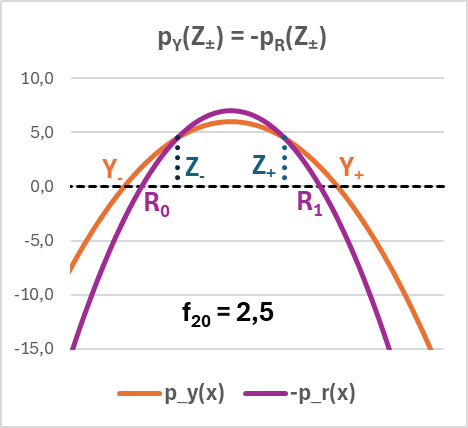}}	&
{\\ \includegraphics[scale=.5]{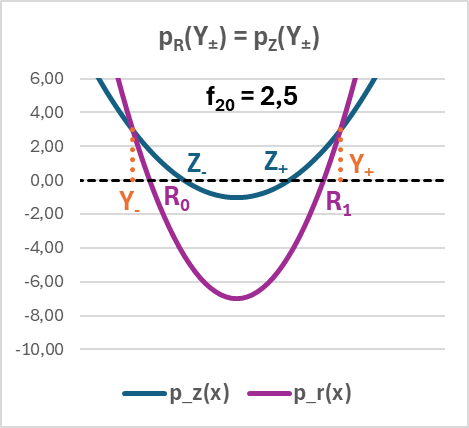}}	
\\
\hline
\end{longtblr}

\begin{remark}
Except for the cases $\dag\dag$ in Table \ref{Tab_graphical}, given the relative ordering of zeros and the sign of the quadratic coefficients of two parabolas, their intersection points  must exist and define the zeros of the third parabola. 
\end{remark}

\end{document}